\documentclass[journal, twoside]{IEEEtran}

\usepackage{amsmath, amssymb, amsfonts}
\usepackage{algorithmic}
\usepackage{enumitem}
\usepackage{graphicx}
\usepackage{textcomp}
\usepackage[dvipsnames, svgnames, x11names]{xcolor}
\usepackage{booktabs}
\usepackage{multirow}
\usepackage{longtable}
\usepackage{array}
\usepackage{url}
\usepackage{tikz}

\usepackage{forest}
\newcommand{\edgebadge}{\textbf{[E]}}
\newcommand{\cloudbadge}{\textbf{[C]}}

\usetikzlibrary{
    shapes,
    arrows.meta,
    positioning,
    trees,
    mindmap,
    calc,
    decorations.pathreplacing,
    shadows,
    backgrounds,
    fit}
\usepackage{acronym}
\usepackage{orcidlink}
\usepackage{balance}
\usepackage{cite}
\usepackage{hyperref}
\usepackage{caption}
\usepackage{tabularx}

\usepackage{wasysym}
\newcommand{\partialc}{\LEFTcircle}

\newcommand{\yes}{\textbf{\Large$\bullet$}}   
\newcommand{\no}{{\Large$\circ$}}             

\begin{document}
\bstctlcite{IEEEexample:BSTcontrol}

\title{AI-Native Closed-Loop Security for 6G-Enabled Cyber-Physical Systems: From Edge Detection to Network-Wide Mitigation}

\author{
    Bilal~Hussain,~\orcidlink{0000-0002-0046-9007}~\IEEEmembership{Member,~IEEE,}
    ~Muhammad~Bilal,~\orcidlink{0000-0003-4221-0877}~\IEEEmembership{Senior~Member,~IEEE,}
    ~Tan~Li,~\orcidlink{0000-0001-6129-4792}~\IEEEmembership{Member,~IEEE,}
    ~Haris~Pervaiz,~\orcidlink{0000-0002-8364-4682}~\IEEEmembership{Member,~IEEE,}
    ~Xiao~Tang,~\orcidlink{0000-0001-8971-5413}~\IEEEmembership{Member,~IEEE,}   ~Qinghe~Du,~\orcidlink{0000-0002-4992-9166}~\IEEEmembership{Member,~IEEE,}
    ~Fawad~Ahmad,~\orcidlink{0000-0001-8271-1073}~\IEEEmembership{Member,~IEEE,}
    ~Muhammad~Azhar,~\orcidlink{0000-0003-3687-0270}
    ~and Jun~Zhang,~\orcidlink{0000-0002-5222-1898}~\IEEEmembership{Fellow,~IEEE}
\thanks{Bilal Hussain and Fawad Ahmad are with the Division of Science, Engineering, and Health Studies, School of Professional Education and Executive Development, The Hong Kong Polytechnic University, Hong Kong SAR, China (e-mails: \{bilal.hussain, fawad.ahmad\}@cpce-polyu.edu.hk).}
\thanks{Muhammad Bilal is with the School of Computing and Communications, Lancaster University, United Kingdom (e-mail: m.bilal8@lancaster.ac.uk).}
\thanks{Tan Li is with Department of Computer Science, The Hang Seng University of Hong Kong, Hong Kong SAR, China (e-mail: tanli@hsu.edu.hk).}
\thanks{Haris Pervaiz is with the School of Computer Science and Electronic Engineering, University of Essex, CO4 3SQ, United Kingdom (e-mail: haris.pervaiz@essex.ac.uk).}
\thanks{Xiao Tang and Qinghe Du are with the School of Information and Communication Engineering, Xi'an Jiaotong University, Xi'an 710049, China (e-mails: tangxiao@xjtu.edu.cn, duqinghe@mail.xjtu.edu.cn).}
\thanks{Muhammad Azhar is with the Department of Applied Data Science, Hong Kong Shue Yan University, Hong Kong SAR, China (e-mail: azhar@hksyu.edu).}
\thanks{Jun Zhang is with the Department of Electronic and Computer Engineering, Hong Kong University of Science and Technology, Hong Kong (e-mail: eejzhang@ust.hk).}
}

\IEEEoverridecommandlockouts
\maketitle
\begin{abstract}
In sixth-generation (6G) networks, billions of cyber-physical systems (CPSs)---autonomous vehicles, smart grids, industrial robots, and remote-surgical equipment---will run over ultra-reliable low-latency slices, collapsing the gap between a remote breach and physical harm to milliseconds, a budget perimeter firewalls and centralised security operations centres cannot meet.
This survey reframes 6G CPS security as a \emph{closed-loop, AI-native pipeline} that senses at the multi-access edge computing (MEC) tier, using minute-scale call-detail records (CDRs) for baseline learning and slow-rate campaigns and sub-millisecond Radio Access Network (RAN)/Open-RAN (O-RAN) telemetry for the latency-critical path. 
The pipeline decides locally with compressed deep models, mitigates network-wide via software-defined networking (SDN), network function virtualization (NFV), and O-RAN controllers, and retrains through federated learning (FL) and digital-twin (DT) replay.
We formalise a per-slice, tail-bounded latency contract on the sense, detect, and mitigate stages, enforced at a slice-dependent tail percentile ($p_{99}$ for safety-critical URLLC slices) as a conservative sum-of-stage bound that prior autonomic and network control-loop models leave undefined.
Organising 128 peer-reviewed studies (2017--2026) under a PRISMA~2020 protocol, we (i)~map the 6G/CPS threat surface to MITRE ATT\&CK and a CDR-observable feature space;
(ii)~unify edge anomaly detection and DDoS classification across twelve datasets and statistical, graph, and transformer models;
(iii)~synthesise SDN/NFV/O-RAN primitives into one closed-loop reference architecture;
(iv)~treat FL, large language models (LLMs), DT, post-quantum cryptography (PQC), zero-trust architecture (ZTA), and explainable AI as cross-cutting enablers, not parallel pillars; and
(v)~consolidate open problems into five directions spanning data, latency, trust, standardisation, and evaluation.
\end{abstract}


\begin{table*}[!t]
\centering
\caption{List of Abbreviations}
\label{tab:abbreviations}
\small
\scriptsize
\setlength{\tabcolsep}{3pt}
\begin{tabular}{@{}p{1.2cm}p{6.0cm}p{1.2cm}p{7.6cm}@{}}
\toprule
\textbf{Acronym} & \textbf{Definition} & \textbf{Acronym} & \textbf{Definition} \\
\midrule
3GPP  & Third Generation Partnership Project & LSTM   & Long Short-Term Memory \\
5G    & Fifth-Generation Mobile Network & MEC    & Multi-access Edge Computing \\
6G    & Sixth-Generation Mobile Network & ML     & Machine Learning \\
AI    & Artificial Intelligence & mMTC   & Massive Machine-Type Communication \\
AMF   & Access and Mobility Management Function & NAS    & Non-Access Stratum \\
aNB   & AI-native NodeB (6G RAN node) & NFV    & Network Function Virtualization \\
API   & Application Programming Interface & NWDAF  & Network Data Analytics Function \\
CDR   & Call Detail Record & O-RAN  & Open Radio Access Network \\
CNN   & Convolutional Neural Network & PQC & Post-Quantum Cryptography \\
CPS   & Cyber-Physical System & PRISMA & Preferred Reporting Items for Systematic Reviews and Meta-Analyses \\
DDoS  & Distributed Denial of Service & RIC & RAN Intelligent Controller \\
DL    & Deep Learning & RRC & Radio Resource Control \\
DT    & Digital Twin & SDN & Software-Defined Networking \\
FL    & Federated Learning & SOC & Security Operations Center \\
GAN   & Generative Adversarial Network & URLLC  & Ultra-Reliable Low-Latency Communication \\
GNN   & Graph Neural Network & V2X & Vehicle-to-Everything \\
GTP   & GPRS Tunneling Protocol & VAE & Variational Autoencoder \\
IDS   & Intrusion Detection System & XAI & Explainable Artificial Intelligence \\
IoT   & Internet of Things & xApp & Extended Application (O-RAN) \\
LLM   & Large Language Model & ZTA & Zero Trust Architecture \\
\bottomrule
\end{tabular}
\end{table*}

\begin{figure*}[!t]
  \centering
  \includegraphics[width=1\textwidth]{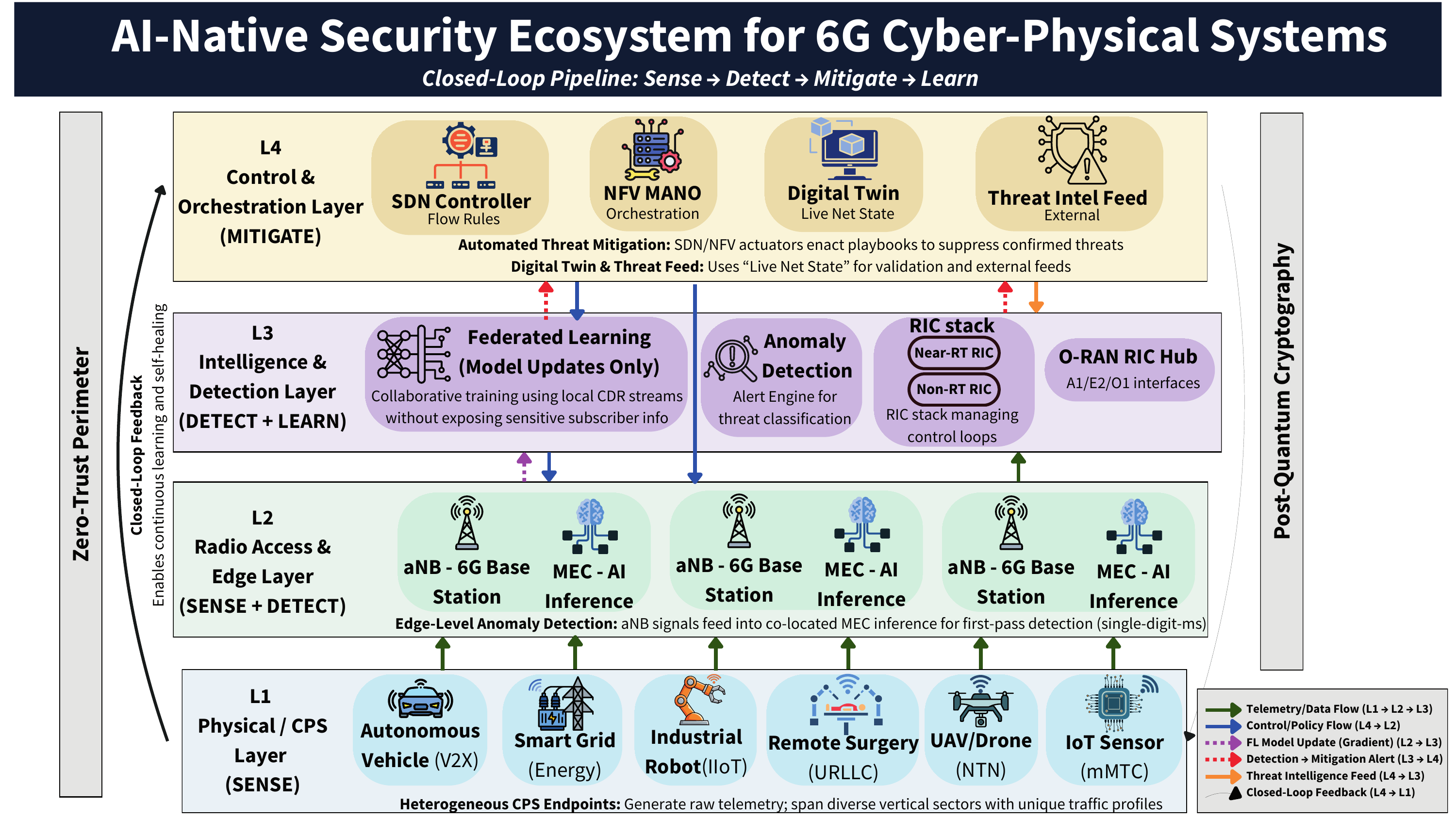}
  \captionsetup{skip=2pt, belowskip=-8pt}
\caption{End-to-end AI-native security ecosystem for 6G cyber-physical systems (CPSs), modelled as a closed control loop whose four stages---sense $\rightarrow$ detect $\rightarrow$ mitigate $\rightarrow$ learn---are realised across four functional planes (physical/CPS, radio access and edge, intelligence and detection, control and orchestration); the intelligence/detection plane hosts the per-MEC detectors and the operator-core FL aggregator that consolidates their gradient updates. 
Zero-trust access (left) and post-quantum cryptography (right) span all planes as cross-cutting substrates, while the coloured arrows denote the inter-plane flows---telemetry, control/policy, federated-learning gradient updates, the detection-to-mitigation alert, and the threat-intelligence feed that sharpens Layer-3 detection---as defined in the legend. 
The bold curved arrow closes the loop from \emph{mitigate} back to \emph{sense}, enabling continuous learning and self-healing. 
The per-plane mechanisms are detailed in Sec.~\ref{sec:intro_ai_native}.}
\label{fig:ecosystem}
\vspace{-5pt}
\end{figure*}

\section{Introduction}
\label{sec:intro}

\IEEEPARstart{S}{ixth-generation} (6G) wireless networks will not merely add bandwidth; they will fuse the digital and the physical until breach latency and physical-harm latency become the same number. 
Industry roadmaps target peak data rates above $1$\,Tb/s, end-to-end latencies below $1$\,ms, connection densities of $10^7$ devices/km$^2$, and ubiquitous on-device intelligence~\cite{wang2023survey_6g, singh2025towards6gevolution, saad2020vision6g, you2021towards_6g}.
These capabilities exist to serve Cyber-Physical Systems (CPSs): autonomous and aerial vehicles, smart grids, Industry~4.0 robotic cells, remote surgery, Unmanned Aerial Vehicle (UAV) swarms, and tactile telepresence~\cite{humayed2017cyber}. 
Across these domains, every packet is effectively a physical event:
an unmitigated attack can move a robot, mis-dose a patient, or bring
down a substation.

The shift from 5G to 6G is not incremental on the security axis. 
5G introduced service-based architecture, network slicing, and edge computing, but its security model still leaned on perimeter-style controls inherited from 4G Evolved Packet Core (EPC) and on a handful of standardised functions such as the Authentication Server Function (AUSF), Security Anchor Function (SEAF), and Access and Mobility Management Function (AMF), which authenticate users and protect the signalling plane~\cite{3gpp_sa3_security,ahmad2019network, porambage2021survey}.
6G dissolves this perimeter. 
The Radio Access Network (RAN) is disaggregated and software-defined~\cite{polese2023oran, bonati2021oran, oran_alliance_spec}; intelligence migrates from the core into the RAN and onto the device~\cite{li_2025_ainativeran}; and the same physical infrastructure simultaneously hosts Ultra-Reliable Low-Latency Communication (URLLC) vehicular slices, massive Machine-Type Communication (mMTC) industrial Internet-of-Things (IoT) slices, and enhanced Mobile Broadband (eMBB) extended-reality slices.
Each of these design choices opens a new attack surface and shortens the chain between an exploit and a physical-world consequence.

The consequence for CPS whose control loops traverse a 6G network---hereafter referred to by the shorthand ``6G CPS''---is best appreciated as a budget.
A Vehicle-to-Everything (V2X) collision-avoidance message has at most a few milliseconds to travel from a vehicle, through the radio, into a remote application server, and back to a peer vehicle that must brake; an autonomous robotic surgeon has perhaps ten milliseconds before a control-loop deviation becomes a physiological event; an industrial protective relay has a few cycles before a fault propagates and trips a substation~\cite{lee2008cyber, rajkumar2010cyber}. 
In all three cases, the attacker does not need to break confidentiality or alter payload---merely delaying or dropping the message is enough to materialise harm. 
Security in 6G CPS is therefore inseparable from latency, and the architectural question is where in the network the defender can decide and act fast enough.

\begin{figure*}[!t]
\centering
\resizebox{\linewidth}{!}{%
\begin{tikzpicture}[
  root/.style={rectangle, rounded corners=5pt,
               draw=blue!60!black, fill=blue!20, very thick,
               text width=4.6cm, minimum height=1.0cm,
               font=\bfseries\small, align=center},
  secA/.style={rectangle, rounded corners=3pt,
               draw=teal!70!black, fill=teal!14, thick,
               text width=3.1cm, minimum height=0.8cm,
               font=\footnotesize\bfseries, align=center},
  secB/.style={rectangle, rounded corners=3pt,
               draw=orange!80!black, fill=orange!14, thick,
               text width=3.1cm, minimum height=0.8cm,
               font=\footnotesize\bfseries, align=center},
  secC/.style={rectangle, rounded corners=3pt,
               draw=purple!60!black, fill=purple!12, thick,
               text width=3.1cm, minimum height=0.8cm,
               font=\footnotesize\bfseries, align=center},
  secD/.style={rectangle, rounded corners=3pt,
               draw=red!60!black, fill=red!10, thick,
               text width=3.1cm, minimum height=0.8cm,
               font=\footnotesize\bfseries, align=center},
  secE/.style={rectangle, rounded corners=3pt,
               draw=violet!60!black, fill=violet!10, thick,
               text width=3.1cm, minimum height=0.8cm,
               font=\footnotesize\bfseries, align=center},
  secF/.style={rectangle, rounded corners=3pt,
               draw=gray!55!black, fill=gray!10, thick,
               text width=3.1cm, minimum height=0.8cm,
               font=\footnotesize\bfseries, align=center},
  sub/.style={font=\scriptsize, align=left,
              text width=2.9cm, inner sep=0pt, outer sep=0pt},
  conn/.style={draw=black!55, thick},
  carr/.style={draw=black!55, thick, -{Stealth[length=2mm]}},
]

\node[root] (root) at (0,0)
  {AI-Native Security for 6G CPS\\[2pt]
   \normalfont\footnotesize(This Survey)};

\draw[conn] (root.south) -- (0,-1.0);
\draw[conn] (-8.25,-1.0) -- (10.0,-1.0);

\node[secA] (s2) at (-8.25,-1.85) {Sec.~\ref{sec:background}\\Background};
\node[secA] (s3) at (-2.75,-1.85) {Sec.~\ref{sec:methodology}\\Methodology};
\node[secB] (s4) at ( 2.75,-1.85) {Sec.~\ref{sec:threats}\\Threat Landscape};
\node[secC] (s5) at ( 8.25,-1.85) {Sec.~\ref{sec:edge_detection}\\Edge Detection};

\draw[carr] (-8.25,-1.0) -- (s2.north);
\draw[carr] (-2.75,-1.0) -- (s3.north);
\draw[carr] ( 2.75,-1.0) -- (s4.north);
\draw[carr] ( 8.25,-1.0) -- (s5.north);

\node[sub, anchor=north] (s2t) at (-8.25,-2.75)
  {\textbullet~6G/CPS Architectures\\[1pt]
   \textbullet~MEC \& O-RAN Substrate\\[1pt]
   \textbullet~CDR Feature Space};
\draw[conn] (s2.south) -- (s2t.north);

\node[sub, anchor=north] (s3t) at (-2.75,-2.75)
  {\textbullet~PRISMA 2020 Protocol\\[1pt]
   \textbullet~Comparison with Prior Surveys\\[1pt]
   \textbullet~Closed-Loop Taxonomy};
\draw[conn] (s3.south) -- (s3t.north);

\node[sub, anchor=north] (s4t) at ( 2.75,-2.75)
  {\textbullet~6G CPS Attack Surface\\[1pt]
   \textbullet~Cellular DDoS Vectors\\[1pt]
   \textbullet~MITRE ATT\&CK Alignment};
\draw[conn] (s4.south) -- (s4t.north);

\node[sub, anchor=north] (s5t) at ( 8.25,-2.75)
  {\textbullet~CDR-Driven DL Models\\[1pt]
   \textbullet~MEC Deployment \& Benchmarks\\[1pt]
   \textbullet~Sub-10\,ms Inference};
\draw[conn] (s5.south) -- (s5t.north);

\draw[conn] (10.0,-1.0) -- (10.0,-4.9);

\draw[conn] (-8.25,-4.9) -- (10.0,-4.9);

\node[secD] (s6) at ( 8.25,-5.75) {Sec.~\ref{sec:mitigation}\\Network Mitigation};
\node[secE] (s7) at ( 2.75,-5.75) {Sec.~\ref{sec:enablers}\\Cross-Cutting Enablers};
\node[secF] (s8) at (-2.75,-5.75) {Sec.~\ref{sec:open_challenges}\\Open Challenges};
\node[secF] (s9) at (-8.25,-5.75) {Sec.~\ref{sec:conclusion}\\Conclusion};

\draw[carr] ( 8.25,-4.9) -- (s6.north);
\draw[carr] ( 2.75,-4.9) -- (s7.north);
\draw[carr] (-2.75,-4.9) -- (s8.north);
\draw[carr] (-8.25,-4.9) -- (s9.north);

\node[sub, anchor=north] (s6t) at ( 8.25,-6.65)
  {\textbullet~SDN/NFV Orchestration\\[1pt]
   \textbullet~O-RAN Closed-Loop Control\\[1pt]
   \textbullet~Cross-Tier Reference Architecture};
\draw[conn] (s6.south) -- (s6t.north);

\node[sub, anchor=north] (s7t) at ( 2.75,-6.65)
  {\textbullet~FL, LLM, Digital Twin\\[1pt]
   \textbullet~Zero-Trust \& PQC\\[1pt]
   \textbullet~Explainable AI};
\draw[conn] (s7.south) -- (s7t.north);

\node[sub, anchor=north] (s8t) at (-2.75,-6.65)
  {\textbullet~Data \& Benchmark Gaps\\[1pt]
   \textbullet~Standardisation \& Governance\\[1pt]
   \textbullet~CPS-Impact Metrics};
\draw[conn] (s8.south) -- (s8t.north);

\node[sub, anchor=north] (s9t) at (-8.25,-6.65)
  {\textbullet~Key Findings Summary\\[1pt]
   \textbullet~Call to Action};
\draw[conn] (s9.south) -- (s9t.north);

\node[font=\scriptsize\itshape, teal!70!black]
  at (-5.50,-0.78) {Foundation};
\node[font=\scriptsize\itshape, orange!80!black]
  at ( 2.75,-0.78) {Threat};
\node[font=\scriptsize\itshape, purple!60!black]
  at ( 8.25,-0.78) {Detection};
\node[font=\scriptsize\itshape, red!60!black]
  at ( 8.25,-4.68) {Mitigation};
\node[font=\scriptsize\itshape, violet!60!black]
  at ( 2.75,-4.68) {Enablers};
\node[font=\scriptsize\itshape, gray!65!black]
  at (-5.50,-4.68) {Synthesis};

\end{tikzpicture}%
}
\captionsetup{skip=2pt, belowskip=-8pt}
\caption{Survey roadmap. The paper is structured around eight sections (Secs~\ref{sec:background}--\ref{sec:conclusion}) that collectively cover the end-to-end AI-native security pipeline for 6G cyber-physical systems, from foundational background and systematic methodology through edge detection and network-wide mitigation to cross-cutting enablers, consolidated open challenges, and concluding synthesis.}
\label{fig:roadmap}
\vspace{-5pt}
\end{figure*}
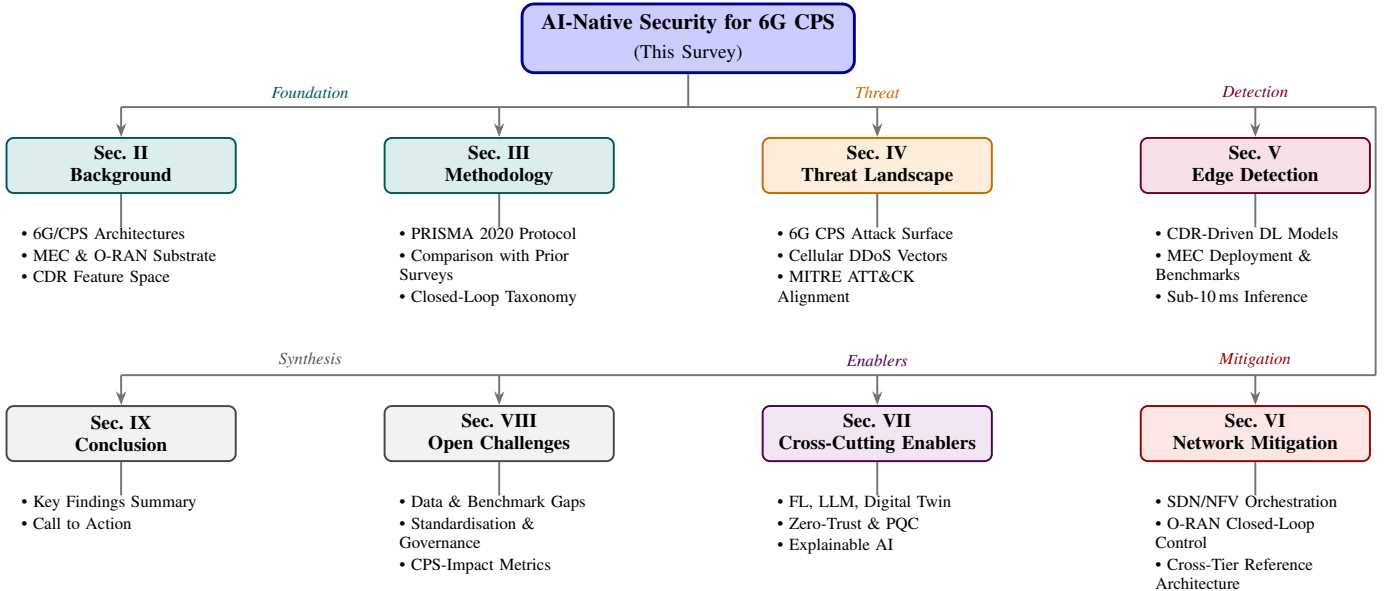

\vspace{-12pt}
\subsection{Why Legacy Security Cannot Meet the 6G CPS Budget?}
\label{sec:intro_legacy}
Four structural gaps make perimeter-based defences and the Security Operations Center (SOC) model unsuitable for 6G CPS: a volume gap, a latency gap, a telemetry gap, and a trust-boundary gap~\cite{ahmad2019network,porambage2021survey, nguyen2022survey}.
First, the \emph{volume gap}: a single 6G artificial-intelligence (AI)-native Node B (AI-native Node B, or aNB) will emit gigabit-rate signaling, mobility, and slice telemetry that no human-supervised SOC can triage in real time~\cite{nguyen2022survey, porambage2021survey}.
Second, the \emph{latency gap}: backhauling traffic to a cloud detector costs $80$--$120$\,ms~\cite{deng2020edge_intelligence,wang2020convergence_edge_ai}, but URLLC slices for V2X or telesurgery have $1$--$10$\,ms end-to-end budgets, so by the time a cloud rule fires, the unsafe action has already happened. 

Third, the \emph{telemetry gap}: legacy intrusion detection systems (IDS) were tuned to packet headers, not to the long-range, multi-modal cellular signals that reveal attacks on cellular infrastructure.
Such signals include Non-Access Stratum (NAS) attach storms, anomalous Radio Resource Control (RRC) setup ratios, GPRS Tunnelling Protocol User-plane (GTP-U) utilisation, and bursts of calls and Short Message Service (SMS) traffic visible in Call Detail Records (CDRs) ~\cite{ahmad2019network, ferrag2020deep_learning, hussain2021ddos}.
Rule-based IDSs over-fit to known signatures and generalise poorly to the polymorphic, slice-aware Distributed Denial of Service (DDoS) variants now reported in the wild~\cite{hoque2025ddos5g}.

A fourth, less-discussed mismatch is the \emph{trust-boundary gap}. 
The 5G perimeter assumed a hard inside/outside split between operator core and external internet; 6G dissolves that split because (i)~multi-access edge computing (MEC) hosts execute third-party applications adjacent to the user-plane data path~\cite{etsi_mec,  etsi2026_mec062, sabella2019mec, abbas2018mec}, (ii)~Open Radio Access Network (O-RAN) extended applications (xApps) and RAN applications (rApps) are sourced from heterogeneous vendors and dynamically loaded into the RAN Intelligent Controllers (RICs)~\cite{polese2023oran, oran_alliance_spec, bonati2020oran_survey, abdalla2022oran_security, wen2024oran}, and (iii)~network slices are rented to verticals, and the operator cannot audit the security posture of those tenants~\cite{alwis2026, shi2024ims}. 
The implication is that there is no longer a meaningful ``inside'' to defend; every interface must be authenticated and every model decision must be interpretable, which is the operational case for zero-trust architecture (ZTA)~\cite{nahar_2024_ztasurvey6g, agarwal2026} and explainable AI (XAI)~\cite{salmi2026xai} as native, not optional, components of the security stack.

These four gaps compound. 
A volumetric control-plane signalling storm originating from a slice-tenant's compromised IoT fleet~\cite{zhang2023signalingstorm, nguyen2025rrc} crosses the trust boundary at the MEC tier, demands a sub-millisecond response on a URLLC slice, and is invisible to an IDS that inspects only packet headers and does not parse Third Generation Partnership Project (3GPP) signalling.

The closed-loop framing returns to these effects later as explicit, CPS-aware mitigation metrics and gives the slice tenant a runtime control surface over them (Sec.~\ref{sec:mitigation_metrics}).
A complete list of abbreviations used throughout this survey is provided in Table~\ref{tab:abbreviations}.

\subsection{The AI-Native, Closed-Loop Alternative}
\label{sec:intro_ai_native}
The remedy emerging in the literature is to make security a \emph{native function of the network fabric itself}, executed close to the data source and continuously trained~\cite{hussain2019mec, hussain2020mec, polese2023oran, blika2024fl_6g, alwis2026}. We unify these threads under a single thesis:

\medskip
\noindent\fbox{\parbox{0.97\linewidth}{\textbf{Thesis.} In 6G CPS, security must be reframed as a \emph{closed-loop, AI-native pipeline}: \textbf{sense} on rich edge telemetry (CDR + RAN/O-RAN signals), \textbf{detect} locally at the MEC tier with latency-bounded, compressed deep models, \textbf{mitigate} network-wide through Software-Defined Networking (SDN)/Network Function Virtualization (NFV)/O-RAN actuators, and \textbf{learn} continuously via Federated Learning (FL) and Digital Twin (DT) replay---all under a zero-trust, post-quantum substrate.\\[4pt]
The novelty is not the four-stage decomposition---which is, in shape, shared with Monitor–Analyze–Plan–Execute over a shared Knowledge base (MAPE-K) and Observe–Orient–Decide–Act (OODA)---but the \emph{per-slice, tail-bounded design contract} that binds the loop to each slice's CPS safety budget, making slice-admissibility a falsifiable property rather than a qualitative aspiration (formalised in Sec.~\ref{sec:closed_loop}).}}
\medskip

The loop is treated here as an operational pipeline with measurable telemetry, latency, and actuation stages, each realised as a concrete tier rather than an abstraction.
Fig.~\ref{fig:ecosystem} renders each verb as a concrete operational tier with measurable latency, identifiable telemetry, and surveyed empirical evidence: \emph{sense} on the MEC-resident CDR/RAN/Network Data Analytics Function (NWDAF) telemetry streams~\cite{barlacchi2015multi, christopoulou_2024_ddos5gnwdaf}; 
\emph{detect} via lightweight deep models running at single-digit-millisecond latency on commodity MEC hardware~\cite{doriguzzi2020lucid, franco2026, sabella2019mec}; 
\emph{mitigate} through SDN/NFV/O-RAN actuators composed by an AI-native orchestrator~\cite{moore2025, allaw2025cross_layer, polese2025airanarch, etsi2026nfv}; and 
\emph{learn} through FL aggregation and DT replay that closes the loop back to sense~\cite{mcmahan2017fedavg, blika2024fl_6g, alwis2026}, while external threat-intelligence feeds are distributed to the Layer-3 detection models to sharpen anomaly classification against known indicators of compromise.
Two cross-cutting substrates bracket the loop: a zero-trust perimeter that authenticates every sense--detect--mitigate transaction, and post-quantum cryptography that protects the telemetry, gradient, and policy flows against harvest-now-decrypt-later adversaries, motivating the integration of post-quantum cryptography (PQC) with FL aggregation~\cite{alwis2026, etsi2026nfv}.
The contribution of this survey is not any single verb but their explicit composition into a continuously-running pipeline whose end-to-end metrics are the binding constraints---and whose composition surfaces the five open challenges catalogued in Sec.~\ref{sec:open_challenges}.
The loop framing matters because the pillars share a latency budget, a telemetry substrate, and a control plane: coupling problems that per-pillar surveys structurally cannot surface (Sec.~\ref{sec:open_challenges}).

We scope the survey to \emph{6G-tethered CPS}---excluding generic
enterprise IT and pure-IoT settings---and frame our contributions around the
\emph{sense\,$\rightarrow$\,detect\,$\rightarrow$\,mitigate\,$\rightarrow$\,learn} closed loop introduced in Sec.~\ref{sec:intro_ai_native}, so that each contribution maps to one stage of a single pipeline rather than standing alone:
\begin{itemize}[leftmargin=*]
    \item \textbf{One closed-loop reference architecture} (Fig.~\ref{fig:ecosystem}) that binds the four stages below into a single, latency-contracted pipeline and resolves the long-standing split between the ``edge-detection'' and ``network-mitigation'' literatures.
    
    \item \textbf{Sense---CDR as a first-class security signal.} We ground a CDR-driven sensing layer in 3GPP-observable features and 5G/6G NWDAF interfaces, with explicit MITRE ATT\&CK alignment (Sec.~\ref{sec:threats}, Table~\ref{tab:mitre}), establishing the telemetry substrate the loop senses on.
    
    \item \textbf{Detect---deployment-aware AI at the MEC.} We consolidate CDR-driven anomaly detection and DDoS classification under one taxonomy and one accuracy-and-latency table segregated by deployment tier (Sec.~\ref{sec:edge_detection}, Table~\ref{tab:anomaly_benchmark}), removing the duplication earlier surveys inherited from the SDN-IDS and cellular-anomaly strands.

    \item \textbf{Mitigate---composable network-wide actuation.} We treat SDN, NFV, O-RAN xApps, and AI orchestration as composable actuators of one control loop rather than competing schools (Sec.~\ref{sec:mitigation}, Table~\ref{tab:mitigation_comparison}).

    \item \textbf{Learn---enablers as services to the loop.} We position FL, large language models (LLMs), DT, PQC, ZTA, and XAI as continuous-learning and trust services that sustain the loop (Sec.~\ref{sec:enablers}), not as a parallel ``emerging paradigms'' silo.

    \item \textbf{A loop-level research roadmap} (Sec.~\ref{sec:open_challenges}) of five composition-level challenges---latency, auditability, federated governance, standardisation, and CPS-specific benchmarking---each carrying a falsifiable 2027/2028 target and a named standardisation vehicle.
\end{itemize}

\subsection{Survey Organisation}
\label{sec:intro_org}
Sec.~\ref{sec:background} establishes the 6G/CPS/MEC/CDR substrate that defines the sensing surface. 
Sec.~\ref{sec:methodology} presents our Preferred Reporting Items for Systematic Reviews and Meta-Analyses (PRISMA)-based \cite{page2021prisma} selection protocol and contrasts the survey with prior work.
Sec.~\ref{sec:threats} characterises the threats that the loop must defend against. 
Sec.~\ref{sec:edge_detection} surveys edge-based detection.
Sec.~\ref{sec:mitigation} surveys network-wide mitigation and instantiates the closed loop. 
Sec.~\ref{sec:enablers} covers the cross-cutting enablers. 
Sec.~\ref{sec:open_challenges} consolidates the five open challenges and Sec.~\ref{sec:conclusion} concludes. 
The roadmap of how sections compose into one narrative is in Fig.~\ref{fig:roadmap}: Secs.~\ref{sec:background}--\ref{sec:threats} frame the loop and the \emph{sense} stage (threat surface and CDR-observable feature space); Sec.~\ref{sec:edge_detection} treats the \emph{detect} stage; Sec.~\ref{sec:mitigation} the \emph{mitigate} stage; and Sec.~\ref{sec:enablers} the \emph{learn} stage together with the cross-cutting substrates (PQC, ZTA, XAI) that span all four stages; Secs.~\ref{sec:open_challenges}--\ref{sec:conclusion} close the survey.

\section{Background: 6G CPS, MEC, and CDR}
\label{sec:background}
This section fixes the substrate on which the rest of the survey runs. We summarise (i)~the 6G architectural pillars that define the new attack surface, (ii)~how CPSs are formally bound to 6G slices, (iii)~why MEC is the right enforcement plane, and (iv)~why CDRs are the most under-exploited security signal in cellular networks. 
AI/Machine Learning (ML) preliminaries are not repeated here; we assume readers are familiar with Convolutional Neural Networks (CNNs), Recurrent Neural Networks (RNNs)/Long Short-Term Memory networks (LSTMs), autoencoders, Graph Neural Networks (GNNs), and transformers~\cite{lecun2015deep, vaswani2017attention, wu2021gnn_survey} (see also \cite[Ch.~1]{sarker2024ai} for an applied AI/ML primer in a security context) and introduce architecture-specific notation in situ in Sec.~\ref{sec:edge_detection}.
Throughout this survey, $\mathbf{x}\in\mathbb{R}^d$ denotes a $d$-dimensional feature vector extracted from cellular telemetry; $f_\theta:\mathbb{R}^d\!\to\!\mathbb{R}^K$ a $K$-class classifier with parameters $\theta$; $g_\theta$ an autoencoder; $s(\mathbf{x})\in\mathbb{R}_{\geq 0}$ an anomaly score; $\theta^{(t)}_k$ the model parameters of MEC client $k$ at FL round $t$; $\Delta\theta^{(t)}_k = \theta^{(t)}_k - \theta^{(t-1)}_{\text{global}}$ the local update; $\Delta t_{\text{infer}}$ per-decision inference latency; $\Delta t_{\text{mit}}$ time from alert to enacted mitigation; and $\tau_{\max}$ the slice's safety budget.

\subsection{6G Architectural Pillars and Their Security Implications}
\label{sec:6g_pillars}
Six pillars distinguish 6G from 5G and each opens a new attack surface. 
\emph{AI-native RAN} replaces hand-engineered protocol stacks with learned ones~\cite{li_2025_ainativeran, bonati2020oran_survey, huang2025_ailcm_ran}, exposing the radio scheduler, beam manager, and Hybrid Automatic Repeat reQuest (HARQ) policy to adversarial ML; the AI-as-a-Service (AIaaS) exposure surface introduced by operator-side RAN AI service APIs~\cite{li_2025_ainativeran, polese2025airanarch} further widens the attack surface to include model-theft and capability-abuse vectors. 
\emph{Open RAN (O-RAN)} disaggregates the RAN into Open Radio Unit (O-RU), Open Distributed Unit (O-DU), and Open Centralised Unit (O-CU) components controlled by Near-Real-Time (Near-RT) and Non-Real-Time (Non-RT) RICs hosting xApps and rApps~\cite{polese2023oran, bonati2021oran, oran_alliance_spec,agarwal2026}; openness multiplies vendor diversity and supply-chain risk while creating rogue-xApp and Application Programming Interface (API)-exploitation vectors. 
\emph{Network slicing} carves logically isolated eMBB/URLLC/mMTC virtual networks over shared physical resources but leaves cross-slice interference and resource-starvation attacks~\cite{thantharate2020slice}. 

\emph{Non-Terrestrial Networks (NTNs)} extend coverage with Low Earth Orbit (LEO) satellites and High-Altitude Platform Stations (HAPS) but inherit limited onboard security and are open to eavesdropping~\cite{singh2025towards6gevolution}. 
\emph{Reconfigurable Intelligent Surfaces (RISs)} offer passive beamforming with new pilot-contamination and beam-hijacking risks~\cite{wu2021ris_survey, hassouna2026_riscs}. 
\emph{Integrated Sensing and Communication (ISAC)} reuses radio waveforms for environmental sensing, creating adversarial-sensing and surveillance hazards~\cite{etsi2026ISAC,singh2025towards6gevolution}. 
Table~\ref{tab:6g_pillars_security} summarises this map, linking each pillar's key capability to the primary security concern it introduces; the concrete exploitation path for each concern is developed in Sec.~\ref{sec:attack_surface}.

\begin{table}[!t]
\centering
\caption{6G Architectural Pillars and Associated Security Implications}
\label{tab:6g_pillars_security}
\renewcommand{\arraystretch}{1.15}
\scriptsize
\setlength{\tabcolsep}{3pt}
\begin{tabular}{|p{2.0cm}|p{2.6cm}|p{2.6cm}|}
\hline
\textbf{Pillar} & \textbf{Key Capability} & \textbf{Primary Security Concern} \\
\hline
AI-Native RAN   & Learned protocol stack, autonomous optimisation & Adversarial ML, model poisoning/inversion \\
\hline
O-RAN / RIC & Open interfaces, xApps/rApps, multi-vendor & Rogue xApps, API exploitation, supply-chain risk \\
\hline
Network Slicing & eMBB/URLLC/mMTC virtual networks & Cross-slice leakage, resource starvation \\
\hline
NTN (LEO/HAPS)  & Global ubiquitous coverage & Eavesdropping, limited onboard security \\
\hline
RIS & Passive beamforming, coverage extension & Beam hijacking, pilot contamination \\
\hline
ISAC & Joint sensing and communication & Adversarial sensing, unauthorised surveillance \\
\hline
\end{tabular}
\vspace{-12pt}
\end{table}

\subsection{Cyber-Physical Systems over 6G}
\label{sec:cps_def}
A CPS is a tight coupling of computation, networking, and physical processes~\cite{lee2008cyber, rajkumar2010cyber}; over 6G, that coupling is realised by binding each CPS workload class (defined by its latency, reliability, and density profile) to a network slice. 
URLLC slices serve V2X and Industry~4.0 motion control with $\leq 1$\,ms one-way latency and $10^{-9}$ packet error~\cite{wang2023survey_6g, saad2020vision6g, you2021towards_6g, nguyen2022survey}; eMBB slices serve Extended Reality (XR) and holographic telepresence; mMTC slices serve smart-metering and large-scale IoT. 
Security therefore inherits the slice's Key Performance Indicator (KPI) envelope: an IDS that needs $80$\,ms to decide is not an IDS for a URLLC slice.

Four CPS classes recur throughout this survey because they collectively span the 6G slice envelope, anchoring distinct corners of the latency--reliability--throughput--density design space~\cite{humayed2017cyber}, \cite[Ch. 8]{sarker2024ai}.
\emph{(1) Connected and autonomous mobility} (vehicular platoons, urban autonomy, UAV swarms) couples per-vehicle perception/control loops with V2X cooperative perception and is most sensitive to URLLC latency violations and to localisation tampering; a delayed actuator command becomes a collision.
\emph{(2) Smart grid and substation automation} couples wide-area measurement with protective relaying over sub-millisecond-sensitive International Electrotechnical Commission (IEC)~61850 Generic Object Oriented Substation Event (GOOSE) and Sampled Values (SV) traffic; late or dropped messages cause a real-world trip or out-of-step condition.
\emph{(3) Industrial automation and Industry~4.0 robotic cells} couples Programmable Logic Controllers (PLCs) and motion controllers with Time-Sensitive Networking (TSN)-style deterministic transport---the canonical converged Information Technology (IT) / Operational Technology (OT) case in which a network breach moves physical actuators (e.g., a faulty weld).
\emph{(4) Remote healthcare and telesurgery} couples haptic and visual feedback with operator-side actuation and demands not only latency bounds but also explainability of any defensive action that interrupts a clinical session, since a missed defibrillation has the same cost as a missed attack~\cite{lee2008cyber, rajkumar2010cyber, humayed2017cyber}.
The cross-class invariant is that a successful attack does not need to read or alter payload---it only needs to delay, drop, or deflect packets long enough to push a control loop outside its safety envelope~\cite{kim2022cps_resilient}.

This framing has two implications for security. First, confidentiality, integrity, and availability must be re-prioritised: for safety-critical slices, availability and timely integrity dominate confidentiality, and the cost of a false positive (blocking legitimate control traffic) can exceed the cost of a missed attack. 
Second, the closed-loop pipeline of Sec.~\ref{sec:intro_ai_native} must produce \emph{type-aware} mitigation: a generic ``alert$\rightarrow$rate-limit'' reflex that is appropriate for an mMTC SMS flood is catastrophic for a URLLC vehicular slice, where the same primitive can starve a safety message. 
Sec.~\ref{sec:mitigation} returns to this point.

\subsection{Multi-Access Edge Computing as the Enforcement Plane}
\label{sec:mec_arch}
The European Telecommunications Standards Institute (ETSI) MEC reference architecture~\cite{etsi_mec, sabella2019mec, bonati2020oran_survey, mao2017survey_mec} co-locates compute and storage with the base station, delivering single-digit millisecond detection latencies versus $80$--$120$\,ms cloud round trips~\cite{deng2020edge_intelligence, wang2020convergence_edge_ai}.
A typical MEC host offers $8$--$32$ CPU cores, $32$--$128$\,GB RAM, and an optional GPU---enough for moderate-complexity CNN/LSTM detectors, but not for uncompressed LLMs~\cite{zhou2019edge_intelligence}. 
Deploying a deep model at the MEC therefore requires aggressive compression~\cite{abbas2018mec, wang2020convergence_edge_ai}: 8-bit integer quantisation, structured pruning, knowledge distillation from a larger cloud teacher, and---where supported---hardware acceleration through Neural Processing Units (NPUs), Field-Programmable Gate Arrays (FPGAs), or vendor-specific inference chips. 
Even after compression, the host time-shares its compute among the security model, application workloads, and orchestration agents, so model design must account for tail-latency excursions and not just average throughput. 
Edge-intelligence surveys~\cite{deng2020edge_intelligence, wang2020convergence_edge_ai, zhou2019edge_intelligence} converge on a layered pattern that we recover throughout Sec.~\ref{sec:edge_detection}: a small ``cheap'' model triages every flow, a larger model re-evaluates a sampled or alerted subset, and a cloud or core model owns long-horizon model lifecycle, FL aggregation, and policy synthesis.

The compute envelope is only one half of the case for MEC; the other half is telemetry. 
The MEC tier is the only point in the
6G stack with simultaneous low-latency access to four security-relevant modalities, each over its native interface: 
(i) the operator's CDR pipeline streamed from the Charging Function (CHF) over the \emph{Nchf} reference point~\cite{cdr3gpp,abbas2018mec,sabella2019mec}; 
(ii) 3GPP analytics streams---the Network Data Analytics Function (NWDAF) in 5G and its 6G successor still under standardisation---exposed to consumers as event analytics over the Service-Based Architecture (SBA) \emph{Nnwdaf} service~\cite{christopoulou_2024_ddos5gnwdaf};
(iii) O-RAN Near-RT RIC \emph{E2~Key Performance Measurement (KPM)} measurement feeds~\cite{Zadeh2025data, azkaei2025AD, li_2025_ainativeran}; and 
(iv) MEC-local user-plane probes~\cite{mao2017survey_mec}.

This four-interface vantage point---hereafter referred to as the \emph{MEC telemetry vantage}---is the architectural reason detectors that fuse CDR with RAN telemetry can be operationally deployed at the MEC but rarely elsewhere~\cite{li_2025_ainativeran, polese2025airanarch}. 
The price of this proximity is local visibility: each host sees only its attached cells, so distributed attacks demand hierarchical aggregation (discussed in Sec.~\ref{sec:mec_detection_tier}, Sec.~\ref{sec:edge_synthesis}).

These capabilities are conditional on a property that is frequently under-stated: MEC is also the \emph{trust boundary} between the operator's administrative control and third-party application workloads. 
Hosts run operator-managed platform software alongside tenant workloads from Content-Delivery Networks (CDNs), enterprise customers, and vendor-supplied O-RAN functions, with separation enforced by container isolation, per-tenant namespaces, and role-based API access~\cite{abbas2018mec, sabella2019mec, polese2025airanarch}. 
Co-tenancy is two-edged: the operator gains fine-grained observability over every workload---the enabling condition for MEC-tier detection---but a compromise of the platform software exposes the security model's parameters, its training data, and cross-tenant traffic observations. 
Surveys of MEC security~\cite{abbas2018mec, porambage2021survey} agree that platform-integrity attestation (secure boot, measured-launch, runtime integrity monitoring) is a prerequisite for treating the MEC as the detection tier; without it, the detector runs on a substrate whose trustworthiness has not been established, and the closed loop inherits that uncertainty~\cite{etsi_mec, etsi2026nfv}.

\subsection{Call Detail Records as Security Telemetry}
\label{sec:cdr}
Call Detail Records (CDRs) are per-event records of calls, SMS messages, data sessions, and handovers, written in near-real-time by the 5G Charging Function (CHF) under the Converged Charging System (CCS)~\cite{cdr3gpp}. 
Operators retain them for billing, performance management, and lawful interception~\cite{cdr3gpp}, and typically expose them to analytics consumers either as the raw per-event stream or as per-cell aggregates over $5$--$15$\,minute reporting windows~\cite{barlacchi2015multi}. 
CDRs are not packet captures; they cannot reveal payload, but they reliably reveal volumetric, mobility, and call-pattern anomalies that are the dominant signature of cellular DDoS attacks~\cite{hussain2017, hussain2018semi, hussain2018globecom, sultan2018cdr, aziz2024kmeans}.
Crucially, CDRs are already collected, already privacy-pseudonymised, and already streamed to operator data lakes. While the minute-scale aggregation horizon makes them unsuitable as a sub-millisecond actuator signal, they are the operationally cheapest source of evidence for the volumetric and behavioural attack regimes (signalling storms, sustained DDoS events, slow-rate campaigns) that develop over seconds to minutes; the closed loop therefore composes CDR-class telemetry with sub-ms RAN/E2 KPM probes, each at its native timescale.

Four properties make CDRs particularly well-suited to closed-loop, MEC-tier security.
\begin{description}[leftmargin=*,nosep]
\item[Granularity.] Although individual records aggregate over minutes, per-cell streams expose the spatio-temporal structure that distinguishes a flash crowd from a coordinated signalling storm~\cite{hussain2017, hussain2018semi, hussain2018globecom, hussain2021ddos}.
\item[Mobility-awareness.] CDRs encode handover sequences and dwell-time distributions, essential for detecting impossible-trajectory or SIM-swap behaviour~\cite{sultan2018cdr}.
\item[Cross-generation continuity.] The CDR schema has evolved but not been replaced across 2G/3G/4G/5G, so detectors trained on legacy data generalise to current networks with modest re-engineering~\cite{barlacchi2015multi, aziz2024kmeans, aziz2024tmc}; their continuity into 6G is widely expected but not yet standardised.
\item[Privacy posture.] CDRs do not contain payload and operator pipelines pseudonymise subscriber identifiers, so secondary use for security is closer to the regulatory comfort zone than packet capture or full Deep Packet Inspection (DPI).
\end{description}

The limitations are equally important. 
CDRs miss short-lived, low-volume attacks that complete inside a single reporting window; they cannot resolve protocol-level fields needed to attribute a NAS-attach storm to a specific compromised UE class; and labelled attack data on real CDR is essentially unobtainable outside operator R\&D groups~\cite{duan2025, Stoian2026}.
The operational pattern that has emerged is therefore to use CDR telemetry for first-level detection and triage at the MEC, with packet/RAN telemetry pulled in selectively for confirmation~\cite{xylouris_2025_predictiveddos}---a pattern motivating Sec.~\ref{sec:edge_detection}'s benchmark comparing methods across both CDR and packet-based corpora.

The CDR feature space that recurs in the surveyed literature is conveniently small for ML deployment. 
Base features include per-cell call attempt count and success rate, per-cell SMS volume in inbound/outbound directions, internet session counts and aggregate bytes, handover counts in/out of the cell, and the inter-event distribution of these counts within the reporting window~\cite{barlacchi2015multi, hussain2017, hussain2018semi, sultan2018cdr, hussain2021ddos}.
Engineered features add ratios (success/attempt for both calls and attaches), rolling baselines (z-scores against the same hour-of-day median over a reference period), and spatial features (neighbour-cell correlation, hot-spot indicator, mobility-graph centrality). 
The result is a feature vector of $50$ to $200$ scalar features per cell per window---small enough to feed any of the model families in Sec.~\ref{sec:edge_method_families} and large enough to expose the cross-modality structure that makes CNN, GNN, and transformer architectures useful. 
These features are already computed inside the operator's billing pipeline, so the marginal compute cost of using them for security is dominated by the model's inference cost rather than by feature extraction.

The CDR feature surface also aligns naturally with the NWDAF event-exposure interface that 5G has standardised and that 6G will extend. 
The NWDAF Analytics Logical Function (AnLF) consumes event streams from each Network Function (NF) and exposes derived analytics; the CDR features above can be expressed as NWDAF analytics in straightforward ways, so a CDR-driven detector transfers cleanly to a 6G NWDAF-hosted detector once the appropriate event-exposure profile is standardised (Challenge~\ref{sec:c4}). 
Christopoulou \textit{et al.}~\cite{christopoulou_2024_ddos5gnwdaf} demonstrates exactly this transition for a DDoS detector running inside a 5G NWDAF; the CDR-trained detectors of~\cite{hussain2017, hussain2018semi, hussain2019mec, hussain2020mec, hussain2021ddos} are direct architectural antecedents.

\section{Survey Methodology and Comparison with Prior Surveys}
\label{sec:methodology}

\subsection{PRISMA-Based Literature Protocol}
\label{sec:prisma}
We followed the PRISMA~2020 systematic review protocol~\cite{page2021prisma}. 
The search strategy decomposed the keyword matrix \{6G, 5G, O-RAN, MEC, CDR, slicing\} $\times$ \{security, intrusion detection, anomaly detection, DDoS\} $\times$ \{machine learning, deep learning, federated learning, LLM, digital twin\} into five consolidated Boolean queries that match the five thematic strands of this survey: (Q1)~5G/6G/O-RAN security baseline; (Q2)~MEC and edge-intelligence security; (Q3)~CDR and cellular-telemetry-driven detection; (Q4)~federated learning for cellular security; (Q5)~LLM and digital-twin approaches to network security.
Each query was executed in April~2026 against IEEE Xplore, Scopus, Google Scholar, the ACM Digital Library, and arXiv, and restricted to English-language peer-reviewed venues, top-tier conferences, and arXiv preprints with $\geq$\,5 citations or active follow-up.

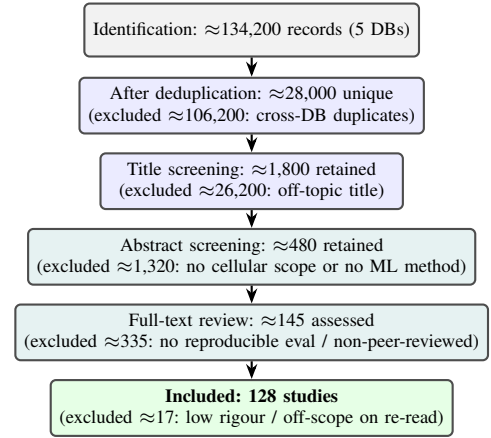
\begin{figure}[!t]
\centering
\begin{tikzpicture}[
    every node/.style={font=\scriptsize},
    box/.style={rectangle, draw=black!70, fill=white, rounded corners=2pt,
                minimum height=0.7cm, minimum width=4.0cm, align=center, thick},
    arr/.style={-{Stealth[length=2mm]}, thick},
]
\node[box, fill=gray!10]   (id)  at (0, 4.0) {Identification: $\approx$134{,}200 records (5 DBs)};
\node[box, fill=blue!8]    (sc)  at (0, 3.0) {After deduplication: $\approx$28{,}000 unique\\\scriptsize(excluded $\approx$106{,}200: cross-DB duplicates)};
\node[box, fill=blue!8]    (ts)  at (0, 2.0) {Title screening: $\approx$1{,}800 retained\\\scriptsize(excluded $\approx$26{,}200: off-topic title)};
\node[box, fill=teal!10]   (as)  at (0, 1.0) {Abstract screening: $\approx$480 retained\\\scriptsize(excluded $\approx$1{,}320: no cellular scope or no ML method)};
\node[box, fill=teal!10]   (ft)  at (0, 0.0) {Full-text review: $\approx$145 assessed\\\scriptsize(excluded $\approx$335: no reproducible eval / non-peer-reviewed)};
\node[box, fill=green!10]  (in)  at (0,-1.0) {\textbf{Included: 128 studies}\\\scriptsize(excluded $\approx$17: low rigour / off-scope on re-read)};
\draw[arr] (id) -- (sc); \draw[arr] (sc) -- (ts); \draw[arr] (ts) -- (as);
\draw[arr] (as) -- (ft); \draw[arr] (ft) -- (in);
\end{tikzpicture}
\caption{PRISMA~2020 flow of the systematic literature review, yielding the final corpus of 128 peer-reviewed studies, augmented by a small number of very recent studies via citation snowballing. 
A further 14 standardisation/grey-literature and three supplementary references are cited outside the protocol ($128{+}14{+}3=145$).}
\label{fig:prisma_flow}
\vspace{-10pt}
\end{figure}

Inclusion required (i)~explicit treatment of cellular network security or (ii)~ML/DL methodology directly applicable to CDR or RAN telemetry; exclusion eliminated pure cryptography, pure physical-layer security, and non-cellular IoT-only studies that did not generalise.
Across the five queries, IEEE Xplore returned 10{,}992 hits (Q1--Q5 combined), Scopus 12{,}473, Google Scholar 102{,}900, the ACM Digital Library 6{,}778, and arXiv 1{,}090. 
As illustrated by the complete PRISMA~2020 flow diagram in Fig.~\ref{fig:prisma_flow}, these initial hits were rigorously filtered down to the final corpus. The figure details the per-stage breakdown of identification, screening, and inclusion, demonstrating how our explicit focus on latency-bounded edge capability eliminated off-topic studies.
DOI- and title-based deduplication followed by sequential title, abstract, and full-text screening against the inclusion criteria yielded the final corpus of 128 peer-reviewed studies. 
The bibliography additionally cites 14 standardisation and grey-literature references (3GPP, ETSI, O-RAN, NIST, and MITRE) and three supplementary references, all of which sit outside the PRISMA protocol.

Quality appraisal followed standard systematic-review practice: each candidate was rated for venue quality, methodological rigour (clear problem statement, reproducible evaluation, baseline comparison), domain relevance, and recency. 
We additionally tracked whether each study reported on real operator data, on a 5G/O-RAN testbed, or on a public-only benchmark~\cite{hoque2025ddos5g}, because the gap between curated-benchmark accuracy and operational-CDR accuracy is one of the main quantitative findings of the survey (Sec.~\ref{sec:edge_benchmark}). 
Forward and backward citation tracking added a small number of very recent O-RAN security studies not yet indexed by all databases at query time, and inter-rater consistency on the abstract-screening stage was checked on a 10\% sample by a second author with disagreements resolved by discussion.

\subsection{Taxonomy and Comparison with Prior Surveys}
\label{sec:taxonomy}
We organise the corpus into the closed-loop taxonomy of Fig.~\ref{fig:roadmap}: \emph{threats} feed \emph{detection} (edge-based), which feeds \emph{mitigation} (network-wide), sustained by \emph{enablers} (FL, LLM, DT, PQC, ZTA, XAI). This loop-centred view, rather than the pillar list of prior cellular-security surveys, is the organising principle of this work.
Table~\ref{tab:survey_comparison} contrasts our work with the most recent and most cited related surveys across dimensions spanning 5G/6G scope, CPS integration, MEC deployment, CDR-driven detection, DDoS detection, network-wide mitigation, ML/DL and FL methodology, LLM coverage, and---uniquely---whether detection and mitigation are bound into one closed loop and whether downstream CPS impact is evaluated.
Competing surveys each populate only a subset of these dimensions: Hoque \textit{et al.}~\cite{hoque2025ddos5g} reach CDR-driven detection and Kumar \textit{et al.}~\cite{kumar2026} reach CPS, but always in isolation, and no other entry treats LLM-based methods. Ours is the only fully filled row; the contribution is therefore the joint coverage under a single closed-loop framing, not any individual dimension.
We structure the rest of this subsection around five axes on which prior surveys differ from ours: clustering and closed-loop framing, benchmarks, CPS-impact metrics, open challenges, and standardisation alignment.

\paragraph{Clustering and the closed-loop gap} The closest prior surveys split into two clusters. 
The detection-centric cluster~\cite{pang2021deep, thakkar2022survey_ddos, chandola2009anomaly} catalogues ML/DL detectors but says little about how alerts become actions and does not treat CDR as a first-class signal, while broader 5G-security surveys~\cite{ahmad2019network} chart the architecture-and-technology landscape without addressing detector-to-mitigator composition.
The mitigation/architecture-centric cluster~\cite{kumar2026, nguyen2022survey,porambage2021survey, nahar_2024_ztasurvey6g, hoque2025ddos5g,alwis2026, blika2024fl_6g,mothukuri2021survey_fl} treats mitigation primitives or FL/ZTA as enablers but stops short of binding them to a concrete edge-detection benchmark.
Within the ranked comparison of Table~\ref{tab:survey_comparison}, neither cluster's entries combine CDR-driven detection with O-RAN-driven mitigation under a single closed-loop framing, and neither poses its open problems at the level of cross-stage composition.

\paragraph{Benchmarks} The detection-centric cluster reports per-method accuracy on public benchmarks (CICDDoS2019~\cite{sharafaldin2019cicddos}, UNSW-NB15~\cite{moustafa2015unswnb15}, Bot-IoT~\cite{koroniotis2019towards_botiot}) without distinguishing the deployability of the underlying model on a MEC tier; high-accuracy methods that require centralised training on tens of millions of records and tens of seconds of inference are reported on equal footing with compressed quantised models that meet a sub-10\,ms MEC budget. 
The mitigation/architecture cluster reports primitives and reference architectures without binding them to a measured detector performance. 
Our consolidated comparison (Sec.~\ref{sec:edge_benchmark}) addresses both gaps by reporting accuracy and latency together on the same reference benchmarks, segregated by deployment tier (MEC vs.~core vs.~cloud), and Sec.~\ref{sec:reference_impl} sketches an integrated implementation that closes the loop end-to-end.

\paragraph{CPS-impact metrics} The detection-centric cluster evaluates detectors on accuracy, precision, recall, and F1 against labelled benchmark datasets; the mitigation/architecture cluster evaluates primitives on block-effectiveness, false-block rate, and recovery time. 
Neither cluster measures the downstream physical impact---the attack-induced delay to defibrillation in a tele-health slice, the missed V2X hazard warning in a vehicular slice, the unscheduled relay trip in a smart-grid slice---that is the ultimate object of concern for 6G CPS. 
Our treatment (Sec.~\ref{sec:mitigation_metrics} and Challenge~C5) elevates CPS-impact metrics to the same standing as detection accuracy and proposes that future benchmarks carry per-incident impact labels produced by operator-CPS-vendor partnerships or by digital-twin replay (Sec.~\ref{sec:dt}). 
Detection accuracy and mitigation block-effectiveness are necessary but not sufficient criteria for closed-loop viability; without CPS-impact metrics, the loop optimises the wrong objective.

\paragraph{Open challenges}
Prior surveys typically enumerate ten to twenty isolated sub-problems (drift detection, FL non-IID heterogeneity, xApp lifecycle, PQC migration, dataset scarcity) without indicating which of them compose into a working closed loop. 
We compress this enumeration to five challenges (Sec.~\ref{sec:open_challenges}) chosen for cross-stage leverage: each challenge sits at the intersection of detection, mitigation, and at least one enabler, so that progress on a single challenge advances the loop as a whole rather than improving one stage in isolation.
Five is sufficient rather than arbitrary because the criterion admits exactly one challenge per functional bottleneck the loop must clear (detection viability, trustworthiness, cross-operator training, standards-conformant deployment, and honest evaluation) so finer sub-problems fold into whichever of the five they serve; Sec.~\ref{sec:open_challenges} develops this closed-agenda argument in full.

\paragraph{Standardisation alignment}
Prior surveys treat 3GPP SA3, the O-RAN Security Working Group, and ETSI MEC specifications as background references~\cite{porambage2021survey,ahmad2019network}; our treatment binds each Challenge (Sec.~\ref{sec:open_challenges}) to the specific standardisation venue at which closure would have the largest operational effect. 
This is the methodological complement of the closed-loop framing and the precondition for turning the research agenda into deployable infrastructure rather than academic prototype.

\begin{table*}[!t]
\centering
\caption{Comparison of This Survey with Existing Related Surveys}
\label{tab:survey_comparison}
\renewcommand{\arraystretch}{1.1}
\setlength{\tabcolsep}{3pt}
\footnotesize
\begin{tabular}{|p{2.7cm}|c|c|c|c|c|c|c|c|c|c|c|c|}
\hline
\textbf{Reference} & \textbf{Year} & \textbf{5G/6G} & \textbf{CPS} & \textbf{MEC} & \textbf{CDR} & \textbf{DDoS Det.} & \textbf{Net. Mitig.} & \textbf{ML/DL} & \textbf{FL} & \textbf{LLM} & \textbf{Closed-Loop\textsuperscript{$\dagger$}} & \textbf{CPS-Impact\textsuperscript{$\ddagger$}} \\
\hline
Ahmad \textit{et al.}~\cite{ahmad2019network}          & 2019 & \yes & \no  & \yes  & \no  & \yes & \no  & \yes & \no  & \no  & \no  & \no  \\
\hline
Pang \textit{et al.}~\cite{pang2021deep}               & 2021 & \no  & \no  & \no  & \no  & \no  & \no  & \yes & \no  & \no  & \no  & \no  \\
\hline
Porambage \textit{et al.}~\cite{porambage2021survey}   & 2021 & \yes & \no  & \yes & \no  & \yes & \no  & \yes & \yes & \no  & \no  & \no  \\
\hline
Mothukuri \textit{et al.}~\cite{mothukuri2021survey_fl} & 2021 & \no  & \no  & \no  & \no  & \no  & \no  & \yes & \yes & \no  & \no  & \no  \\
\hline
Nguyen \textit{et al.}~\cite{nguyen2022survey}         & 2022 & \yes & \no  & \yes & \no  & \no  & \no  & \yes & \yes & \no  & \no  & \no  \\
\hline
Thakkar \& Lohiya~\cite{thakkar2022survey_ddos}         & 2022 & \no  & \no  & \no  & \no  & \yes & \no  & \yes & \no  & \no  & \no  & \no  \\
\hline
Blika \textit{et al.}~\cite{blika2024fl_6g}             & 2024 & \yes & \no  & \yes & \no  & \yes & \no  & \yes & \yes & \no  & \no  & \no  \\
\hline
Nahar \textit{et al.}~\cite{nahar_2024_ztasurvey6g}      & 2024 & \yes & \no  & \no  & \no  & \no  & \yes & \no  & \no  & \no  & \no  & \no  \\
\hline
Hoque \textit{et al.}~\cite{hoque2025ddos5g}           & 2025 & \yes & \no  & \yes & \yes & \yes & \yes & \yes & \no  & \no  & \no  & \no  \\
\hline
De Alwis \textit{et al.}~\cite{alwis2026}              & 2026 & \yes & \no  & \yes & \no  & \yes & \yes & \yes & \yes  & \no  & \no  & \no  \\
\hline
Agarwal \textit{et al.}~\cite{agarwal2026}             & 2026 & \yes & \no  & \yes & \no  & \no  & \yes & \no  & \no  & \no  & \no  & \no  \\
\hline
Kumar \textit{et al.}~\cite{kumar2026}                 & 2026 & \yes & \yes & \yes & \no  & \no  & \yes & \yes & \yes & \no  & \no  & \no  \\
\hline
\textbf{This Survey}                                   & 2026 & \yes & \yes & \yes & \yes & \yes & \yes & \yes & \yes & \yes & \yes & \yes \\
\hline
\end{tabular}
\vspace{2pt}
\begin{minipage}{0.98\textwidth}
\scriptsize
A filled circle ($\bullet$) is awarded where a survey provides defensible coverage of the dimension (dedicated discussion, taxonomy, comparative analysis, or substantive treatment); an open circle ($\circ$) indicates the dimension is absent or appears only in passing.
\textbf{5G/6G}: substantive 5G/6G security-architecture coverage.
\textbf{CPS}: cyber-physical or safety-critical control systems treated as the protected asset, beyond generic IoT.
\textbf{MEC}: edge/MEC analysed as an enforcement or detection tier.
\textbf{CDR}: call-detail-record or equivalent cellular-telemetry features used as a security signal.
\textbf{DDoS Det.}: DDoS/flooding detection methods.
\textbf{Net. Mitig.}: network-wide mitigation (SDN/NFV/O-RAN actuation), not detection alone.
\textbf{ML/DL}: ML/DL methodology surveyed comparatively.
\textbf{FL}: federated learning as a security-relevant training paradigm.
\textbf{LLM}: large-language-model methods as a distinct topic.
\textsuperscript{$\dagger$}\textbf{Closed-Loop (Det.$\rightarrow$Mitig.)}: filled only where the survey treats detection and network-wide mitigation as one bound feedback loop (alerts$\rightarrow$actuation$\rightarrow$retraining), not as separate chapters.
\textsuperscript{$\ddagger$}\textbf{CPS-Impact}: filled only where the survey evaluates downstream physical/safety impact (e.g., missed hazard warning, relay trip, control-loop deviation), beyond accuracy/F1 or block-rate metrics.
\end{minipage}
\vspace{-12pt}
\end{table*}

Having situated this work against twelve prior surveys, we now turn to the threat landscape itself: the attack classes that the closed-loop architecture must detect, contain, and recover from.

\section{Threat Landscape for 6G CPS}
\label{sec:threats}
This section answers \emph{what} the closed loop must defend against. 
We expand the 6G CPS attack surface (Sec.~\ref{sec:attack_surface}), enumerate the cellular DDoS vectors that dominate empirical reports (Sec.~\ref{sec:ddos_vectors}), formalise an attacker model (Sec.~\ref{sec:attacker_model}), survey the public dataset landscape that detectors are evaluated on (Sec.~\ref{sec:datasets}), and align the picture with MITRE ATT\&CK so that downstream detection and mitigation map cleanly onto a common adversarial vocabulary (Sec.~\ref{sec:mitre}).

\begin{figure}[!t]
\centering
\includegraphics[width=1\columnwidth]{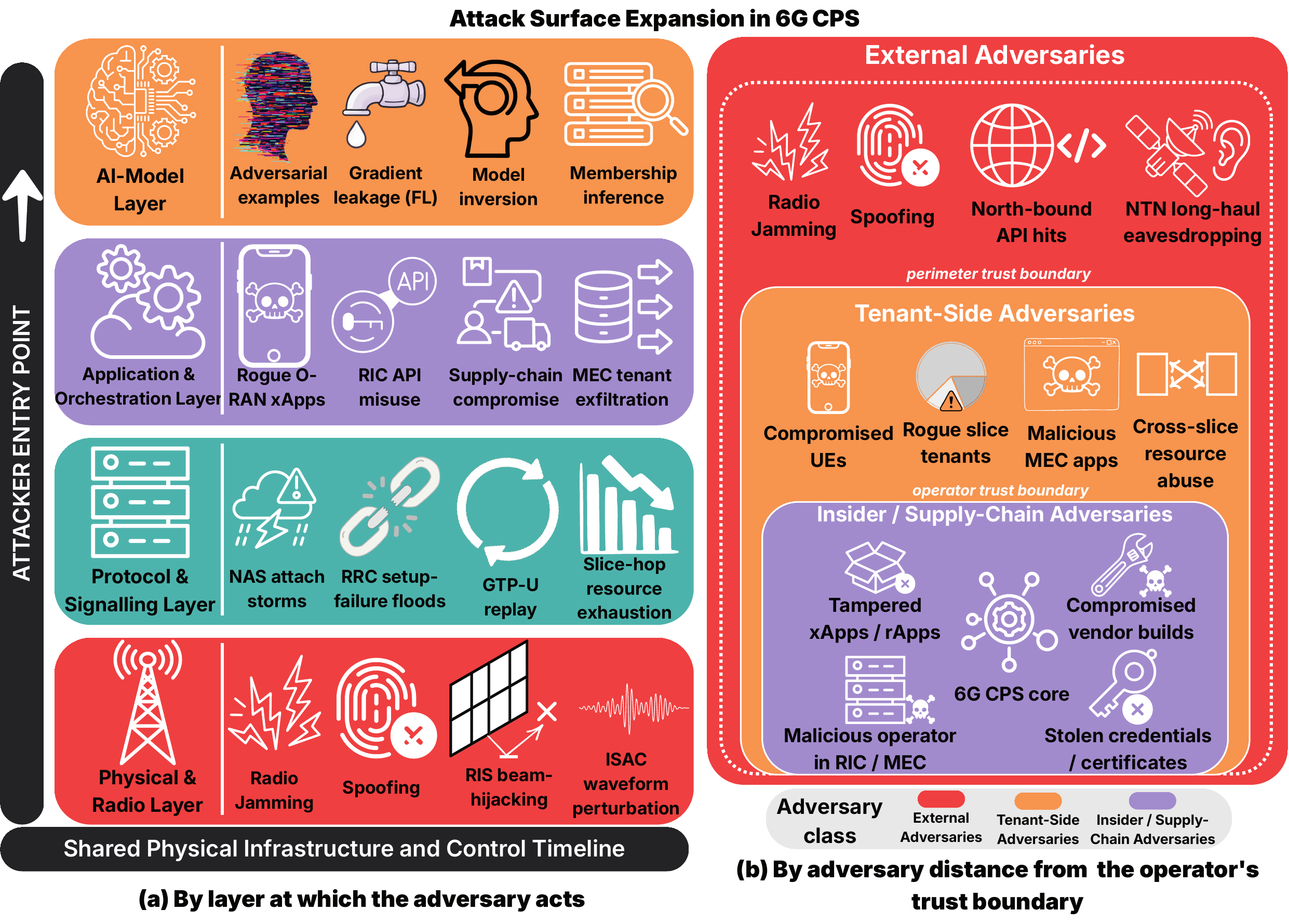}
\caption{Attack surface of 6G CPS along two orthogonal axes: \textbf{(a)} the protocol layer at which the adversary acts (physical/radio up to the AI-model layer, with upward pivot), and \textbf{(b)} the adversary's distance from the operator's trust boundary (external, tenant-side, insider/supply-chain). Both axes are detailed in Sec.~\ref{sec:attack_surface}.}
\label{fig:attack_surface}
\vspace{-10pt}
\end{figure}

\subsection{Attack Surface Expansion in 6G CPS}
\label{sec:attack_surface}
The pillars of Sec.~\ref{sec:6g_pillars} together produce an attack surface that no single perimeter can wrap around, which is the operational reason the closed-loop pipeline is necessary rather than optional.
We frame this surface along two orthogonal axes, summarised in Fig.~\ref{fig:attack_surface}.

\textit{By layer at which the adversary acts}
[Fig.~\ref{fig:attack_surface}(a)]. 
Physical and radio layer attacks include jamming, spoofing, RIS beam-hijacking and pilot contamination on the joint active-passive beamforming substrate\cite{wu2021ris_survey}, adversarial sensing---including waveform-level perturbations against ISAC and unauthorised surveillance via reused sensing beams---and long-haul eavesdropping and jamming on NTN backhaul\cite{hassouna2026_riscs, etsi2026ISAC}.
Protocol and signalling layer attacks include NAS attach storms, RRC setup-failure floods, GTP-U replay, cross-slice resource exhaustion (including slice-hopping variants), isolation bypass, and cross-RAT downgrade via EPS-fallback or rogue base stations~\cite{zhang2023signalingstorm, nguyen2025rrc, abdulqadder2020sdn5g, thantharate2020slice, cui2022voice}. 

Application and orchestration layer attacks include rogue O-RAN xApps, RIC API misuse via the E2/A1 interfaces, supply-chain compromise of vendor-supplied components, and exfiltration through MEC-hosted third-party applications~\cite{polese2023oran, oran_alliance_spec, abdalla2022oran_security, wen2024oran}. 
AI-model layer attacks---model poisoning at training time, adversarial examples at inference time, gradient leakage from FL clients, model inversion, and membership inference---turn the defender's own intelligence into a target~\cite{salmi2026xai, Alauthman2026GANIDS,nguyen2023deep_rl_survey}. 
The MEC tier is doubly exposed because it terminates user-plane traffic and runs security models, making it the natural target for both denial-of-service (DoS) and model-extraction attacks. 

\textit{By adversary distance from the operator's trust boundary} [Fig.~\ref{fig:attack_surface}(b)]. External adversaries hit the radio and the public application surface and have been the historical focus of cellular IDSs; tenant-side adversaries leverage compromised UEs or rogue slice tenants and are the dominant operational threat as IoT density grows; insider and supply-chain adversaries operate from inside the MEC or RIC software stack and are the emergent threat that O-RAN openness has materialised.
Defences that compose across these classes (multi-tier sensing, federated cross-operator learning, and attestable xApp lifecycles) are what the closed loop is designed to provide.
Because both decompositions describe the same incidents from different angles, the closed loop has to hold along the layer axis and the trust-boundary axis at the same time: every attack ultimately traverses the same physical infrastructure and the same control timeline, so a defence that closes one axis while leaving the other open does not contain the threat.

\subsection{DDoS Vectors Against 6G CPS}
\label{sec:ddos_vectors}
DDoS dominates the 6G CPS threat landscape because each control-plane procedure is computationally expensive on the operator side, giving the attacker a high amplification factor: a small UE-side cost translates to a large operator-side cost. 
Six attack vectors recur in the literature.

\noindent\textbf{Signaling Storms.} A botnet of compromised IoT devices triggers excessive NAS Attach/Detach or RRC Setup/Release sequences, exhausting gNB RRC contexts and AMF/Session Management Function (SMF) control-plane CPU, and starving URLLC slices, even at modest message rates~\cite{zhang2023signalingstorm, nguyen2025rrc, wen2024oran}.

\noindent\textbf{GTP-U Flooding.} Spoofed or replayed GTP-U tunnels~\cite{christopoulou_2024_ddos5gnwdaf} overwhelm the User Plane Function (UPF), disrupting V2X latency budgets and consequently block safety-of-life messages.

\noindent\textbf{SMS Flooding on mMTC Slices.} SMS remains critical signaling channel for two-factor authentication and emergency alerts, leaving healthcare and similar notification-bound CPS especially exposed~\cite{cui2022voice, shi2024ims, hussain2021ddos}.

\noindent\textbf{Silent Call Attacks.} Calls initiated and dropped after minimal ring time consume RRC and AMF resources without producing billable minutes; the resulting CDRs carry a distinctive short-duration high-frequency signature~\cite{hussain2021ddos}.

\noindent\textbf{Blended/Polymorphic Attacks.} Modern campaigns interleave the above vectors and adapt their mix in response to mitigation, with reinforcement-learning-driven mutation now demonstrated on the operational frontier~\cite{nguyen2023deep_rl_survey}.

\noindent\textbf{Slice-Hop and Cross-Slice DDoS.} An adversary controls UEs that hand over between slices and concentrates load on whichever slice is currently most cost-effective to disrupt; because slice isolation on shared O-RU/O-DU resources is statistical rather than physical, the target slice can be degraded without direct attachment~\cite{alam2026}.
Predictive load-forecasting defences pre-emptively reallocate resources before the attack peaks~\cite{xylouris_2025_predictiveddos}.

\begin{figure*}[!t]
\centering
\setlength{\abovecaptionskip}{2pt}
\setlength{\belowcaptionskip}{0pt}
\begin{tikzpicture}[
    node distance=1.0cm and 1.1cm,
    every node/.style={font=\scriptsize},
    attack/.style={rectangle, draw=red!70!black, fill=red!8,
                   rounded corners=2pt, minimum height=0.7cm,
                   minimum width=1.7cm, align=center, thick,
                   inner sep=2pt},
    defense/.style={rectangle, draw=blue!70!black, fill=blue!8,
                    rounded corners=2pt, minimum height=0.7cm,
                    minimum width=1.7cm, align=center, thick,
                    inner sep=2pt},
    consequence/.style={rectangle, draw=orange!80!black, fill=orange!8,
                        rounded corners=2pt, minimum height=0.7cm,
                        minimum width=1.7cm, align=center, thick,
                        inner sep=2pt},
    arrowstyle/.style={-{Stealth[length=2.0mm]}, thick},
    darrowstyle/.style={-{Stealth[length=2.0mm]}, thick, blue!70!black},
    obscoupling/.style={-{Stealth[length=2.0mm]}, thick, dashed, black!50},
    actcoupling/.style={-{Stealth[length=2.0mm]}, thick, dashed, blue!70!black},
    labelstyle/.style={font=\tiny, midway, above, text=black!70,
                       inner sep=1pt},
]
\node[attack] (attacker) {Attacker\\(Botnet)};
\node[attack,      right=of attacker] (vector)   {Attack Vector\\Selection};
\node[attack,      right=of vector]   (ran)      {5G/6G RAN\\Impact};
\node[consequence, right=of ran]      (slice)    {CPS Slice\\Disruption};
\node[consequence, right=of slice]    (physical) {Physical\\Consequence};

\draw[arrowstyle, red!70!black] (attacker) -- node[labelstyle]{coordinate} (vector);
\draw[arrowstyle, red!70!black] (vector)   -- node[labelstyle]{launch}     (ran);
\draw[arrowstyle, red!70!black] (ran)      -- node[labelstyle]{propagate}  (slice);
\draw[arrowstyle, red!70!black] (slice)    -- node[labelstyle]{cascade}    (physical);

\node[defense, below=1.3cm of attacker] (monitor)  {CDR/KPI\\Monitoring};
\node[defense, right=of monitor]        (detect)   {ML-Based\\Detection};
\node[defense, right=of detect]         (decide)   {Anomaly\\Classification};
\node[defense, right=of decide]         (mitigate) {SDN/NFV\\Mitigation};
\node[defense, right=of mitigate]       (feedback) {Feedback \&\\Adaptation};

\draw[darrowstyle] (monitor)  -- node[labelstyle, text=blue!60!black]{features} (detect);
\draw[darrowstyle] (detect)   -- node[labelstyle, text=blue!60!black]{scores}   (decide);
\draw[darrowstyle] (decide)   -- node[labelstyle, text=blue!60!black]{policy}   (mitigate);
\draw[darrowstyle] (mitigate) -- node[labelstyle, text=blue!60!black]{update}   (feedback);

\draw[obscoupling] (ran)      -- (detect);
\draw[obscoupling] (slice)    -- (decide);
\draw[actcoupling] (mitigate) -- (ran);

\draw[actcoupling]
  (feedback.south) -- ++(0,-0.55) coordinate (Fdown)
  -- (Fdown -| monitor.south) coordinate (Mdown)
  -- (monitor.south);

\node[font=\scriptsize\bfseries, blue!70!black, below=1pt, yshift=-2pt]
  at ($(Fdown)!0.5!(Mdown)$) {Closed-Loop Defence};

\node[above=0.18cm of vector, font=\scriptsize\bfseries, red!70!black]
  {Attack Lifecycle};

\end{tikzpicture}
\caption{DDoS attack lifecycle in 6G CPS.
The upper red chain traces the attacker's progression from vector selection through RAN impact and slice disruption to physical-world consequences.
The lower blue chain mirrors the closed-loop defence: CDR/KPI monitoring feeds ML-based detection, anomaly classification, SDN/NFV mitigation, and feedback-driven adaptation.
Black dashed arrows mark observational coupling (telemetry feeding the detector and classifier); blue dashed arrows mark active coupling (mitigation acting on the RAN and the feedback-driven model-adaptation loop).}
\label{fig:ddos_lifecycle}
\vspace{-10pt}
\end{figure*}
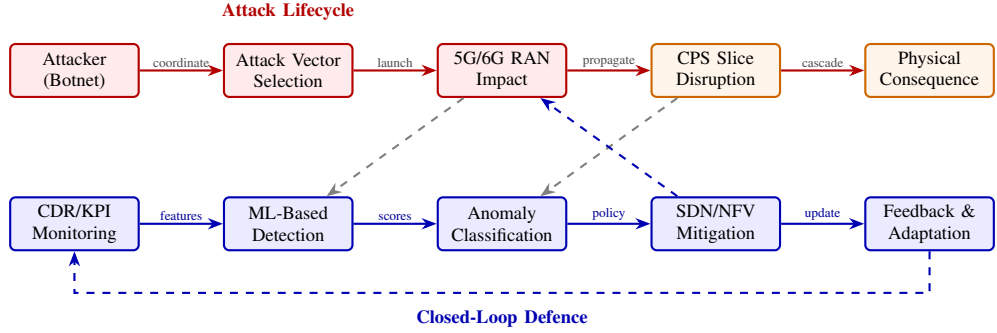

These vectors are normalised in the literature as binary or multi-class classification under class-imbalanced training conditions~\cite{hussain2021ddos}; the methodological treatment is deferred to Sec.~\ref{sec:edge_detection}. 
Fig.~\ref{fig:ddos_lifecycle} renders the attack--defence pairing that motivates the closed-loop pipeline of this survey: the upper chain traces the attacker's path to physical impact, the lower chain previews the detection--mitigation--feedback loop elaborated in Secs.~\ref{sec:edge_detection} and~\ref{sec:mitigation}, and the dashed cross-links mark the telemetry and actuation that bind them.

\subsection{Attacker Model and Profiles}
\label{sec:attacker_model}
We characterise an attacker by the triple $(\kappa, \rho, \pi)$: capability $\kappa\in\{\kappa_0,\kappa_1,\kappa_2\}$ (script-kiddie, organised, nation-state), goal $\rho\in\{\mathcal{G}_D, \mathcal{G}_E, \mathcal{G}_P\}$ (denial, exfiltration, physical impact), and persistence/position $\pi\in\{\pi_{\text{external}},\pi_{\text{edge}},\pi_{\text{core}}\}$~\cite{humayed2017cyber}.
Table~\ref{tab:attacker_profiles} instantiates five representative profiles spanning the easy $(\kappa_0, \mathcal{G}_D, \pi_{\text{external}})$ to the hardest $(\kappa_2, \mathcal{G}_P, \pi_{\text{core}})$ corner of this cube.

\emph{Why even modest attack rates collapse a slice.}
For attack-rate $\lambda$ and per-message control-plane cost $c$, the slice's effective service rate $\mu_{\text{eff}}=\mu - \lambda c$ falls below its required rate when $\lambda > (\mu - \mu_{\text{req}})/c$.
Because cellular signalling traverses long chains of cryptographic and database operations across the AMF, SMF, AUSF, and Unified
Data Management (UDM) functions, $c$ is large and even modest $\lambda$ collapses $\mu_{\text{eff}}$ to zero---which is why detection must observe protocol-level rates rather than byte volumes~\cite{zhang2023signalingstorm, nguyen2025rrc, hussain2021ddos}. 
Reinforcement-learning-driven adversaries~\cite{nguyen2023deep_rl_survey} extend this picture: the attack policy $\pi_{\text{atk}}(s)$, where $s$ is the attacker's observed state including the defender's recent mitigation actions $a_{\text{mit}}$, adapts in response to those actions, so a static threshold-based defence is provably out-evolved.
The closed-loop pipeline of Sec.~\ref{sec:mitigation} is therefore a structural requirement, not a convenience.

\emph{Time horizon stratifies the closed loop's value.}
Three regimes recur. 
\emph{Single-shot} attacks (row 1--2 in Table~\ref{tab:attacker_profiles}) are easy to detect; mitigation latency, not detection accuracy, is the binding constraint. 
\emph{Campaign} attacks (organised criminal DDoS-as-a-service, prolonged signalling-storm campaigns, rows 2 and 5) require drift-aware online learning because the attacker rotates vectors and source pools.
\emph{Persistent} attacks (row 3--4: nation-state Advanced Persistent Threats (APTs), supply-chain compromise of vendor-supplied O-RAN components, insider threats from compromised MEC software) minimise volumetric signal and move laterally through trust boundaries; detection requires multi-modal CDR + RAN + audit-trail fusion plus federated cross-operator correlation, since a single operator's telemetry is rarely sufficient.

\emph{Cross-profile composition.} A single incident may begin as a script-kiddie probe (row 1), be folded into an organised-criminal campaign (row 2 or 5), and provide reconnaissance that a nation-state actor later exploits through a dormant xApp backdoor (row 4). 
A detector that treats each profile in isolation reports three separate incidents and misses the compositional attack. 
The operational response is to maintain a per-source and per-slice reputation vector that persists across detection windows: a rate-limit against a third-time offender is qualitatively different from the same rate-limit against a first-time one. 
Reputation decay, cross-operator reputation sharing, and the integrity of the reputation store against poisoning feed forward into Challenges~C1 and~C3 (Sec.~\ref{sec:open_challenges}).


Adversarial ML threats (evasion, poisoning, model extraction) are documented in the surveyed literature~\cite{salmi2026xai, wan2024survey}, and the agentic-operations literature complements this with prompt-injection, telemetry-manipulation, and unsafe-tool-invocation threats specific to LLM-mediated control loops~\cite{bilal2026netops}; their composition against the closed loop is the subject of Challenge~C1 (Sec.~\ref{sec:c1}).

\begin{table}[!t]
\centering
\caption{Representative Attacker Profiles for 6G CPS}
\label{tab:attacker_profiles}
\renewcommand{\arraystretch}{1.15}
\scriptsize
\setlength{\tabcolsep}{3pt}
\begin{tabular}{|p{1.6cm}|c|c|c|p{1.8cm}|p{1.7cm}|}
\hline
\textbf{Profile} & \textbf{Cap.} & \textbf{Goal} & \textbf{Persist.} & \textbf{Primary Vector} & \textbf{CPS Target} \\
\hline
Script kiddie       & $\kappa_0$ & $\mathcal{G}_D$ & One-shot & Volumetric DDoS              & General service \\
\hline
Botnet operator     & $\kappa_1$ & $\mathcal{G}_D$ & One-shot & Signaling storm, GTP-U flood & V2X, smart grid \\
\hline
Insider threat      & $\kappa_2$ & $\mathcal{G}_E$ & APT      & Data exfil., model poison    & Industrial IoT \\
\hline
Nation-state APT    & $\kappa_2$ & $\mathcal{G}_P$ & APT      & Supply chain, xApp backdoor  & Critical infra. \\
\hline
Cyber-criminal org. & $\kappa_1$ & $\mathcal{G}_E$ & APT      & Ransomware, slice-hop        & Healthcare CPS \\
\hline
\end{tabular}
\vspace{-10pt}
\end{table}

\subsection{Public Dataset Landscape: Threat Evidence and Detection Benchmarks}
\label{sec:datasets}
Public datasets shape what detectors are trained on (Sec.~\ref{sec:edge_detection}), and the cellular-security community remains structurally short of 6G-native data.
Table~\ref{tab:datasets} surveys twelve corpora that recur across the surveyed literature.

Three observations matter for the rest of this article. 
First, only the Telecom Italia Milan \& Trentino CDR set~\cite{barlacchi2015multi} provides authentic operator-grade CDR telemetry, yet it carries no labelled attacks; researchers therefore inject synthetic anomalies into it, biasing benchmark comparability, although recent unsupervised reference-only pipelines~\cite{franco2026} demonstrate that meaningful evaluation against curated event ground-truth (e.g., match schedules, exhibition calendars) is feasible without injected attack labels. 
A recent network-digital-twin evaluation on the same Milan grid reports 96.7\%/98.3\% prediction accuracy and 88--95\% anomaly-prevention rates~\cite{sengendo2026}.
Second, the legacy intrusion corpora that still dominate ML evaluations---KDD'99~\cite{tavallaee2009nslkdd}, NSL-KDD~\cite{tavallaee2009nslkdd}, UNSW-NB15~\cite{moustafa2015unswnb15}, CICIDS2017~\cite{sharafaldin2018}, CICDDoS2019~\cite{sharafaldin2019cicddos}, Bot-IoT~\cite{koroniotis2019towards_botiot} and TON\_IoT~\cite{booij2022ton_iot}---predate 5G and contain no New Radio (NR)/O-RAN signalling, no GTP encapsulation and no slice context. 
Third, the only datasets with native 5G or O-RAN provenance are recent: 5G-NIDD~\cite{siriwardhana2025data} captures real-radio traffic over an operational 5G test network at the University of Oulu, the NANCY corpus~\cite{nancy_oran_2024} covers an O-RAN 5G coverage-expansion testbed, and NetsLab-5GORAN-IDD~\cite{Zadeh2025data} adds synchronised packet-level and radio-telemetry data from a live OAI-based O-RAN deployment. 

None of the three yet covers the full slice-aware, closed-loop scenario that 6G will require---a gap that we revisit as Challenge~C5 in Sec.~\ref{sec:open_challenges}. 
We also note that Generative AI (GenAI)-based synthesis of mobile-network data---spanning per-cell traffic volumes, base-station-association trajectories, and application-usage records---is an emerging but still-private alternative~\cite{goodfellow2014generative, duan2025, kotelnikov2023tabddpm}, with no public release at the time of writing.

\begin{table*}[!t]
\centering
\caption{Public Datasets for Network Intrusion and Anomaly Detection Used in the Surveyed Literature}
\label{tab:datasets}
\renewcommand{\arraystretch}{1.15}
\scriptsize
\setlength{\tabcolsep}{3pt}
\begin{tabular}{|p{2.4cm}|c|p{1.6cm}|p{1.6cm}|p{4.5cm}|p{3.0cm}|c|c|}
\hline
\textbf{Dataset} & \textbf{Year} & \textbf{Type} & \textbf{Net.\ Gen.} & \textbf{Attack Types} & \textbf{Size} & \textbf{Avail.} & \textbf{CDR} \\
\hline
KDD Cup 99~\cite{tavallaee2009nslkdd} & 1999 & Network & Pre-4G & DoS, Probe, R2L, U2R & $\sim$4.9M conn.\ records & Public  & No  \\
\hline
NSL-KDD~\cite{tavallaee2009nslkdd} & 2009 & Network & Pre-4G & DoS, Probe, R2L, U2R & $\sim$148K records & Public  & No  \\
\hline
UNSW-NB15~\cite{moustafa2015unswnb15} & 2015 & Network & General & 9 families: Fuzzers, Analysis, Backdoors, DoS, Exploits, Generic, Reconnaissance, Shellcode, Worms & $\sim$2.54M records & Public  & No  \\
\hline
Telecom Italia Milan \& Trentino CDR~\cite{barlacchi2015multi} & 2015 & Operator CDR (multi-source) & 3G/4G & None (benign telemetry only); multi-source side data: news, weather, social, electricity (Trentino), precipitation & $\sim$2 months (Nov 2013--Jan 2014); two regions: Milan (10,000-cell grid) \& Province of Trentino (6,575-cell grid) & Public  & Yes \\
\hline
CICIDS2017~\cite{sharafaldin2018} & 2017 & Network & General & 8 families: Brute Force, Heartbleed, DoS, DDoS, Web (XSS/SQLi/BForce), Infiltration, Botnet, PortScan & $\sim$2.8M flows & Public & No  \\
\hline
CICDDoS2019~\cite{sharafaldin2019cicddos} & 2019 & DDoS & General & 13 reflection/exploitation DDoS variants: DNS, LDAP, MSSQL, NetBIOS, NTP, PortMap, SNMP, SSDP, SYN, TFTP, UDP, UDP-Lag, WebDDoS & Tens of M records & Public  & No  \\
\hline
Bot-IoT~\cite{koroniotis2019towards_botiot}  & 2019 & IoT & IoT (LAN) & 4 categories / 10 sub-types: Probing, DoS, DDoS, Information Theft & $\sim$73.36M records & Public  & No  \\
\hline
TON\_IoT~\cite{booij2022ton_iot} & 2022 & IoT/IIoT & IoT/IIoT & 9 types: Scanning, DoS, DDoS, Ransomware, Backdoor, Injection, XSS, Password, MITM & $\sim$22M log records & Public  & No  \\
\hline
NANCY O-RAN 5G~\cite{nancy_oran_2024} & 2024 & 5G/O-RAN & 5G O-RAN (SA) & 7 types: Reconnaissance, UDP/TCP-Connect/SYN scan, SYN/ICMP/HTTP flood, Slow-rate DoS & $\sim$588K flow samples (84 features) & Public  & No  \\
\hline
5G-NIDD~\cite{siriwardhana2025data} & 2025 & 5G & 5G NSA & 8 types: ICMP/UDP/SYN/HTTP flood, Slowrate DoS, Torshammer, SYN/TCP-Connect/UDP scan & $\sim$1.22M flows & Public  & No  \\
\hline
NetsLab-5GORAN-IDD~\cite{Zadeh2025data} & 2025 & 5G/O-RAN & 5G O-RAN (SA) & DoS/DDoS (SYN/UDP/TCP/ICMP/HTTP, Slowloris); port \& OS scans; SSH/FTP/HTTP brute force; web (DirBust, SQLi, XSS) & $\sim$28M packets / $\sim$1.72M flows / $\sim$45K radio-telemetry records & Public  & No  \\
\hline
GenAI-synthesised mobile data~\cite{duan2025}  & 2025 & Synthetic mobile-network data & 4G/5G & Synthetic traffic / trajectory / app-usage; no attack labels & Varies (no public release) & Limited & No \\
\hline
\end{tabular}
\vspace{-10pt}
\end{table*}

\subsection{MITRE ATT\&CK Mapping}
\label{sec:mitre}
The MITRE Adversarial Tactics, Techniques, and Common Knowledge (ATT\&CK) framework~\cite{mitreattack},\cite[Ch. 2]{sarker2024ai} assigns each adversary technique a stable identifier that detectors, orchestrators, and SOCs can all key against. 
Three sub-matrices apply to 6G CPS: the Enterprise matrix (T-codes) for IT-layer techniques, the Industrial Control Systems (ICS) matrix (T0-codes) for impact on the physical process layer, and the Adversarial Threat Landscape for AI Systems (ATLAS) catalogue (AML codes) for attacks against deployed ML models~\cite{mitreatlas}. 
We mix all three in Table~\ref{tab:mitre} because a single 6G CPS incident typically chains an Enterprise entry, lateral movement through a MEC host, an Adversarial-ML step that blinds the detector, and a final ICS-layer impact on the controlled process---no individual matrix covers this full chain. 
The resulting vendor-neutral vocabulary is the keying input used by the closed-loop orchestrator of Sec.~\ref{sec:mitigation} to select an appropriate mitigation playbook per detected technique class; for each technique, Table~\ref{tab:mitre} therefore lists the observable telemetry that surfaces it (Sec.~\ref{sec:edge_detection}) and the candidate mitigation primitive that answers it (Sec.~\ref{sec:mitigation}).


\begin{table}[!t]
\caption{MITRE ATT\&CK Mapping for 6G/CPS Threat Landscape, Linked to Closed-Loop Telemetry and Mitigation Primitives}
\label{tab:mitre}
\centering
\scriptsize
\setlength{\tabcolsep}{2.5pt}
\begin{tabular}{|p{1.15cm}|p{0.7cm}|p{1.35cm}|p{1.5cm}|p{1.45cm}|p{1.45cm}|}
\hline
\textbf{Tactic} & \textbf{ID} & \textbf{Technique} & \textbf{6G/CPS Relevance} & \textbf{Observable Telemetry} & \textbf{Candidate Mitigation Primitive} \\
\hline
Impact & T1498 & Network DoS / Flooding & Volumetric DDoS against aNB/MEC & GTP-U utilisation, per-slice load curve & SDN rate-limit / redirect-to-scrubber \\
\hline
Defense Evasion & T1036 & Masquerading & Spoofed UE identity in 6G slices & IMSI-distribution entropy, RRC setup-failure ratio & O-RAN xApp UE suppression \\
\hline
Lateral Movement & T1210 & Exploit Remote Services & MEC-to-core lateral movement & MEC audit-trail, east-west flow logs & ZTA re-authentication, micro-segmentation \\
\hline
Collection & T1040 & Network Sniffing & CDR/telemetry interception at aNB & Interface tap counters, anomalous read patterns & PQC-protected telemetry channel \\
\hline
Impair Process Control & T0836 & Modify Parameter & CPS sensor value manipulation & Sensor-value residuals, control-loop deviation & Slice isolation, control-message integrity check \\
\hline
Inhibit Response & T0881 & Service Stop & Slice isolation bypass / shutdown & Per-slice availability, KPI drop & NFV failover, slice re-instantiation \\
\hline
Initial Access & T1195 & Supply Chain Compromise & O-RAN multi-vendor supply chain & xApp/rApp attestation logs & xApp lifecycle attestation, signed onboarding \\
\hline
ML Attack & AML. T0043 & Physical Adversarial Perturbation & Evasion of MEC-hosted DL models~\cite{salmi2026xai} & Detector confidence drift, input-perturbation score & Adversarial retraining, FL/DT replay \\
\hline
\end{tabular}
\vspace{-12pt}
\end{table}

With the threat surface mapped, the next section examines how edge-side detection (the first arrow of the closed loop) converts CDR and RAN telemetry into actionable alerts.

\section{Edge-Based Detection for 6G CPS}
\label{sec:edge_detection}
This section instantiates the \emph{detect} stage of the closed loop introduced in Sec.~\ref{sec:intro_ai_native}. 
We treat CDR-driven anomaly detection and DDoS classification under one chapter because, in operational 6G, the same MEC host runs the same model families over shared telemetry to separate benign, anomalous, and attack-class flows; the historical split between the two bodies of literature is an artefact of dataset provenance, not of method. 
We organise the survey by \emph{method family} (Secs.~\ref{sec:edge_method_families}--\ref{sec:edge_synthesis}), collect attack-type-specific findings in Sec.~\ref{sec:attack_specific}, and report a unified deployment-aware benchmark in Sec.~\ref{sec:edge_benchmark}.

\subsection{Why MEC Is the Right Detection Tier?}
\label{sec:mec_detection_tier}
Sec.~\ref{sec:mec_arch} established the architectural case for MEC as the enforcement plane and Table~\ref{tab:latency_comparison} consolidates the deployment-tier trade-off the rest of this section optimises against; the recurring pattern is to train large, deploy small, and aggregate hierarchically when local visibility is insufficient. Recent measurements on ARM-SoC platforms~\cite{franco2026} reinforce this case quantitatively: a single Apple M4~Pro completes a full per-cell anomaly-detection pipeline over the Milan CDR dataset $\sim$35\% faster than a 36-core x86 HPC node at $\sim$96\% lower energy, suggesting that ARM-based MEC hosts can sustain CDR-scale analytics at the data source without offloading to the core.

\begin{table}[!t]
\centering
\caption{Detection Deployment Tiers: Latency, Compute, and Visibility}
\label{tab:latency_comparison}
\renewcommand{\arraystretch}{1.15}
\scriptsize
\setlength{\tabcolsep}{3pt}
\begin{tabular}{|l|c|c|c|}
\hline
\textbf{Attribute} & \textbf{On-Device} & \textbf{MEC} & \textbf{Cloud} \\
\hline
Detection latency  & $<$1\,ms        & 1--10\,ms       & 80--120\,ms          \\
\hline
Compute capacity   & Very limited    & Moderate        & Abundant             \\
\hline
Model complexity   & Tiny ML         & Medium DNN/CNN  & Large / ensemble     \\
\hline
Network visibility & Single UE       & Cell/site level & Network-wide         \\
\hline
Energy budget      & Battery         & Mains powered   & Mains powered        \\
\hline
Suitable CPS apps  & Wearables       & URLLC, V2X      & Analytics, forensics \\
\hline
\end{tabular}
\vspace{-12pt}
\end{table}

\subsection{Detection Problem Formulation}
\label{sec:edge_formulation}
Using the notation of Sec.~\ref{sec:background}, the surveyed literature collapses onto three formulations of the detection problem~\cite{chandola2009anomaly}:
\begin{itemize}\setlength{\itemsep}{1pt}
  \item binary detection $f_{\text{bin}}:\mathbb{R}^d\!\to\!\{0,1\}$;
  \item multi-class classification $f_{\text{mc}}:\mathbb{R}^d\!\to\!\{1,\dots,K\}$, which exposes the attack type and is the precondition for type-aware mitigation~\cite{hussain2021ddos};
  \item unsupervised anomaly scoring $s(\mathbf{x})$, typically a reconstruction error from an autoencoder $g_\theta$ trained on benign traffic.
\end{itemize}
A deployed pipeline must additionally meet a streaming budget $\Delta t_{\text{infer}}\!\le\!\tau_{\max}$, with $\tau_{\max}$ in the single-digit-millisecond range for URLLC slices and tens of milliseconds for eMBB and mMTC slices. 
Class imbalance is severe (attack samples typically below $1\%$), so weighted cross-entropy or focal-loss~\cite{lin2017focal_loss,xylouris_2025_predictiveddos} objectives dominate. 
The feature surface fuses CDR-derived signals (per-subscriber call rate, entropy of destination numbers, ratio of failed-to-successful calls, average inter-arrival time, and burstiness indices) with RAN- and NWDAF-side signals (NAS message rate, RRC setup-failure ratio, GTP-U tunnel utilisation, and per-slice resource consumption)~\cite{christopoulou_2024_ddos5gnwdaf}. 
Because these streams arrive over the disjoint native interfaces of the \emph{MEC telemetry vantage} (Sec.~\ref{sec:mec_arch}), multi-modal fusion (Sec.~\ref{sec:multimodal_fusion}) requires explicit time-alignment and trust-domain handling~\cite{li_2025_ainativeran}.

\begin{figure}[!t]
\centering
\scriptsize
\begin{forest}
  for tree={
    grow=east,
    parent anchor=east,
    child anchor=west,
    edge={->, semithick, draw=black!60},
    l sep=10pt,
    s sep=2pt,
    anchor=west,
    align=left,
    font=\scriptsize,
    inner sep=2pt,
    rounded corners=2pt,
    draw,
  }
  [{\textbf{Edge-Side Detection}}, fill=black!75, text=white, font=\scriptsize\bfseries, align=center, minimum width=20mm
    [{\textbf{Federated / PP}}, fill=black!10, font=\scriptsize\bfseries, minimum width=18mm
      [{FedAvg + DP~\edgebadge}, draw=black!40, fill=white]
      [{Split Learning~\edgebadge}, draw=black!40, fill=white]
    ]
    [{\textbf{Deep Learning}}, fill=black!10, font=\scriptsize\bfseries, minimum width=18mm
      [{Autoencoder (recon.)~\edgebadge}, draw=black!40, fill=white]
      [{LSTM (temporal)~\edgebadge}, draw=black!40, fill=white]
      [{Transformer (attn.)~\cloudbadge}, draw=black!40, fill=white]
      [{GNN (topological)~\cloudbadge}, draw=black!40, fill=white]
    ]
    [{\textbf{Classical ML}}, fill=black!10, font=\scriptsize\bfseries, minimum width=18mm
      [{$k$-means (unsup.)~\edgebadge}, draw=black!40, fill=white]
      [{Isolation Forest~\edgebadge}, draw=black!40, fill=white]
      [{Random Forest (sup.)~\edgebadge}, draw=black!40, fill=white]
    ]
    [{\textbf{Statistical}}, fill=black!10, font=\scriptsize\bfseries, minimum width=18mm
      [{Z-score / EWMA~\edgebadge}, draw=black!40, fill=white]
      [{Holt--Winters~\edgebadge}, draw=black!40, fill=white]
    ]
  ]
\end{forest}
\vspace{2pt}
{\footnotesize\textbf{\\Legend:} \edgebadge{} = edge-deployable ($\leq 8$\,GB RAM, $\leq 4$\,W);\quad \cloudbadge{} = cloud-only.}
\caption{Taxonomy of edge-side detection method families surveyed in Sec.~\ref{sec:edge_detection}. Edge-deployable leaves (\edgebadge{}) are those validated on $\leq 8$\,GB RAM / $\leq 4$\,W power envelopes; cloud-only leaves (\cloudbadge{}) currently exceed these envelopes on commodity MEC hardware.}
\label{fig:method_family_tree}
\vspace{-10pt}
\end{figure}
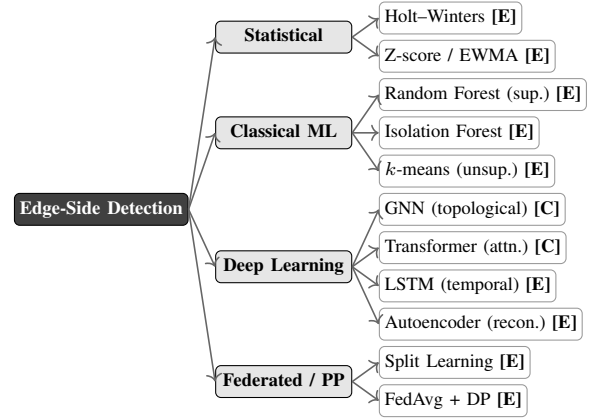

\subsection{Method Families}
\label{sec:edge_method_families}

The MEC host runs every detector family on the same CDR/RAN telemetry, so the families differ not in the data they consume but in the inductive bias each encodes and---decisively for this survey---in whether that bias fits a latency-, memory-, and label-constrained edge tier. We therefore compare families along four deployment axes rather than re-deriving their mechanics (assumed in Sec.~\ref{sec:background}): inference latency, memory/power footprint, labelled-data requirement, and interpretability. Fig.~\ref{fig:method_family_tree} organises the families by mathematical framework and annotates the edge-deployable (\edgebadge{}, $\leq 8$\,GB RAM / $\leq 4$\,W) versus cloud-bound (\cloudbadge{}) boundary; the surveyed evidence per family follows.

\noindent\textbf{Statistical and Classical ML.} Fixed-threshold rules and Principal Component Analysis (PCA)-based residual methods provide interpretable but linear baselines~\cite{chandola2009anomaly, buczak2016survey_ids}. 
Franco-Valiente \textit{et al.}~\cite{franco2026} learn per-cell truncated-Singular Value Decomposition (SVD) subspaces from stable reference weeks and score anomalies as squared residuals with a MAD-based dimensionless severity. Density-estimation approaches~\cite{hussain2017} fit a Gaussian model to normal-profile CDR features and flag low-likelihood grids, achieving $\sim$93\% accuracy on the Milan dataset. Clustering with $k$-means or Density-Based Spatial Clustering of Applications with Noise (DBSCAN) has been used by Sultan \textit{et al.}~\cite{sultan2018cdr} (90\%) and Aziz \textit{et al.}~\cite{aziz2024kmeans} (96\% on 5G CDR). Hussain \textit{et al.}~\cite{hussain2018semi} introduced a semi-supervised cluster-then-classify pipeline at $\sim$92\% accuracy, and Shone \textit{et al.}~\cite{shone2018deep} proposed a non-symmetric deep autoencoder as a complementary deep-shallow hybrid.

This family persists in production for three deployment reasons~\cite{buczak2016survey_ids}: tree ensembles and PCA residuals run in microseconds per record on commodity MEC hardware; a per-feature threshold is auditable in a way a deep model is not, which matters under the lawful-interception and slice-service-level agreements (SLAs) regimes operators face; and these methods stabilise on markedly smaller corpora than the more data-hungry deep models~\cite{ahmad2021benchmark_ids, ferrag2020deep_learning, buczak2016survey_ids}. 
The pattern recovered in operational telco SOCs is therefore a layered ensemble---a statistical front-line filter at the MEC with deep models invoked only on flagged segments, the operational realisation of which is a classical-plus-deep hybrid pipeline~\cite{shone2018deep}.

\noindent\textbf{Deep Feedforward and Convolutional Networks.} An $L$-layer Multilayer Perceptron (MLP) on Milan CDR achieved 94.6\% accuracy and 1.7\% FPR~\cite{hussain2018globecom}, and a deep CDR-based anomaly detector reached 98.8\%/0.44\% on a real operational CDR set~\cite{hussain2019mec}. CNNs exploit the spatial correlation between a cell and its geographic neighbours, which flat MLPs discard; the spatially-aware CDR CNN of~\cite{hussain2020mec} attained 96\% on real operator data, while ResNet-50 and a custom Deep Rudimentary CNN (DRC) model~\cite{he2016resnet, hussain2021ddos} pushed multi-class DDoS classification past 91\% on real CDRs. 
Lightweight CNNs (LUCID~\cite{doriguzzi2020lucid})  reach 99\% on CICDDoS2019 at parameter counts compatible with MEC deployment. 

The trade-off is now well-characterised: deeper feedforward and CNN models gain roughly $1$--$3$ percentage points of accuracy over shallow baselines on identical features at $3$--$10\times$ inference latency~\cite{ferrag2020deep_learning}, a cost that quantisation-aware training and structured pruning recover to a sub-percentage-point accuracy loss, enabling the sub-10\,ms inference reported by MEC-deployed detectors~\cite{doriguzzi2020lucid}. The empirical lesson, recovered across studies, is that the binding constraint at deployment time is feature engineering and class balance, not network depth; the dominant 5G-CPS designs accordingly reuse deep architectures from the encrypted-traffic lineage~\cite{aceto2024traffic} rather than reinventing them.

\noindent\textbf{Recurrent and Hybrid Spatio-Temporal Models.} Recurrent and convolutional-recurrent hybrids matter because cellular attacks have temporal structure on multiple scales: a signalling storm builds over seconds, a silent-call campaign over minutes, and a slow-rate slice-hop attack over hours. Roopak \textit{et al.}~\cite{roopak2019deep} reported 97\%+ on CICIDS2017 with a CNN--LSTM hybrid; Elsayed \textit{et al.}~\cite{elsayed2020ddos_rnn} combined an RNN--autoencoder with softmax (DDoSNet) at 99\% on CICDDoS2019; and ConvLSTM with transfer learning~\cite{noonari2024convlstm} achieved 99\% on a public benchmark, with the typical gain over single-modality baselines at 2--5 points on multi-class DDoS data~\cite{roopak2019deep, noonari2024convlstm, pang2021deep, alam2026}. The one clearly edge-validated case is the LSTM--RNN autoencoder hosted as a Near-RT RIC xApp on a real over-the-air O-RAN testbed~\cite{moore2025}, which delivered 97.5\% accuracy at sub-second closed-loop mitigation latency. The cost is sequence-handling latency, which on commodity MEC hardware bounds the practical sequence length to a few hundred timesteps; longer-horizon attacks are typically captured by hierarchical models that summarise short windows before feeding a slower recurrent layer.

\noindent\textbf{Self-Supervised, Autoencoder, and Generative Adversarial Network (GAN)-Augmented Detection.} The labelled-CDR scarcity problem is structural and operational~\cite{hussain2018semi, yang2023ssl, buczak2016survey_ids}: production networks emit terabytes of unlabelled CDRs daily, but labelling demands scarce expertise. Plain autoencoders trained on benign traffic~\cite{rey2022fl_anomaly} learn normal manifolds and flag high-reconstruction-error inputs; Mirsky \textit{et al.}~\cite{mirsky2018kitsune} (Kitsune) demonstrated online unsupervised ensemble autoencoders running at line rate. Conditional GANs and tabular diffusion models~\cite{goodfellow2014generative, xu2019modeling, kotelnikov2023tabddpm} synthesise minority-class attack samples to improve recall on rare attack types under bounded re-identification risk (surveyed in~\cite{Stoian2026}); Duan \textit{et al.}~\cite{duan2025} additionally use diffusion to denoise coarse cellular trajectories into realistic mobility-aware traffic, and GAN-based detectors export the trained discriminator as the operational classifier~\cite{Alauthman2026GANIDS, goodfellow2014generative}. Contrastive learning and pseudo-labelling progressively expand the labelled pool from a small seed~\cite{yang2023ssl}. The attractive deployment property across this family is that no attack labels are needed and the compressed autoencoder form is edge-feasible (\edgebadge{}); the cost is a non-trivial threshold-tuning problem, since the reconstruction-score distribution is heavy-tailed even on benign traffic, and GAN generators remain a training-time aid rather than an inference-time edge detector.

\noindent\textbf{Graph Neural Networks.} GNNs exploit the network's structural information that flat features discard~\cite{wu2021gnn_survey, rahmani2026gnn}, supporting both per-flow detection (nodes are flows or endpoints, capturing multi-hop attack patterns invisible to single-flow features) and per-cell detection (nodes are aNBs, capturing spatial diffusion of an attack across the radio plane). Lo \textit{et al.}~\cite{lo2022graphsage_ids} ported GraphSAGE to IDS and gained 3--5 percentage points over flat baselines on identical features. The decisive deployment property is inductivity: 6G sees new UEs and new cells arrive continuously, so the inductive GraphSAGE setting Lo \textit{et al.} demonstrate is workable in production whereas any transductive method that requires retraining on each topology change is not. Survey work by Rahmani \textit{et al.}~\cite{rahmani2026gnn} catalogues the design choices (GCN vs.~GraphSAGE vs.~GAT). On current commodity MEC hardware GNN variants remain cloud-bound (\cloudbadge{}) pending compression advances.

\noindent\textbf{Transformers and Attention.} Kim and Thulasiraman~\cite{kim_2024_5gtransformer} used a transformer autoencoder on 5GAD-2022 flow features, reaching F1~$=$~0.882 and AUC~$=$~0.90 without outlier removal; Liu \textit{et al.}~\cite{liu2023attention_ids} introduced multi-domain (time-plus-frequency) multi-head attention with a packet-level time-series feature meter (TS-NFM) extractor for early-stage NIDS, classifying intrusions from the first ten packets at 50{,}000$\times$ earliness over conventional flow-based detectors; and Umer \textit{et al.}~\cite{Umer2025} combined wrapper-based feature selection with a multi-head attention transformer (and SMOTE rebalancing) to reach 93\% on UNSW-NB15 while exposing which features and time positions drove each decision. Adjewa \textit{et al.}~\cite{adjewa_2024_bertfedids} ported BERT to a federated NIDS for 5G, showing a BERT-style backbone can be trained across distributed edge clients without sharing raw traffic and remain deployable through linear quantisation; Houssel \textit{et al.}~\cite{houssel_2024_explainablenids} combined LLMs with explainability for human-readable NIDS reports; and on the synthesis side, Duan \textit{et al.}~\cite{duan2025} fine-tune a Llama-2 backbone (AppSyn) to generate operator-realistic CDR-like records under DCR-balanced privacy. Two deployment consequences recur: attention exposes \emph{which} feature positions and time-steps drove a decision, providing an XAI primitive without an external explainer~\cite{Umer2025}, and multi-head attention captures simultaneous patterns at different temporal scales, the right inductive bias for blended/polymorphic attacks~\cite{vaswani2017attention}. The cost is quadratic complexity in sequence length~\cite[Table~1]{vaswani2017attention}, which bounds practical context windows on MEC hardware to a few hundred CDR samples; sliding-window and linear-attention variants are the standard remedies, and ViT-style image-of-traffic representations are still maturing.


\noindent\textbf{Native Cellular Interfaces (NWDAF, P4, Cross-Layer).} Three native-interface variants close out the method-family inventory and feed directly into the multi-modal fusion treatment of Sec.~\ref{sec:multimodal_fusion}: NWDAF-resident detectors~\cite{christopoulou_2024_ddos5gnwdaf}, P4 data-plane detectors~\cite{abubakar_2024_5gdad}, and cross-layer SDN/NFV/AI architectures spanning RAN, transport, and core~\cite{allaw2025cross_layer, yang_2025_zerotouchsec}.

\subsection{Attack-Type-Specific Detection}
\label{sec:attack_specific}
Six task formulations recur in the literature; each is a specialisation of the method families above, and Table~\ref{tab:attack_specific} compares them along the dimensions that determine MEC deployability.

\begin{table*}[!t]
\centering
\caption{Attack-type-specific detection task formulations, with discriminative telemetry features, representative methods, best reported performance (as cited), and MEC deployability.}
\label{tab:attack_specific}
\renewcommand{\arraystretch}{1.15}
\scriptsize
\setlength{\tabcolsep}{4pt}
\begin{tabular}{|p{2.6cm}|p{5.7cm}|p{4.5cm}|p{2.2cm}|p{1.2cm}|}
\hline
\textbf{Task} & \textbf{Discriminative features} & \textbf{Representative methods} & \textbf{Best reported performance} & \textbf{MEC-deployable} \\
\hline
Sleeping-cell detection & Per-cell reconstruction error over window $T$; absence of alarm; spatial neighbour state & Statistical/Gaussian~\cite{hussain2017}; MLP~\cite{hussain2018globecom}; CDR-DNN~\cite{hussain2019mec}; spatial CNN~\cite{hussain2020mec} & 93.0--98.8\% acc. & \checkmark \\
\hline
Traffic-surge / congestion (DDoS vs.\ flash crowd) & Volumetric rate, IAT, src-IP entropy, slice-aware load curves & ARIMA $\to$ classifier~\cite{hussain2018semi}; k-means + NN forecaster\cite{sultan2018cdr}; energy-aware ARM subspace pipelines (event-side only)~\cite{franco2026} & 92--96\%$^\dagger$ acc. & \checkmark \\
\hline
Fault localisation / root-cause attribution & Causal-DAG residuals; Grad-CAM/SHAP saliencies over cell grid & Granger / causal discovery; XAI attribution~\cite{Umer2025} & N/A (ranking task) & \checkmark \\
\hline
Signalling-storm detection \& mitigation & NAS rate per cell; RRC setup-failure ratio (Msg5/Msg3); Msg5/Msg4 ratio; mean connection hold & RRC-storm xApp~\cite{nguyen2025rrc}; 5G-SPECTOR~\cite{wen2024oran}; blockchain-5GAKA~\cite{zhang2023signalingstorm} & 90\,ms detect.\ lat.\ /\ 1.2\% CPU~\cite{nguyen2025rrc}$^\ddagger$ & \checkmark \\
\hline
SMS-flooding / silent-call detection & Per-subscriber send rate; destination-entropy; message-length dist.; short-call bursts & ResNet/CNN~\cite{hussain2021ddos}; dilated CNN & 91--94\% acc. & \checkmark \\
\hline
Slice-specific / blended / polymorphic DDoS & Per-slice flow features; cross-vector correlations; wrapper-FS attributions & Slice-aware AI-IDS~\cite{thantharate2020slice, xylouris_2025_predictiveddos, amaizu2021composite, Umer2025}; mitigation taxonomy~\cite{hoque2025ddos5g} & 93--96\%+ acc. & \checkmark \\
\hline
Trust management / collaborative detection & Trust-weighted alert aggregation; multi-node SDN votes; reputation tokens & Trust-CIDN survey~\cite{li2022trustcid_survey}; SDN-coop~\cite{yan2016sdnsec_survey} & N/A (consensus task) & Partial (blockchain latency) \\
\hline
\end{tabular}

\vspace{2pt}
\noindent\begin{minipage}{\textwidth}
\scriptsize $^\dagger$Hussain~\textit{et~al.}~\cite{hussain2018semi} additionally report a 40\% relative reduction in false positives over single-stage baselines. $^\ddagger$Nguyen~\textit{et~al.}~\cite{nguyen2025rrc} report 90\,ms detection latency at 132 Msg3s/sec attack rate, with 60\,ms mitigation window before gNB overload, 1.2\% CPU and $\sim$0\% memory overhead on an OAI+FlexRIC testbed; reduces gNB unavailability (95\% under unmitigated attack) by enabling early xApp-triggered response. 5G-SPECTOR~\cite{wen2024oran} and blockchain-5GAKA~\cite{zhang2023signalingstorm} report distinct performance dimensions (layer-3 attack-class detection and registration-storm mitigation, respectively).
\end{minipage}
\vspace{-15pt}
\end{table*}

\subsection{Unified Comparison of Edge-Based Detectors}
\label{sec:edge_benchmark}
Table~\ref{tab:anomaly_benchmark} consolidates the representative numbers from the literature into a single chronological view that subsumes both the cellular-anomaly and the DDoS-IDS evidence bases.
To keep these figures from being read out of context, each row carries an evidence tier (E1--E4, defined in the table footnote) that records how the result was validated, from real operator data (E1) down to simulation or synthetic labels only (E4). 
Read this way, the table exposes a recurring pattern: the near-ceiling accuracies concentrate at E3 on saturated public benchmarks such as CICDDoS2019, where several models report 99\% or above, whereas the E1 operator-CDR studies tend to report more modest, operationally faithful figures—typically in the low-to-high nineties—examined in Challenge~\ref{sec:c5}. 
The contrast is one of tendency rather than a clean partition, since validation realism, not headline accuracy alone, governs how each figure should be weighed.

\begin{table*}[!t]
\centering
\caption{Unified comparison of edge-based detection methods for 5G/6G CPS, consolidating the cellular-anomaly and DDoS-classification literature into a single deployment-aware view.
Headline figures are reproduced from each cited work; we do not re-execute them.
These figures should be read within, not across, the evidence tiers defined in the footnote.}
\label{tab:anomaly_benchmark}
\renewcommand{\arraystretch}{1.12}
\scriptsize
\setlength{\tabcolsep}{3pt}
\begin{tabular}{|l|c|p{2.7cm}|l|c|c|c|c|c|c|c|c|}
\hline
\textbf{Ref.} & \textbf{Year} & \textbf{Method} & \textbf{Dataset} & \textbf{Acc.\ (\%)} & \textbf{F1} & \textbf{FPR (\%)} & \textbf{Latency} & \textbf{Ev.} & \textbf{Edge} & \textbf{CDR} & \textbf{Domain} \\
\hline
\cite{hussain2017} & 2017 & Semi-sup.\ statistical (Gaussian) & Milan CDR & 92.8 & 0.94 & 14.1 & Near-RT & E3 & & \checkmark & \\
\hline
\cite{hussain2018semi} & 2018 & Semi-sup.\ statistical (Gaussian) & Milan CDR & 92.0 & 0.94 & 14.1 & Near-RT & E3 & & \checkmark & \\
\hline
\cite{hussain2018globecom} & 2018 & $L$-layer DNN (MLP) & Milan CDR & 94.6 & -- & 1.7 & Near-RT & E3 & & \checkmark & N \\
\hline
\cite{sultan2018cdr} & 2018 & $k$-means + Neural Network & CRAWDAD, Nodobo, Milan CDR & -- & -- & -- & Batch & E3 & & \checkmark & \\
\hline
\cite{mirsky2018kitsune} & 2018 & Autoencoder ensemble (Kitsune) & Real traffic (9 datasets) & -- & -- & -- & RT & E3 & \checkmark & & \\
\hline
\cite{koroniotis2019towards_botiot} & 2019 & SVM/RNN/LSTM-RNN baselines (Bot-IoT release) & Bot-IoT & 99.99 & 1.00 & -- & -- & E3 & & & N,\,P \\
\hline
\cite{roopak2019deep} & 2019 & CNN--LSTM hybrid & CICIDS2017 & 97.1 & -- & -- & -- & E3 & & & \\
\hline
\cite{sharafaldin2019cicddos} & 2019 & Dataset release + ID3/RF baselines & CICDDoS2019 & -- & 0.78 & 0.69 & -- & E3 & & & \\
\hline
\cite{hussain2019mec} & 2019 & MEC-deployed DNN & Real CDR (operator) & 98.8 & -- & 0.44 & RT & E3 & \checkmark & \checkmark & N \\
\hline
\cite{doriguzzi2020lucid} & 2020 & Lightweight CNN (LUCID) & CICDDoS2019 & 99.0 & 0.99 & -- & RT & E3 & \checkmark & & \\
\hline
\cite{hussain2020mec} & 2020 & MEC CNN (spatial grid) & Real CDR & $\leq$96 & -- & -- & RT & E1 & \checkmark & \checkmark & N \\
\hline
\cite{abdulqadder2020sdn5g} & 2020 & SDN+NFV+AI multi-layer & NS-3 simulation & 96.1 & -- & -- & RT & E4 & & & N \\
\hline
\cite{elsayed2020ddos_rnn} & 2020 & RNN--Autoencoder + softmax (DDoSNet) & CICDDoS2019 & 99.0 & 0.99 & -- & -- & E3 & & & \\
\hline
\cite{hussain2021ddos} & 2021 & CNN (ResNet-50 / DRC) & Milano CDR + synth. attacks (4 types) & 91--97 & -- & -- & -- & E3 & & \checkmark & P \\
\hline
\cite{amaizu2021composite} & 2021 & Composite DNN (MLP+PCC) & CICDDoS2019 & 99.7 & -- & -- & -- & E3 & & & N \\
\hline
\cite{cil2021detection} & 2021 & Feedforward DNN (MLP) & CICDDoS2019 & 94.6 & -- & -- & RT & E3 & & & \\
\hline
\cite{booij2022ton_iot} & 2022 & RF (also GBM, MLP) & TON\_IoT (vs Aposemat IoT-23) & 98.1 & 0.97 & -- & -- & E3 & & & \\
\hline
\cite{lo2022graphsage_ids} & 2022 & E-GraphSAGE (edge-feature GNN) & NF-ToN-IoT / NF-BoT-IoT & -- & 1.00 & -- & -- & E3 & & & \\
\hline
\cite{liu2023attention_ids} & 2023 & Multi-Domain Transformer (MD-MHA, TS-NFM) & SCVIC-TS-2022 (CICIDS2017 PCAPs) & 99.7 & 0.84 & -- & Early-RT & E3 & & & N \\
\hline
\cite{aziz2024kmeans} & 2024 & $k$-means + deseasonalisation & 14~M one-year operator CDR (5G-deployed) & 96.7 & 0.60 & -- & Batch & E1 & & \checkmark & N \\
\hline
\cite{aziz2024tmc} & 2024 & DBSCAN / IF (deseasonalised; GMM, MS for profiling) & 37~M one-year operator CDR & 98.1 & 0.76 & -- & Batch & E1 & & \checkmark & N \\
\hline
\cite{noonari2024convlstm} & 2024 & Multi-Scale ConvLSTM + transfer learning & Public Kaggle KPI (train) $\to$ KPI (fine-tune) & 99.0 & 0.99 & -- & -- & E3 & & & N \\
\hline
\cite{kim_2024_5gtransformer} & 2024 & Transformer autoencoder & 5GAD-2022 (free5GC) & -- & 0.88 & -- & Offline & E4 & & & N \\
\hline
\cite{Umer2025} & 2025 & Wrapper-FS + multi-head attn.\ Transformer & UNSW-NB15 & 93.0 & 0.92 & -- & Offline & E3 & & & \\
\hline
\cite{moore2025} & 2025 & LSTM--RNN autoencoder (xApp) & O-RAN over-the-air testbed & 97.5 & 0.99 & 2.2 & RT & E2 & \checkmark & & N,\,P \\
\hline
\cite{allaw2025cross_layer} & 2025 & RF/XGB on UNSW-NB15 (cross-layer arch.) & UNSW-NB15 & 86--87 & $\sim$0.86 & -- & & E3 & & & N,\,P \\
\hline
\cite{yang_2025_zerotouchsec} & 2025 & SH-CASH AutoML (ARF/SRP + ADWIN/EDDM) & CICIDS2017 + Oracle RF fingerprint & 99.4 & 0.98 & -- & RT & E3 & & & N \\
\hline
\cite{alam2026} & 2026 & Multi-layer FL + CNN--LSTM (trust scoring) & Mininet-WiFi & 97.6 & 0.97 & 6.5 & RT & E4 & \checkmark & & N \\
\hline
\end{tabular}

\vspace{2pt}
\noindent\begin{minipage}{\textwidth}
\scriptsize \textbf{Evidence strength (Ev.):} E1\,=\,real operator data, privately accessed / not publicly released; E2\,=\,real 5G/O-RAN testbed or over-the-air; E3\,=\,public real-traffic benchmark; E4\,=\,simulation or synthetic labels only. \textbf{Domain:} N\,=\,5G/6G-relevant; P\,=\,explicit CPS setting. ``--'' denotes a value not reported in the cited work.
\end{minipage}
\vspace{-10pt}
\end{table*}

Three trends bear directly on the closed-loop thesis of this survey.
First, headline accuracy on curated benchmarks is now saturated above $97\%$, but performance on real operational CDR data sits in the $91$--$98\%$ band; this gap is a benchmark artefact, not a methodological gain (Challenge~C5).
Second, edge-deployable methods are now common, vindicating the MEC-tier thesis of Sec.~\ref{sec:mec_detection_tier}, and the most recent contributions~\cite{moore2025, christopoulou_2024_ddos5gnwdaf, abubakar_2024_5gdad} are increasingly evaluated on native 5G/O-RAN testbeds rather than legacy public benchmarks.
Third, methods evaluated on native O-RAN hardware~\cite{moore2025} report sub-second \emph{detect-to-mitigate} latency, the first end-to-end empirical confirmation that the loop-time argument of Sec.~\ref{sec:intro_ai_native} is achievable with current technology.

Reporting practice, however, remains uneven: the community standardised on accuracy and F1 in the late 2010s, but three deployment-relevant metrics remain under-reported~\cite{Alauthman2026GANIDS}---calibrated false-positive rates, tail-latency percentiles (especially $p99$, which determines MEC deployability), and energy-per-decision under perturbation. 
Dataset descriptors that publish per-class precision/recall plus training and prediction times~\cite{siriwardhana2025data} provide the baseline against which future 6G-native datasets should be benchmarked.
The most operationally-deployable contributions~\cite{moore2025} are precisely those that report end-to-end mitigation latency on a CPS-relevant slice rather than single-stage accuracy; the lesson for future work is that the deployment-relevant figures of merit are loop-time and CPS-impact harm reduction, and the structural fix to cross-corpus incomparability is the consortium-released benchmark proposed in Challenge~C5.

\subsection{End-to-End MEC Detection Pipeline}
\label{sec:edge_pipeline}
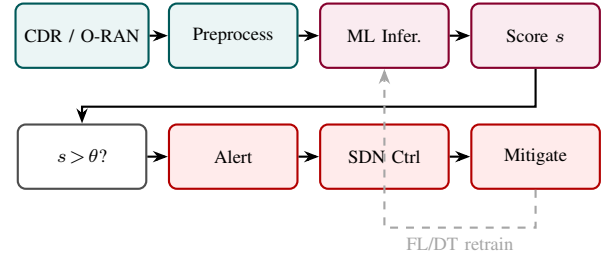
\begin{figure}[!t]
\centering
\begin{tikzpicture}[
    every node/.style={font=\scriptsize},
    block/.style={rectangle, draw=black!70, fill=white, rounded corners=3pt,
                  minimum height=0.85cm, minimum width=1.7cm, align=center, thick},
    datablock/.style={rectangle, draw=teal!70!black, fill=teal!8, rounded corners=3pt,
                      minimum height=0.85cm, minimum width=1.7cm, align=center, thick},
    mlblock/.style={rectangle, draw=purple!70!black, fill=purple!8, rounded corners=3pt,
                    minimum height=0.85cm, minimum width=1.7cm, align=center, thick},
    actionblock/.style={rectangle, draw=red!70!black, fill=red!8, rounded corners=3pt,
                        minimum height=0.85cm, minimum width=1.7cm, align=center, thick},
    arr/.style={-{Stealth[length=2mm]}, thick},
    fb/.style={-{Stealth[length=2mm]}, thick, dashed, gray!70},
]
\node[datablock]   (cdr)        at (0,2.2) {CDR / O-RAN};
\node[datablock]   (preprocess) at (2,2.2) {Preprocess};
\node[mlblock]     (inference)  at (4,2.2) {ML Infer.};
\node[mlblock]     (scoring)    at (6,2.2) {Score $s$};
\node[block]       (threshold)  at (0,0.6) {$s\!>\!\theta$?};
\node[actionblock] (alert)      at (2,0.6) {Alert};
\node[actionblock] (sdn)        at (4,0.6) {SDN Ctrl};
\node[actionblock] (mitigate)   at (6,0.6) {Mitigate};
\draw[arr] (cdr) -- (preprocess); \draw[arr] (preprocess) -- (inference);
\draw[arr] (inference) -- (scoring);
\draw[arr] (scoring.south) -- ++(0,-0.5) -| (threshold.north);
\draw[arr] (threshold) -- (alert); \draw[arr] (alert) -- (sdn); \draw[arr] (sdn) -- (mitigate);
\draw[fb] (mitigate.south) -- ++(0,-0.5) -| (inference.south)
    node[pos=0.25, below, gray!70] {FL/DT retrain};
\end{tikzpicture}
\caption{MEC-based detection pipeline. The \emph{Alert}$\rightarrow$\emph{SDN Ctrl}$\rightarrow$\emph{Mitigate} stages are the detector-to-orchestrator hand-off formalised in Sec.~\ref{sec:handoff}; the dashed feedback path is the closed-loop FL/DT retraining channel of Sec.~\ref{sec:enablers}.}
\label{fig:mec_pipeline}
\vspace{-5pt}
\end{figure}

Fig.~\ref{fig:mec_pipeline} renders the pipeline that each detector of the families above must instantiate end-to-end.
\emph{Ingestion} subscribes to the CDR, RAN E2 KPM, and NWDAF streams at the MEC vantage (Sec.~\ref{sec:mec_arch})~\cite{li_2025_ainativeran};
\emph{feature extraction} computes the engineered signals of Sec.~\ref{sec:cdr};
\emph{inference}---the latency-critical stage---runs the compressed detector to produce per-cell or per-flow scores~\cite{zhou2019edge_intelligence};
and \emph{calibration/thresholding} maps scores to slice-aware decisions $s(\mathbf{x})\!>\!\theta$, trading detection recall against the False Block Rate (FBR) defined in Table~\ref{tab:mitigation_metrics}.
A positive decision is handed off to the orchestrator as an alert whose tuple and per-field action semantics are specified in Sec.~\ref{sec:handoff}.
The end-to-end budget sums these stages and the orchestrator's decision time, which the slice safety envelope must bound; the dashed path in Fig.~\ref{fig:mec_pipeline} closes the loop into FL/DT retraining (Sec.~\ref{sec:enablers}).

\subsection{Multi-Modal Fusion Across CDR, RAN, and Core Telemetry}
\label{sec:multimodal_fusion}
A single telemetry stream rarely contains enough signal to separate sophisticated 6G CPS attacks from benign traffic. 
Drawing on the three operator-facing modalities of the \emph{MEC telemetry vantage} (Sec.~\ref{sec:mec_arch}), each
stream contributes a distinct grain: CDR captures aggregate per-subscriber behaviour at the call-event scale, RAN E2~KPM exposes per-UE RRC state transitions and scheduler decisions, and NWDAF event exposure carries cross-NF correlations from the core.
Each modality has blind spots: CDR cannot see RRC failure cascades that never reach billing, RAN E2 KPM cannot see core-network amplification, and NWDAF events do not yet expose cell-level mobility patterns at fine grain. 
The multi-modal fusion problem is to combine these streams in a way that preserves the latency budget of the MEC tier while extracting cross-modality joint signals~\cite{christopoulou_2024_ddos5gnwdaf}.

Three fusion architectures recur in the surveyed literature. 
\emph{Early fusion} concatenates per-modality features into a single input vector for a unified detector; this is the simplest pattern and gives the strongest gains when the modalities are time-aligned and sampled at compatible rates, which is usually true for CDR and RAN at the MEC. 
Christopoulou \textit{et al.}~\cite{christopoulou_2024_ddos5gnwdaf} provide a representative early-fusion baseline on a native 5G testbed, concatenating UE-, eNB-, and MME-side features into a single input vector consumed by KNN, XGBoost, and Decision-Tree classifiers.
\emph{Late fusion} runs a per-modality detector and combines their scores via a learned aggregator; this scales better when modalities have different sampling rates or live on different hosts, at the cost of losing fine cross-modality interactions. 
\emph{Hybrid fusion} runs per-modality encoders to a shared latent space and then a joint head on the latent representation; this captures cross-modality interactions while controlling the per-host compute footprint and is the dominant pattern in 2024--25 cellular IDS work. 
The remaining gap is standardised cross-modality schema (Challenge~C4) so that the fusion architecture does not have to be re-implemented per operator.

\subsection{Synthesis: Four Open Trade-Offs}
\label{sec:edge_synthesis}
The surveyed evidence converges on four trade-offs that structure the remaining design space.

\emph{Accuracy vs.\ latency:} deeper models gain $1$--$3$ pp accuracy but risk violating URLLC budgets; int8 quantisation, structured pruning, and knowledge distillation typically recover $70$--$90\%$ of the latency overhead at
$0.5$--$1.5$\,pp accuracy loss.

\emph{Local vs.\ global visibility:} MEC-local inference is fast but blind to distributed campaigns; hierarchical aggregation introduces latency.

\emph{Supervised vs.\ unsupervised:} supervised peaks in accuracy but starves for labels; semi-/self-supervised approaches occupy the operationally practical middle.

\emph{Accuracy vs.\ explainability and personalisation:} black-box deep models (deep ensembles, transformer hybrids) score highest on curated benchmarks but are the hardest to explain, while the most interpretable methods (PCA residuals, decision trees) lag in accuracy by 5--15 percentage points; personalised per-slice or per-operator models that incorporate operator-specific priors lose a small amount of headline accuracy but gain robustness to distribution shift and produce explanations that downstream auditors and regulators can act on. 
The trade-off is not a free lunch: explainable models can leak decision-boundary information to adversaries through the explanation surface itself~\cite{salmi2026xai}, and personalised models complicate FL aggregation because per-operator divergence accumulates across rounds. 

The operational compromise motivated by Umer \textit{et al.}~\cite{Umer2025} is layered along three axes simultaneously. 
Along the \emph{accuracy axis}, an interpretable triage layer at the MEC (decision trees, PCA residuals, attention-weight read-outs) scores every alert at low latency, with a black-box deep model invoked only for a second-look pass on flagged segments---recovering deep-family accuracy on the small fraction of traffic that warrants it. 
Along the \emph{explainability axis}, attention-based attribution is emitted at both layers~\cite{Umer2025}, so every decision carries a per-feature rationale, with heavier model-agnostic explainers (SHAP, counterfactuals) reserved for post-incident review at the operator core. 
Along the \emph{personalisation axis}, a federated backbone is shared across operators while each fine-tunes a small operator-specific head on its own non-IID CDR/RAN distribution, capturing local attack patterns without leaking subscriber data or destabilising FL aggregation~\cite{li2020fedprox}.

\subsection{Design Lessons for the Detect Stage of the Closed Loop}
\label{sec:edge_design_lessons}
Four design lessons recur across the literature surveyed in this section and should constrain any future detect-stage design.
\begin{itemize}[leftmargin=*]
\item \emph{Lesson~1 — telemetry first, model second.} Feature engineering on CDR/NWDAF/E2 telemetry has produced larger accuracy deltas than architectural changes within a method family; investment in telemetry pipelines and feature selection out-performs investment in deeper models~\cite{aziz2024tmc, hussain2020mec, hussain2021ddos}.
\item \emph{Lesson~2 — compress as a first-class step, not an afterthought.} Quantisation- and pruning-aware training produces detectors that match unconstrained baselines on commodity MEC hardware; deploying an unconstrained model and then trying to fit it to MEC after the fact rarely succeeds~\cite{zhou2019edge_intelligence, doriguzzi2020lucid, hussain2019mec, hussain2020mec, wang2020convergence_edge_ai}.
\item \emph{Lesson~3 — multi-class, not binary.} Multi-class classifiers~\cite{hussain2021ddos, sharafaldin2019cicddos} produce per-attack-type outputs that the orchestrator can use to select mitigation playbooks; a binary detector forces the orchestrator to default to a generic primitive, which over-blocks legitimate traffic.
\item \emph{Lesson~4 — federate where possible, validate everywhere.} Per-MEC training on local network telemetry plus FL aggregation~\cite{mcmahan2017fedavg, blika2024fl_6g, adjewa_2024_bertfedids, chen2020fl_ids, nguyen2019fl_ddos} is the only path that scales, and DT-based what-if validation is the only way to keep the loop safe under continual model updates~\cite{huang2025_ailcm_ran}.
\end{itemize}

These four lessons converge on a single operational implication: detection performance in isolation is not the binding constraint of the closed loop. Even a perfect detector is operationally useless without the automated, low-latency hand-off into the mitigation machinery that the next section formalises (Sec.~\ref{sec:mitigation}, Tables~\ref{tab:mitigation_metrics} and~\ref{tab:mitigation_comparison}).

\section{Network-Wide Mitigation and the Closed Loop}
\label{sec:mitigation}
This section surveys how the alert produced by the detector (the \emph{Alert} box of Fig.~\ref{fig:mec_pipeline}) is converted into network-wide action.
We treat SDN, NFV, and O-RAN xApps as composable actuators of one closed loop rather than competing schools, define the metrics that matter for CPS (Sec.~\ref{sec:mitigation_metrics}), and present the consolidated comparison of mitigation frameworks (Table~\ref{tab:mitigation_comparison}).

\subsection{From Detection to Response: The Mitigation Gap}
\label{sec:mitigation_gap}

The cellular security literature has historically paid less attention to mitigation than to detection, even though the operational value of an IDS depends on what happens after the alert. 
The mitigation gap is conventionally framed as having three faces: (i)~the policy gap (what action to take), (ii)~the actuation gap (where to enact it), and (iii)~the latency gap (how long it takes). 6G CPS amplifies all three because URLLC slices need millisecond reactions and human-in-the-loop is no longer feasible.

Beyond these three, 6G CPS surfaces a fourth, less-often-named face: the \emph{reversibility gap}. 
Once the orchestrator has throttled a slice or blocked a UE class, undoing that action has its own latency, its own collateral impact, and its own audit burden. Mitigation design in 6G CPS has to treat each action as paired with its inverse---re-admit, un-throttle, restore-handover-weight---and the inverse must be reachable within a budget comparable to the original action or the loop cannot tolerate false positives without breaking safety-of-life workloads. 
This reversibility requirement is what makes graded enforcement (Sec.~\ref{sec:substrates}) and the four-tuple \texttt{ttl} field (Sec.~\ref{sec:handoff}) structurally necessary rather than cosmetic. 
Mitigation systems that are evaluated only by their block-effectiveness~\cite{abdulqadder2020sdn5g} systematically overstate operational viability, because the reversibility cost is invisible to block-effectiveness alone.

The policy gap is the hardest of the three. 
A detector emits a per-flow or per-cell anomaly score; an orchestrator must convert that score into a discrete action drawn from a finite playbook (rate-limit, scrub, isolate, drop, re-authenticate, re-route, slice-rebalance) and parameterise that action against the slice's safety envelope. 
The mapping is not unique: the same anomaly could justify a soft response (rate-limit at 50\%) or a hard response (slice isolation), and the cost of an over-reaction is paid by the very CPS workload the action is supposed to protect~\cite{allaw2025cross_layer, sheibani2024multilayer}.

The actuation gap is the heterogeneity of the actuator surface. 
SDN flow-rule installation, NFV scrubber instantiation, O-RAN xApp scheduler-weight modification, and core-network admission control each have different APIs, latency envelopes, and failure modes; a single playbook must compose across all of them.

The latency gap is what 6G CPS makes binding. 
A URLLC slice with a $1$\,ms end-to-end budget cannot tolerate the sub-minute VNF-instantiation latencies that NFV-only mitigation incurs, even if the eventual mitigation is correct~\cite{fayaz2015bohatei, abdulqadder2020sdn5g}.

The closed loop addresses all three gaps by maintaining a tier-stratified actuator hierarchy: the fastest tier (O-RAN xApp scheduler weights, $10$--$100$\,ms) handles immediate suppression, while slower tiers (SDN re-routing, ${\sim}1$\,s; NFV scrubber instantiation, tens of seconds) sustain the response.

\begin{figure*}[!t]
\centering
\setlength{\abovecaptionskip}{4pt}
\begin{tikzpicture}[
    font=\footnotesize,
    node distance=0.5cm,
    smo/.style={rectangle, draw=black!70, fill=blue!8, thick,
                minimum width=2.2cm, minimum height=0.6cm, align=center, rounded corners=2pt},
    ric/.style={rectangle, draw=black!70, fill=teal!12, thick,
                minimum width=2.0cm, minimum height=0.75cm, align=center, rounded corners=2pt},
    ranbox/.style={rectangle, draw=black!70, fill=gray!10, thick,
                minimum width=1.5cm, minimum height=0.75cm, align=center, rounded corners=2pt},
    hook/.style={rectangle, draw=red!70!black, fill=red!6, thick,
                 minimum width=2.4cm, minimum height=0.9cm, align=center,
                 rounded corners=2pt, font=\scriptsize},
    iflabel/.style={font=\scriptsize\itshape, text=black!70},
    mgmt/.style={dashed, draw=black!55, -{Latex[length=5pt,width=4pt]}, thick, shorten >=2pt},
    dpath/.style={-{Latex[length=6pt,width=5pt]}, very thick, draw=black!85, shorten >=3pt, shorten <=1pt},
    hookline/.style={draw=red!70!black, thick, dotted}
]

\node[smo] (smo) at (0,1.7) {SMO\\\scriptsize(Service Mgmt \& Orchestration)};

\node[ric] (nonrt) at (-6.0,0)  {Non-RT RIC\\\scriptsize(rApps)};
\node[ric] (nearrt) at (-3.0,0) {Near-RT RIC\\\scriptsize(xApps)};
\node[ranbox] (ocu) at (0.0,0)  {O-CU};
\node[ranbox] (odu) at (3.0,0)  {O-DU};
\node[ranbox] (oru) at (6.0,0)  {O-RU};

\draw[dpath] (nonrt) -- node[above,iflabel]{A1} (nearrt);
\draw[dpath] (nearrt) -- node[above,iflabel]{E2} (ocu);
\draw[dpath] (ocu) -- node[above,iflabel]{F1} (odu);
\draw[dpath] (odu) -- node[above,iflabel]{OFH} (oru);

\draw[mgmt] (smo.south) .. controls +(-2,-0.4) and +(0,0.6) .. (nonrt.north)
    node[midway,above,iflabel,xshift=-6pt]{O1};
\draw[mgmt] (smo.south) .. controls +(0,-0.7) and +(0,0.6) .. (ocu.north)
    node[midway,right,iflabel,xshift=2pt,yshift=2pt]{O1};
\draw[mgmt] (smo.south) .. controls +(1.2,-0.4) and +(0,0.6) .. (odu.north)
    node[midway,above,iflabel,xshift=6pt]{O2};

\node[hook] (h1) at (-6.0,-1.6) {\textbf{rApp re-policy}\\$\sim$1\,s\\\textit{network-wide}};
\node[hook] (h2) at (-3.0,-1.6) {\textbf{xApp throttle}\\10--100\,ms\\\textit{cell-group}};
\node[hook] (h3) at (0.0,-1.6)  {\textbf{Slice re-alloc.}\\$\sim$100\,ms\\\textit{per-slice}};
\node[hook] (h4) at (3.0,-1.6)  {\textbf{PRB blocking}\\$\sim$1\,ms\\\textit{single cell}};

\draw[hookline] (nonrt.south) -- (h1.north);
\draw[hookline] (nearrt.south) -- (h2.north);
\draw[hookline] (ocu.south) -- (h3.north);
\draw[hookline] (odu.south) -- (h4.north);

\draw[{Latex[length=5pt,width=4pt]}-{Latex[length=5pt,width=4pt]}, thick, draw=black!60]
    (-7.0,-2.5) -- (4.2,-2.5);
\node[font=\scriptsize, text=black!70] at (-6.0,-2.78) {slower $\cdot$ wider blast radius};
\node[font=\scriptsize, text=black!70] at (3.0,-2.78)  {faster $\cdot$ narrower};

\begin{scope}[shift={(0,-3.4)}, font=\scriptsize]
  \draw[dpath] (-4.2,0) -- (-3.6,0);
  \node[anchor=west] at (-3.6,0) {data-path control};
  \draw[mgmt] (-1.0,0) -- (-0.4,0);
  \node[anchor=west] at (-0.4,0) {O1/O2 mgmt plane};
  \draw[draw=red!70!black, fill=red!6, thick, rounded corners=1.5pt]
       (2.7,-0.12) rectangle (3.0,0.12);
  \node[anchor=west] at (3.05,0) {mitigation hook};
\end{scope}

\end{tikzpicture}
\caption{O-RAN/SDN mitigation hooks mapped onto the disaggregated RAN control plane. Four actuation points trade latency against blast radius: rApp re-policy via A1 ($\sim$1\,s, network-wide), xApp throttle via E2 (10--100\,ms, cell-group), slice re-allocation via F1 ($\sim$100\,ms, per-slice), and PRB blocking at O-DU MAC ($\sim$1\,ms, single cell). The dashed O1/O2 management plane carries telemetry and lifecycle events but is not in the data-path control loop. Faster hooks have narrower blast radius, motivating the closed-loop design in Sec.~\ref{sec:closed_loop}.}
\label{fig:oran_hooks_overlay}
\vspace{-12pt}
\end{figure*}
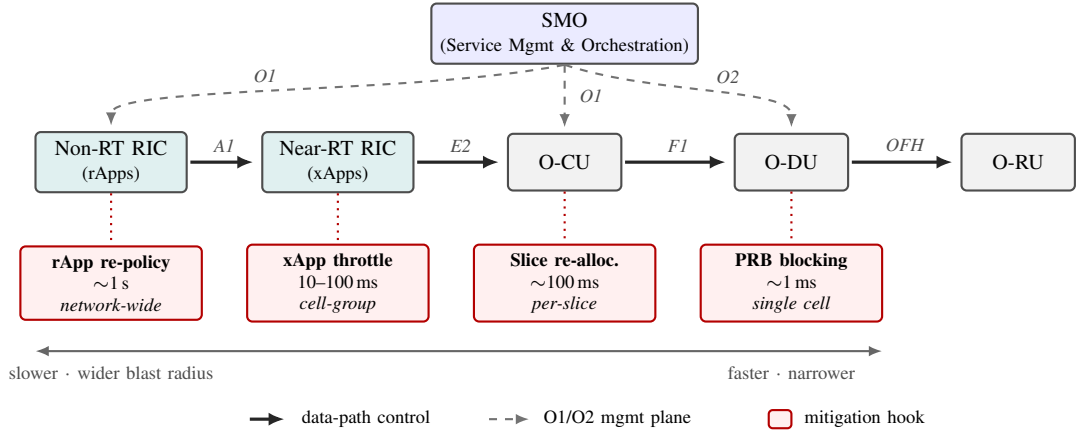

\subsection{Detection-to-Mitigation Hand-Off Protocol}
\label{sec:handoff}
The operational link between Sec.~\ref{sec:edge_detection} and Sec.~\ref{sec:mitigation_primitives} is a hand-off protocol whose properties determine whether the closed loop runs end-to-end~\cite{bilal2026netops}. 
We model the hand-off as a four-tuple $\langle \text{score}, \text{class}, \text{evidence}, \text{ttl}\rangle$ emitted by the detector and consumed by the orchestrator. 
The \emph{score} is the calibrated probability that the observed traffic is malicious. 
The \emph{class} is the MITRE technique label (Table~\ref{tab:mitre}) that selects the playbook. 
The \emph{evidence} carries the per-feature attribution (Sec.~\ref{sec:substrates}) needed for the orchestrator's downstream rationale and for forensic retention. 
The \emph{ttl} bounds the alert's validity window, after which the orchestrator must request a refresh; this is what prevents stale alerts from triggering mitigations after the underlying traffic has already cleared. 
The hand-off protocol is the natural locus of standardisation in Challenge~C4 (Sec.~\ref{sec:c4}); the SDN, NFV, and O-RAN primitives below are organised by the tuple field each consumes. 

\subsection{SDN-, NFV-, and O-RAN-Based Mitigation Primitives}
\label{sec:mitigation_primitives}

\textbf{SDN-based mitigation} centralises forwarding decisions in a controller that can install or revoke flow rules in seconds: blacklisting source IPs, redirecting flows to scrubbing services, or isolating compromised slices~\cite{yan2016sdnsec_survey, scott2016sdn_ddos, allaw2025cross_layer}. 
The strength is global view; the weakness is controller scalability and the new attack surface of the controller itself. Cooperative SDN architectures~\cite{yan2016sdnsec_survey} federate decisions across controllers to extend reach, and flow-rule programming via OpenFlow/P4~\cite{abubakar_2024_5gdad} pushes mitigation as far down the stack as the data-plane switch will permit. 
The mitigation playbook---rate-limit, redirect-to-scrubber, isolate-slice, drop---is selected from the hand-off class field (Sec.~\ref{sec:mitre}) and parametrised with the false-block-rate and slice-availability constraints of Table~\ref{tab:mitigation_metrics}.

\textbf{NFV-based mitigation} virtualises mitigation appliances (firewalls, scrubbers, deep packet inspection) and elastically scales them in response to attack volume. 
Bohatei~\cite{fayaz2015bohatei} demonstrated elastic NFV-based DDoS scrubbing with sub-minute provisioning. Semantic SDN/NFV policy frameworks~\cite{zarca2020semantic_sdn} compose primitives at higher abstraction. 
The operational appeal is that mitigation capacity scales with attack capacity rather than being statically provisioned, but the latency to instantiate a new VNF (typically tens of seconds even with optimisations) is too coarse for URLLC-slice protection on its own; NFV is therefore best paired with a faster SDN/O-RAN reflex layer.

\textbf{O-RAN-based mitigation} is the natively-cellular path: xApps on the Near-RT RIC can suppress malicious UEs, modify scheduler weights, or trigger handovers within the $10$--$100$\,ms RIC budget~\cite{bonati2021oran, polese2023oran, oran_alliance_spec, moore2025, yang_2025_zerotouchsec, nguyen2025rrc, azkaei2025AD}. 
xApps for layer-3 attack detection (5G-SPECTOR~\cite{wen2024oran}, RRC-storm~\cite{nguyen2025rrc}) are now operational on testbeds, and the anomaly detection/traffic-steering xApp pattern is widely documented~\cite{agarwal2026}. 
Because the RIC controls scheduling and admission directly, an xApp can degrade service to a suspicious UE without dropping its connection, preserving the operator's ability to forensically observe the attack while neutralising its impact---an option not available to upstream SDN-only defences. 
The architectural mapping of these O-RAN and SDN mitigation hooks onto the disaggregated RAN control plane is illustrated in Fig.~\ref{fig:oran_hooks_overlay}, which highlights how actuation points natively trade latency against blast radius. 
Surveys of O-RAN architecture and security~\cite{abdalla2022oran_security, bonati2020oran_survey, polese2023oran} caution that the same openness that enables this also expands the supply-chain surface; xApp lifecycle attestation (Sec.~\ref{sec:c4}) is therefore an open standardisation problem.

\textbf{Hybrid composition} is the dominant pattern in 2024--25: SDN steers, NFV scales, O-RAN xApps act at the air-interface, and an AI orchestrator picks the playbook. 
Sheibani \textit{et al.}~\cite{sheibani2024multilayer} formalise the multi-layer defence. 
Allaw et al.~\cite{allaw2025cross_layer} sketch a cross-layer SDN/NFV/AI mitigation framework for 5G/6G slices, with the AI-detection stage validated on UNSW-NB15.
The orchestrator's job is to map a detected MITRE technique to a primitive, choose the actuator tier whose latency fits the slice's budget, and roll back if the false-block rate exceeds a slice-specific threshold. 
Recent work treats the orchestrator itself as a learned policy~\cite{nguyen2023deep_rl_survey}, with reinforcement learning over a discretised action space of (primitive, target-tier, intensity) triples; this lifts the orchestrator out of static playbooks but reintroduces explainability and safety questions that XAI (Sec.~\ref{sec:substrates}) and DT validation (Sec.~\ref{sec:dt}) must answer before production rollout.


\textbf{AI-native orchestration} is the integrating layer. Each MITRE technique class produced by the detector (Sec.~\ref{sec:mitre}) maps to one or more primitives. The orchestrator chooses among primitives by minimising a slice-aware cost
\begin{equation}
J(a) = w_1\,\Delta t_{\text{mit}}(a) + w_2\,\mathrm{FBR}(a) + w_3\,\Delta t_{\text{rec}}(a) + w_4\,\mathrm{cost}(a),
\label{eq:orch_cost}
\end{equation}
where the weights $w_1,\dots,w_4$ are slice-specific (URLLC slices weight $\Delta t_{\text{mit}}$ and FBR much more heavily than mMTC slices). 
This formulation cleanly composes the metrics of Table~\ref{tab:mitigation_metrics}; the multi-layer SDN/NFV defence of~\cite{sheibani2024multilayer} approximates it operationally and the cross-layer architecture of~\cite{allaw2025cross_layer} sketches the same composition at the design level, while the binary-isolation pattern of~\cite{moore2025} is a special case where $w_1$ dominates and the action set collapses to a single primitive.

\subsection{Mitigation Performance Metrics}
\label{sec:mitigation_metrics}
The mitigation literature has historically optimised three quantities---detection accuracy, block-effectiveness, and post-attack recovery time---and reported each in isolation from the others. 
For 6G CPS, this decomposition is insufficient because it omits the downstream physical consequence that is the ultimate object of concern. 
A mitigation that blocks a volumetric flood with $99$\% effectiveness but delays a vehicular collision-warning message by $4$\,ms can still cause harm; a mitigation that reports a $200$\,ms recovery time but collaterally blocks a legitimate telesurgery session is operationally worse than the attack it contains. 
We therefore adopt a CPS-aware metric set, summarised in Table~\ref{tab:mitigation_metrics}, in which every metric is paired with the slice-class envelope against which it must be evaluated.

Five first-order metrics recur across the surveyed closed-loop systems and form the basis of our comparative analysis in Table~\ref{tab:mitigation_comparison}, with full definitions and CPS relevance consolidated in Table~\ref{tab:mitigation_metrics}.
\emph{Mitigation latency} $\Delta t_{\text{mit}}$ is tail-bounded for URLLC, not mean-bounded, since a mitigation that fires in $10$\,ms on average but occasionally spikes to $100$\,ms is unusable for V2X collision avoidance~\cite{abdulqadder2020sdn5g}.
The CPS-weighted \emph{false block rate} (FBR), unlike the detector-level false-positive rate (FPR), is weighted by the traffic's CPS relevance, so a false block on a meter reading and a false block on a cardiac telemetry stream enter the metric with different costs~\cite{sheibani2024multilayer}.

\emph{Slice availability} under attack is negotiated into the CPS slice tenant's SLA and ultimately optimised by the orchestrator in Eq.~\ref{eq:orch_cost}.
\emph{Recovery time} $\Delta t_{\text{rec}}$ includes the time to release rate-limits, re-admit blocked sources, and restore scheduler weights (the reversibility cost of Sec.~\ref{sec:mitigation_gap}).
Finally, \emph{resource overhead} bounds the operator's ability to scale the defence to network-wide attacks.

Beyond these five first-order metrics, two compound metrics are emerging as the credible operational targets of closed-loop systems. 
The \emph{CPS-impact harm score} fuses mitigation latency, FBR, and slice availability through a slice-specific cost function that encodes the physical consequence of each outcome (a missed V2X warning, an unscheduled relay trip, a delayed defibrillation command); no public dataset labels this quantity today, which is why Challenge~C5 (Sec.~\ref{sec:c5}) names its construction as a priority. 
The \emph{closed-loop drift indicator} captures the rate at which the detector's score distribution on production telemetry diverges from the distribution on the FL-aggregated training set; when the indicator crosses a slice-aware threshold, the orchestrator must either escalate to a larger reference model~\cite{agarwal2026}, trigger a targeted DT replay (Sec.~\ref{sec:dt})~\cite{rahmani2026gnn}, or degrade gracefully to a conservative fallback policy. 
Both compound metrics are operational refinements of the five first-order metrics rather than replacements, and they are the natural reporting surface for the benchmark agenda of Sec.~\ref{sec:c5}.

An important methodological caveat is that the metrics in Table~\ref{tab:mitigation_metrics} are not independent: an orchestrator that minimises $\Delta t_{\text{mit}}$ by firing the fastest available actuator will typically raise the FBR (because the fastest actuator has the coarsest granularity), and an orchestrator that minimises FBR by waiting for high-confidence evidence will typically raise $\Delta t_{\text{mit}}$. 
The Pareto frontier of these trade-offs is slice-specific, and the operational pattern is to expose the slice tenant's preferred point on the frontier as a per-slice configuration rather than as a global operator choice. 
This preference encoding is what lets the same closed loop serve a URLLC vehicular slice (FBR-tolerant, latency-intolerant) and an mMTC metering slice (latency-tolerant, FBR-intolerant) from the same detector and the same actuator pool, and it is the metric-level realisation of the graded-enforcement principle we return to in Sec.~\ref{sec:substrates}.

\begin{table}[!t]
\centering
\caption{Mitigation Performance Metrics and CPS Relevance}
\label{tab:mitigation_metrics}
\renewcommand{\arraystretch}{1.15}
\scriptsize
\setlength{\tabcolsep}{3pt}
\begin{tabular}{|p{2.0cm}|p{2.7cm}|p{2.5cm}|}
\hline
\textbf{Metric} & \textbf{Definition} & \textbf{CPS Relevance} \\
\hline
Mitigation latency $\Delta t_{\text{mit}}$ & Time from alert to attack suppression & Critical for URLLC (V2X, surgery) \\
\hline
False Block Rate (FBR) & Fraction of legitimate traffic blocked & Safety hazard if CPS control traffic blocked \\
\hline
Slice availability & Up-time of target slice during attack & Direct service-level metric \\
\hline
Recovery time $\Delta t_{\text{rec}}$ & Time from suppression to nominal QoS & Bounds CPS downtime \\
\hline
Resource overhead & Compute/bandwidth used by mitigation & Limits scaling to network-wide attacks \\
\hline
\end{tabular}
\vspace{-17pt}
\end{table}

\noindent\textbf{Operational takeaways.} Five lessons recur across the deployed frameworks:
\begin{itemize}[leftmargin=*, nosep]
\item \emph{Latency is bounded by the slowest tier:} commit the fast tier (O-RAN xApp) first; use slower tiers (SDN, NFV) only for sustained mitigation.
\item \emph{False-positive cost is slice-asymmetric:} treat slice class as a first-class input to the orchestrator cost (Eq.~\ref{eq:orch_cost}).
\item \emph{Rollback is as important as forward action:} a loop without a budgeted rollback path is operationally incomplete.
\item \emph{Cross-tier feedback dominates per-tier tuning:} post-mitigation telemetry is the only direct evidence that the chosen action was correct.
\item \emph{Operator-specific tuning is unavoidable:} deployed systems report a 2--6~month tuning period (implications for C3 and C5).
\end{itemize}

\subsection{Comparative Analysis}
\label{sec:mitigation_comparison}

Table~\ref{tab:mitigation_comparison} compares the most cited mitigation frameworks across the SDN/NFV/O-RAN axes and along the AI/automation axis.
The trajectory is clear: early work is either survey-level or rule-driven~\cite{yan2016sdnsec_survey, scott2016sdn_ddos, fayaz2015bohatei}, whereas from 2020 onward SDN/NFV and O-RAN frameworks increasingly close the detect-then-act loop with AI~\cite{abdulqadder2020sdn5g, zarca2020semantic_sdn, polese2023oran, sheibani2024multilayer, allaw2025cross_layer, moore2025}.
This shift toward AI-orchestrated hybrid stacks is the empirical basis for the closed-loop thesis of this survey, yet the empty CPS column shows that none of these frameworks treats cyber-physical impact as a first-class goal---the gap this survey targets.

\begin{table*}[!t]
\centering
\caption{Comparative Analysis of Network-Wide DDoS Mitigation Frameworks.}
\label{tab:mitigation_comparison}
\renewcommand{\arraystretch}{1.1}
\scriptsize
\setlength{\tabcolsep}{4pt}
\begin{tabular}{|c|c|l|c|c|c|c|c|c|c|c|}
\hline
\textbf{Reference} & \textbf{Year} & \textbf{Strategy} & \textbf{SDN} & \textbf{NFV} & \textbf{O-RAN} & \textbf{Latency} & \textbf{CPS} & \textbf{5G/6G} & \textbf{Auto} & \textbf{AI} \\
\hline
\cite{fayaz2015bohatei} & 2015 & Elastic NFV defense (Bohatei) & \yes & \yes & \no & Low & \no & \no & \partialc & \no \\
\hline
\cite{yan2016sdnsec_survey} & 2016 & SDN-based DDoS defense (survey) & \yes & \no & \no & -- & \no & \no & \no & \no \\
\hline
\cite{scott2016sdn_ddos} & 2016 & SDN security (survey) & \yes & \no & \no & -- & \no & \no & \no & \no \\
\hline
\cite{abdulqadder2020sdn5g} & 2020 & SDN+NFV multi-layer & \yes & \yes & \no & Low & \no & \yes & \partialc & \yes \\
\hline
\cite{zarca2020semantic_sdn} & 2020 & Semantic SDN/NFV policy & \yes & \yes & \no & Medium & \no & \no & \yes & \yes \\
\hline
\cite{polese2023oran} & 2023 & O-RAN architecture \& security & \partialc & \partialc & \yes & Low & \no & \yes & \yes & \yes \\
\hline
\cite{sheibani2024multilayer} & 2024 & Multi-layer defence & \yes & \yes & \no & Medium & \no & \yes & \yes & \yes \\
\hline
\cite{allaw2025cross_layer} & 2025 & Cross-layer arch. (SDN/NFV/AI) & \yes & \partialc & \no & -- & \no & \yes & \yes & \yes \\
\hline
\cite{moore2025} & 2025 & xApp closed loop (over-the-air) & \no & \no & \yes & Low & \no & \yes & \yes & \yes \\
\hline
\end{tabular}
\noindent\begin{minipage}{\textwidth}
\footnotesize
\textit{Note:} $\bullet$ = supported; \partialc = partial / rule-based rather than fully AI-driven; $\circ$ = not supported or not reported.
\emph{Latency} is the qualitative end-to-end mitigation budget reported by each framework (Low~$\le$~100~ms, Medium~$\le$~1~s).
\end{minipage}
\vspace{-15pt}
\end{table*}

\subsection{The Unified Closed-Loop Reference Architecture}
\label{sec:closed_loop}
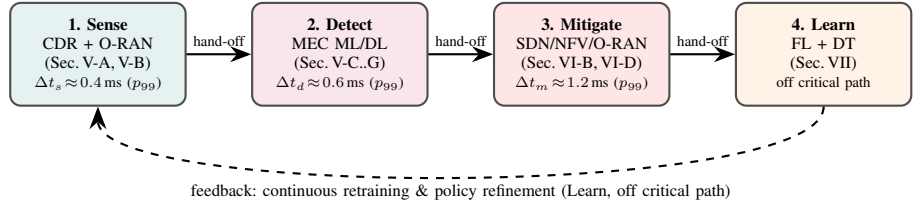
\begin{figure*}[!t]
\centering
\begin{tikzpicture}[
    scale=0.8, transform shape,
    every node/.style={font=\footnotesize},
    stage/.style={rectangle, draw=black!70, rounded corners=4pt,
                  minimum height=1.7cm, minimum width=2.9cm, align=center, thick},
    arr/.style={-{Stealth[length=2.8mm]}, thick},
    bigarr/.style={-{Stealth[length=3mm]}, thick, dashed, black},
]
\useasboundingbox (-0.4,-1.0) rectangle (16.6,3.2);
\node[stage, fill=teal!10]   (sense)   at ( 1.6, 1.6) {\textbf{1.\ Sense}\\CDR + O-RAN\\(Sec.\,V-A,\,V-B)\\{\scriptsize $\Delta t_{s}\!\approx\!0.4$\,ms ($p_{99}$)}};
\node[stage, fill=purple!10] (detect)  at ( 5.6, 1.6) {\textbf{2.\ Detect}\\MEC ML/DL\\(Sec.\,V-C..G)\\{\scriptsize $\Delta t_{d}\!\approx\!0.6$\,ms ($p_{99}$)}};
\node[stage, fill=red!10]    (orch)    at ( 9.6, 1.6) {\textbf{3.\ Mitigate}\\SDN/NFV/O-RAN\\(Sec.\,VI-B,\,VI-D)\\{\scriptsize $\Delta t_{m}\!\approx\!1.2$\,ms ($p_{99}$)}};
\node[stage, fill=orange!10] (learn)   at (13.6, 1.6) {\textbf{4.\ Learn}\\FL + DT\\(Sec.\,VII)\\{\scriptsize off critical path}};
\draw[arr] (sense)  -- node[above, font=\scriptsize, black] {hand-off} (detect);
\draw[arr] (detect) -- node[above, font=\scriptsize, black] {hand-off} (orch);
\draw[arr] (orch)   -- node[above, font=\scriptsize, black] {hand-off} (learn);
\draw[bigarr] (learn.south)
    .. controls (13.6,-0.8) and (1.6,-0.8) ..
   (sense.south)
    node[pos=0.5, below, black] {feedback: continuous retraining \& policy refinement (Learn, off critical path)};
\end{tikzpicture}
\caption{Unified closed-loop reference architecture for AI-native 6G CPS security. 
The architecture replaces the five-pillar list of prior surveys with one continuously-running loop. 
Per-stage annotations are p99 stage latencies $\Delta t_s, \Delta t_d, \Delta t_m$ for the safety-critical V2X worked example of Sec.~\ref{sec:closed_loop_contract} (the percentile mandated for URLLC);
their slice-admissibility is governed by Eq.~\ref{eq:latency_contract}, not by their unweighted sum. 
The Learn stage is off the critical path.}
\label{fig:detection_mitigation}
\vspace{-10pt}
\end{figure*}

Fig.~\ref{fig:detection_mitigation} instantiates the four-stage thesis of Sec.~\ref{sec:intro_ai_native} as a concrete reference architecture.
Sensing is on CDR/O-RAN telemetry at the aNB and MEC tier; detection runs the model families of Sec.~\ref{sec:edge_method_families}; mitigation composes SDN/NFV/O-RAN actuators per detected MITRE technique class (Sec.~\ref{sec:mitre}); learning is FL- and DT-driven retraining (Sec.~\ref{sec:enablers}). 
The closure of the loop---explicit, in-protocol, and continuous---is what the prior literature has been missing.

The loop has three operational properties.
\emph{End-to-end latency budgeting}: each stage is annotated with its target latency, and the sum bounds the slice's safety envelope; a violation in any stage propagates to the next, so optimisation must be loop-wide rather than stage-local~\cite{bonati2021oran}. 
\emph{Telemetry-as-feedback}: the orchestrator's actions become part of the next sense step, so the detector continuously sees how the system responds to its alerts; this is what makes drift detection and on-line model selection feasible at the MEC tier. 
\emph{Federated learning as the lifecycle bus}: per-MEC detectors are continuously refined by FL aggregation of local gradients, and DT replay supplies adversarial scenarios that the operational telemetry does not contain~\cite{mcmahan2017fedavg, alwis2026, blika2024fl_6g}.

The operational evidence for the loop is now sufficient to claim feasibility. 
Moore \textit{et al.}~\cite{moore2025} report a complete sense-detect-mitigate cycle on an over-the-air O-RAN testbed with sub-second mitigation latency; Allaw \textit{et al.}~\cite{allaw2025cross_layer} sketch a cross-layer SDN+NFV+AI architecture and validate the AI-detection stage on UNSW-NB15; Christopoulou \textit{et al.}~\cite{christopoulou_2024_ddos5gnwdaf} show that the NWDAF can host the detection step inside a standardised cellular interface.

\noindent\textbf{Differentiator: a latency-contracted, falsifiable closed loop.} \label{sec:closed_loop_contract}
The four-stage loop above is, in shape, isomorphic to MAPE-K \cite{kephart2003autonomic}, OODA, and the O-RAN RIC closed-loop hierarchy~\cite{polese2023oran}. 
The survey's contribution is therefore \emph{not} the four-stage decomposition itself but a falsifiable contract that ties the loop to the URLLC slice's safety envelope.

\paragraph{Latency contract.} Let $\Delta t_s$, $\Delta t_d$, $\Delta t_m$ denote the per-stage latencies of Sense, Detect and Mitigate measured at a slice-dependent tail percentile $p$, and let $\tau_{\max}$ be the slice's end-to-end safety envelope ($1$--$5$\,ms for URLLC, $10$--$100$\,ms for eMBB, seconds for mMTC).
We set $p=p_{99}$ for safety-critical slices (URLLC vehicular, telesurgery, and grid protection), where a one-in-twenty budget violation is not tolerable, and relax to $p=p_{95}$ for latency-tolerant eMBB/mMTC classes; $p_{99.9}$ is the appropriate target where a slice's reliability requirement ($10^{-9}$ packet error for URLLC) dominates, at the cost of requiring substantially more telemetry to estimate the tail reliably.
The closed loop is \emph{slice-admissible} iff
\begin{equation}
\Delta t_s + \Delta t_d + \Delta t_m \;\le\; \tau_{\max} - \Delta t_{\text{ho}}
\quad\text{at}\;p,
\label{eq:latency_contract}
\end{equation}
where $\Delta t_{\text{ho}}$ is the four-tuple hand-off serialisation cost
(Sec.~\ref{sec:handoff}).
The tail-percentile choice distinguishes the contract from a mean-latency claim, since tail excursions---not the average---are the operational failure mode.
Two caveats make the contract honest.
First, summing per-stage tail latencies does not in general equal the end-to-end tail: $\sum_i \Delta t_i\,@\,p$ equals the chain's true $p$-percentile only when the stage latencies are perfectly positively correlated, and otherwise upper-bounds it.
The contract is therefore analytic rather than empirical: Eq.~\eqref{eq:latency_contract} is a conservative, sufficient admission test derived from per-stage tail bounds, not a measurement of the end-to-end distribution. 
A satisfied bound certifies admissibility; a violated bound mandates direct end-to-end percentile measurement before the slice is rejected.
Second, the percentile is admission-time conservative by design, because a false admission of a safety-critical slice is costlier than a false rejection.
The Learn stage is off the critical path.
Making the contract \emph{per-slice} and \emph{tail-bounded} is what MAPE-K and OODA leave undefined and what the O-RAN RIC closed-loop specification leaves operator-discretionary.

\paragraph{Worked V2X example ($\tau_{\max}=3$\,ms collision-warning slice).}
Because this is a safety-critical URLLC slice, the contract is evaluated at $p_{99}$.
\textbf{Sense:} the MEC host pulls KPMs over E2 with $\Delta t_s\!\approx\!0.4$\,ms.
\textbf{Detect:} a quantised LUCID-class detector~\cite{doriguzzi2020lucid} scores a $32$-flow batch in $\Delta t_d\!\approx\!0.6$\,ms.
\textbf{Mitigate:} the orchestrator emits a scheduler-weight modification to the gNB MAC scheduler via E2 RIC-control, taking $\Delta t_m\!\approx\!1.2$\,ms including Remote Procedure Call (RPC).
With $\Delta t_{\text{ho}}\!\approx\!0.1$\,ms, the summed $p_{99}$ bound consumes $2.3$\,ms of the $3$\,ms budget; because this bound is conservative, the true end-to-end $p_{99}$ is also within budget and the loop is admissible.
If the same slice instead required SDN flow-rule installation ($\Delta t_m\!\approx\!50$\,ms~\cite{abdulqadder2020sdn5g}), Eq.~\ref{eq:latency_contract} fails by an order of magnitude and the loop must fall back to the O-RAN xApp tier or admission-reject the load.
This is a falsifiable architectural property, not an empirical observation.

\subsection{Reference Implementation Sketch}
\label{sec:reference_impl}
To make the closed-loop architecture concrete, we sketch the components that would compose a minimal but realistic instantiation on commodity hardware. 
The MEC tier hosts a lightweight detector (a quantised CNN for spatial CDR features and a compressed LSTM for temporal sequences) co-located with an orchestrator agent that consumes the four-tuple hand-off (Sec.~\ref{sec:handoff}) and selects from a finite playbook. 
The Near-RT RIC hosts an xApp registered with the detector's MITRE technique vocabulary (Table~\ref{tab:mitre}), exposing a small RPC surface for scheduler-weight modification, UE suppression, and handover triggers~\cite{bonati2021oran, polese2023oran, moore2025}. 
The transport tier exposes an SDN controller that accepts authenticated flow-rule installations from the orchestrator, with a P4-programmable data-plane fast-path for high-rate drop and rate-limit primitives~\cite{abubakar_2024_5gdad}. 
The core network exposes the NWDAF event surface for cross-MEC visibility~\cite{christopoulou_2024_ddos5gnwdaf}, and an FL aggregator runs in a privacy-isolated enclave to consume gradient updates from the per-MEC detectors~\cite{mcmahan2017fedavg, blika2024fl_6g}.

The sketch is deliberately conservative: every component above is empirically demonstrated in the surveyed literature. 
Doriguzzi-Corin \textit{et al.}~\cite{doriguzzi2020lucid} and Franco-Valiente \textit{et al.}~\cite{franco2026} show that lightweight CNN and energy-aware ARM-SoC pipelines achieve sub-10\,ms inference on commodity edge-class hardware, and Hussain \textit{et al.}~\cite{hussain2019mec, hussain2020mec} establish the deep CDR-detector accuracy on real operator data that such edge runtimes must reproduce;
Moore \textit{et al.}~\cite{moore2025} demonstrate the xApp orchestrator path on an over-the-air O-RAN testbed; Christopoulou \textit{et al.}~\cite{christopoulou_2024_ddos5gnwdaf} exhibit NWDAF-hosted detection inside a standardised cellular interface; Allaw \textit{et al.}~\cite{allaw2025cross_layer} report a cross-layer SDN+NFV orchestrator. 
The novelty here is the integration: previous work has demonstrated each piece in isolation; the closed loop's value is the composition. 
The five outstanding integration questions---xApp attestation (C4), cross-operator FL governance (C3), audit-grade XAI under adversarial probing (C2), CPS-impact harm metrics (C5), and certified sub-millisecond detection under model drift (C1)---are exactly the consolidated open challenges of Sec.~\ref{sec:open_challenges}.

\section{Cross-Cutting Enablers}
\label{sec:enablers}
The closed loop of Fig.~\ref{fig:detection_mitigation} is sustained by three operational services---FL, LLMs, and DTs (Secs.~\ref{sec:fl}--\ref{sec:dt})---and a three-layer substrate of PQC, ZTA, and XAI (Secs.~\ref{sec:substrates}--\ref{sec:substrate_synthesis}). 
The services produce training data, summaries, and counterfactuals for the loop; the substrate makes the loop's transactions confidential, trusted, and auditable. We treat each below in its operational role inside the loop.

\subsection{Federated Learning: Training Across Operators Without Sharing CDRs}
\label{sec:fl}
\begin{figure}[!t]
\centering
\begin{tikzpicture}[
    every node/.style={font=\scriptsize},
    server/.style={rectangle, draw=blue!70!black, fill=blue!8, rounded corners=2pt,
                   minimum height=0.7cm, minimum width=2.2cm, align=center, thick},
    client/.style={rectangle, draw=teal!70!black, fill=teal!8, rounded corners=2pt,
                   minimum height=0.6cm, minimum width=1.4cm, align=center, thick},
    arr/.style={-{Stealth[length=2mm]}, thick},
]
\node[server] (s) at (3,2.2) {Aggregator (telco core)};
\node[client] (c1) at (0,0.6) {MEC$_1$};
\node[client] (c2) at (2,0.6) {MEC$_2$};
\node[client] (c3) at (4,0.6) {MEC$_3$};
\node[font=\large] (dots) at (5.2,0.6) {$\cdots$};
\node[client] (cn) at (6.4,0.6) {MEC$_n$};
\foreach \c in {c1,c2,c3,cn} {
  \draw[arr] (\c.north) -- (s.south);
}
\node[font=\scriptsize, gray, anchor=west] at (6.8,1.4) {$\Delta\theta$ only};
\end{tikzpicture}
\caption{Federated learning topology for distributed 6G CPS security.
The $n$ MEC clients ($\text{MEC}_1,\ldots,\text{MEC}_n$) each train locally; CDRs never leave the MEC host, and only gradient updates $\Delta\theta$ are exchanged with the aggregator at the telco core.}
\label{fig:fl_architecture}
\vspace{-10pt}
\end{figure}
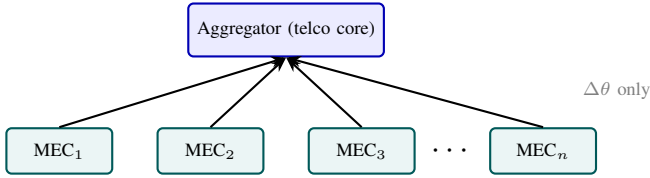

FL is the operational answer to the labelled-CDR scarcity problem identified in Sec.~\ref{sec:edge_method_families}: each MEC node trains a local model on its own CDRs (labelled by the self-supervised/pseudo-labelling means of Sec.~\ref{sec:edge_method_families}, not manual annotation) and shares only model updates with a central aggregator (Fig.~\ref{fig:fl_architecture})~\cite{mcmahan2017fedavg, kairouz2021advances_fl,you2021towards_6g, nguyen2022survey}. 
The architectural fit with MEC is exact---one client per host, one round per training window, secure aggregation across hosts.


\textbf{Algorithm variants.} FedAvg~\cite{mcmahan2017fedavg} is the baseline; FedProx~\cite{li2020fedprox} stabilises training under client heterogeneity; FedAGRU (attention-GRU FL)~\cite{chen2020fl_ids} reaches $\sim$97\% on KDDCup99/CICIDS2017/WSN-DS. 
Federated GRU per device type~\cite{nguyen2019fl_ddos} achieves $>$95\% on a 33-device IoT testbed.

\textbf{Byzantine resilience.} Single-server secure aggregation with multi-Krum-style outlier removal~\cite{so2021byzantine_fl, wan2024survey} bounds the influence of malicious clients while preserving per-client privacy; empirical evaluations on non-IID device-partitioned IoT-malware data~\cite{rey2022fl_anomaly} show that standard FedAvg averaging is broken by a single Byzantine client (gradient-factor and model-cancelling attacks reduce the model to a constant predictor) while coordinate-wise median aggregation tolerates one or two malicious clients but still collapses at three out of eight; this matters operationally because compromised MEC hosts are an in-scope threat~\cite{wan2024survey} (Sec.~\ref{sec:attacker_model}).

\textbf{Privacy and communication efficiency.} Secure aggregation~\cite{bonawitz2019fedlearn} and differentially private FedAvg~\cite{wei2020differential_fl} reduce information leakage; gradient compression and asynchronous aggregation reduce communication overhead~\cite{zhang2025gccd}. 
Federated autoencoders and MLP classifiers detect IoT-botnet (Mirai/BASHLITE) malware on N-BaIoT at $\sim$99.9\% supervised accuracy and $\sim$99.98\% TPR (92--97\% TNR) unsupervised, including on devices unseen during training, but standard FedAvg averaging is broken by a single Byzantine client~\cite{rey2022fl_anomaly}; non-IID heterogeneity is the central practical headache~\cite{zhao2018noniid_fl, wan2024survey}.

\textbf{For 6G CPS specifically.} BERT-based federated IDS for 5G~\cite{adjewa_2024_bertfedids}, FL-enhanced cyber-security~\cite{blika2024fl_6g}, and the full FL-for-6G survey~\cite{alwis2026}, together with the focused FL security and privacy taxonomy of~\cite{mothukuri2021survey_fl}, confirm both feasibility and the need for new aggregation protocols when participants are competing operators across regulatory domains. 
Table~\ref{tab:fl_comparison} consolidates the empirical evidence.

\textbf{Convergence under non-IID CDR distributions.} The convergence guarantees of FedAvg~\cite{mcmahan2017fedavg} are weakened when client data distributions are heterogeneous~\cite{kairouz2021advances_fl}, which is the default regime for cellular CDRs: traffic on an urban MEC host differs structurally from a rural host, a tourist-heavy coastal cell, or an industrial campus slice. 
FedProx~\cite{li2020fedprox} adds a proximal regulariser $\tfrac{\mu}{2}\|\theta_k - \theta^{(t)}\|^2$ to each client's local objective to bound local drift, and its convergence analysis under bounded dissimilarity is the formal result that justifies its empirical robustness on non-IID FEMNIST and derivative cellular workloads. 
The practical knobs for 6G CPS deployments are (i)~the proximal strength $\mu$, tuned per-round from measured client-drift variance; (ii)~the local-epoch count $E$, which trades communication cost against local-drift accumulation; and (iii)~the client-sampling strategy, where stratified sampling across operator-role archetypes (urban/rural/industrial/transit) materially reduces round-to-round variance compared to uniform sampling. 
Empirically, non-IID CDR federations converge in 3--10$\times$ the rounds of IID baselines~\cite{zhao2018noniid_fl, li2020fedprox, chen2020fl_ids}, and the operational cost is the control-plane bandwidth consumed by additional aggregation rounds rather than per-round compute. 
The research direction in Sec.~\ref{sec:c3} is to combine FedProx-style drift control with Byzantine-robust aggregation~\cite{so2021byzantine_fl} and PQC-protected secure aggregation~\cite{bonawitz2019fedlearn, fips204_dilithium} inside a single protocol whose convergence, privacy budget, and Byzantine tolerance are all analysed jointly rather than bolted on one axis at a time.

\begin{table*}[!t]
\centering
\caption{Federated Learning Methods for 6G CPS Security}
\label{tab:fl_comparison}
\renewcommand{\arraystretch}{1.15}
\scriptsize
\setlength{\tabcolsep}{3pt}
\begin{tabular}{|l|c|p{3.2cm}|l|l|c|c|c|c|}
\hline
\textbf{Ref.} & \textbf{Year} & \textbf{FL Algorithm} & \textbf{Security Task} & \textbf{Dataset} & \textbf{Acc.\ (\%)} & \textbf{Privacy} & \textbf{Comm.\ Eff.} & \textbf{Non-IID} \\
\hline
\cite{mcmahan2017fedavg} & 2017 & FedAvg & General classification & MNIST/CIFAR & 99.4/85.0 & \Circle & \CIRCLE & \CIRCLE \\
\hline
\cite{zhao2018noniid_fl} & 2018 & FedAvg + data sharing & Non-IID mitigation & CIFAR-10 & $\sim$78 & \Circle & \Circle & \CIRCLE \\
\hline
\cite{nguyen2019fl_ddos} & 2019 & Federated GRU per device & IoT device-type anomalies & Real IoT (33 dev.) & $>$95 & \Circle & \CIRCLE & \Circle \\
\hline
\cite{bonawitz2019fedlearn} & 2019 & SecAgg + FedAvg & Mobile keyboard prediction & Production & -- & \CIRCLE & \CIRCLE & \Circle \\
\hline
\cite{wei2020differential_fl} & 2020 & NbAFL (DP-FedAvg) & Privacy-preserving FL (general) & MNIST & -- & \CIRCLE & \Circle & \Circle \\
\hline
\cite{chen2020fl_ids} & 2020 & FedAGRU (Attn-GRU) & Wireless-edge IDS & KDDCup99/CICIDS/WSN-DS & $\sim$97 & \Circle & \CIRCLE & \CIRCLE \\
\hline
\cite{li2020fedprox} & 2020 & FedProx & Heterogeneous FL & FEMNIST & -- & \Circle & \Circle & \CIRCLE \\
\hline
\cite{so2021byzantine_fl} & 2021 & BREA (single-server multi-Krum + VSS + RS-aggregation) & Byzantine-resilient secure FL & MNIST / CIFAR-10 & -- & \Circle & \Circle & \CIRCLE \\
\hline
\cite{rey2022fl_anomaly} & 2022 & Federated AE + MLP (FedAvg) & IoT-botnet detection & N-BaIoT & 99.9 & \CIRCLE & \CIRCLE & \Circle \\
\hline
\cite{blika2024fl_6g} & 2024 & FL-enhanced security survey & 5G/6G network security & Multiple & -- & \CIRCLE & \CIRCLE & \CIRCLE \\
\hline
\cite{alwis2026} & 2026 & Survey & FL for 6G security & -- & -- & \CIRCLE & -- & \CIRCLE \\
\hline
\multicolumn{9}{l}{\begin{minipage}{0.98\textwidth}\vspace{2pt}\scriptsize \textbf{Capability indicators:} \CIRCLE\ =\ the method explicitly provides or addresses the property; \Circle\ =\ it does not, or is not designed to. \textbf{Privacy:} a formal privacy mechanism beyond baseline FL (e.g., differential privacy, secure aggregation, or encryption). \textbf{Comm.\ Eff.:} the method is designed to reduce communication overhead from exchanging model updates. \textbf{Non-IID:} the method explicitly handles heterogeneous (non-independent-and-identically-distributed) client data. ``--'' denotes a value not reported in the cited work.\end{minipage}} \\
\end{tabular}
\vspace{-12pt}
\end{table*}

\subsection{Large Language Models Should Explain and Query the Loop, Not Control It}
\label{sec:llm}
Two roles for LLMs are emerging in cellular security~\cite{bilal2026netops}. \emph{Telecom-specific foundation models}, pre-trained on protocol specifications, RFCs, and operator runbooks, can ingest raw logs and emit structured anomaly hypotheses, replacing bespoke feature engineering~\cite{ferrag2026}. \emph{Natural-language SOC interfaces} let operators query the loop in English (``which slices saw RRC anomalies in the past hour?'') and receive answers grounded in the supporting telemetry. Houssel \textit{et al.}~\cite{houssel_2024_explainablenids} show that LLMs can produce human-interpretable NIDS reports that surface the features behind a classifier's decision, while also documenting their twin weaknesses---unreliable detection and fabricated facts---that fix the role boundary below.

\paragraph{Latency.}
The first reason to confine LLMs to explain-and-query rather than decide-and-act is timing. Reported inference runs from roughly one second per query on hosted lightweight models to tens of seconds on self-hosted 30\,B+ models~\cite{ferrag2026}, and Houssel \textit{et al.} measure a single 8\,B-model NIDS inference at about three orders of magnitude slower than a lightweight detector---too slow for real-time use~\cite{houssel_2024_explainablenids}. Against the URLLC envelope of Eq.~\ref{eq:latency_contract} ($1$--$5$\,ms), this rules the LLM off the sense--detect--mitigate path entirely. It also dictates placement: the MEC host sized in Sec.~\ref{sec:mec_arch} for a quantised CNN--LSTM detector cannot hold a foundation model, so the LLM lives in the core or a private cloud and is queried by MEC agents, consistent with the clustered-GPU serving that latency-sensitive operation requires~\cite{ferrag2026}. The result is a clean separation of timescales: the millisecond loop runs on the compressed detector and the xApp actuator, while the LLM works the human-supervised SOC timescale of seconds to minutes, where its latency is immaterial---and where streaming first-token output, not full completion, is the figure of merit.

\paragraph{Reliability.}
The second reason is reliability, which differs in kind from detector reliability. A detector fails as a miscalibrated score that calibration and the FBR metric of Table~\ref{tab:mitigation_metrics} already bound; an LLM fails as a fluent fabrication carrying no confidence signal, most likely on the rare out-of-distribution incident the operator can least afford to miss. 
Houssel \textit{et al.} observe exactly this---confident invention of geographic and protocol facts and confusion of protocol numbers with port numbers~\cite{houssel_2024_explainablenids}---which we generalise as protocol, telemetry, and stale-context hallucination. 
None is removed by a larger model; each is bounded only by architecture: retrieval grounding against the operator's own telemetry and runbook corpus, telecom-tuned pre-training, and a hard refusal that returns \texttt{INSUFFICIENT\_EVIDENCE} rather than guessing~\cite{ferrag2026}. The decisive choice is the action boundary---the LLM may summarise, rank, and propose but never command an actuator---so an undetected hallucination degrades to a bad \emph{suggestion} that the hand-off protocol of Sec.~\ref{sec:handoff} and a human or policy-decision-point still gate, not a bad \emph{mitigation} on a live slice. 
Quantifying the residual hallucination rate under distribution shift, and certifying that the refusal envelope holds, is what Challenge~C2 (Sec.~\ref{sec:c2}) names as bounded LLM risk.

We distil from this literature a four-field prompt pattern that binds, rather than eliminates, hallucination: \textbf{telemetry scope} (slice, cell set, window); \textbf{evidence requirement} (at least one telemetry record cited per factual claim); \textbf{refusal envelope} (return \texttt{INSUFFICIENT\_EVIDENCE} with the missing-field list when support is absent); and \textbf{action boundary} (proposals enter the hand-off protocol of Sec.~\ref{sec:handoff} as suggestions gated by a human or policy-decision-point, never as direct orchestrator commands). The two further roles reported in the literature---\emph{policy synthesis} (turning operator intent into orchestrator actions parameterised against Eq.~\ref{eq:orch_cost})~\cite{ferrag2026} and \emph{retrieval-grounded reasoning} over the operator's telemetry and incident history---inherit the same discipline, leaving the open question of composing telecom-tuned models with retrieval under hard refusal envelopes, i.e. bounded LLM risk (Sec.~\ref{sec:c2}).

\subsection{Digital Twins Are the Counterfactual Workbench for the Loop}
\label{sec:dt}
A digital twin (DT) is a synchronised executable replica of a physical CPS that allows the security stack to (i)~test detector and mitigation policies against synthetic adversarial scenarios before deployment, (ii)~run what-if analyses on candidate mitigation playbooks before committing to them on the live network, and (iii)~replay observed incidents to refine the FL training set~\cite{sengendo2026}. 
The DT therefore plays the role of a safety net for the closed loop: it is the place where model drift can be detected by replaying recent traffic and measuring divergence between live-detector and twin-detector outputs. 
The DT closes back to FL for retraining.

For 6G CPS, three DT concepts are converging. 
The \emph{network DT} models the radio, transport, and core; it is the natural place to evaluate orchestrator playbooks for slice availability and recovery time~\cite{sengendo2026}. 
The \emph{CPS DT} models the physical asset (vehicle dynamics, grid topology, robotic-arm kinematics); it is where the harm consequence of a delayed mitigation can be quantified. 
The \emph{security DT} layers attack scenarios on top of both. 
Operationally, the security DT is the only place where Challenge~C5 (CPS-impact harm metrics) can be operationalised today, because no public dataset labels harm; the DT generates harm labels by execution. 
Recent work increasingly co-designs DT pipelines with FL-aggregated update streams so that scenarios deemed high-risk by the DT trigger targeted FL rounds focused on the affected slice, bringing labelled-CDR scarcity (Sec.~\ref{sec:fl}) closer to a workable equilibrium~\cite{sengendo2026}. 
The DT is therefore not a sidecar but a first-class participant in the closed loop's learn stage.

Four operational uses recur: (i)~\emph{pre-deployment validation} against a curated scenario library before any change touches the production network; (ii)~\emph{counterfactual replay} of observed incidents under alternative orchestrator policies; (iii)~\emph{synthetic adversary generation}, which pairs the DT with reinforcement-learning attackers~\cite{nguyen2023deep_rl_survey} or diffusion-based traffic synthesis~\cite{Stoian2026, duan2025} to produce attack scenarios the operational telemetry has not yet seen; and (iv)~\emph{drift attribution}, which disambiguates benign distribution shift from adversarial campaign by replaying recent traffic against a controlled adversary model. No off-the-shelf platform yet integrates all four (Challenge~C5).

\subsection{Substrate-Level Enablers: PQC, ZTA, XAI}
\label{sec:substrates}
Three cross-cutting substrates underpin the loop and are best treated jointly because each is operationally constrained by the same URLLC envelope that bounds the detector and the orchestrator~\cite{you2021towards_6g, salmi2026xai}. 
We restrict attention here to the loop-specific consequences and refer the reader to~\cite{rose2020nist_zta, salmi2026xai, kumar2026} for the underlying primitives.

\emph{Post-quantum cryptography vs.\ the URLLC slice envelope.} The standardisation of Kyber (ML-KEM~\cite{nist2024fips203}) and Dilithium (ML-DSA) by NIST in 2024 settles the algorithmic question; the binding constraint for 6G CPS is the latency overhead at the AMF/AUSF and on the E2 control plane~\cite{3gpp_sa3_security}.
A Dilithium-3 signature is ${\sim}3.3$\,kB per ML-DSA-65 in NIST FIPS~204~\cite{fips204_dilithium}; on reference AVX2 x86 implementations the verification cost is ${\sim}0.1$\,ms and $0.3$--$0.5$\,ms on ARM Cortex-A cores, comfortably within a $1$\,ms slice envelope, with signing the dominant cost at ${\sim}0.3$--$0.6$\,ms. 
Even so, per-message PQC on the sub-millisecond E2 fast path remains marginal once signature transport ($\sim$3.3\,kB) and jitter are accounted for, so session-setup or per-batch granularity remains the safer operating point; harvest-now-decrypt-later threats still motivate PQC-protected FL aggregation and orchestrator control channels, where the retention horizon dominates the latency cost. 
Hybrid classical/PQC handshakes are the dominant transitional pattern in 3GPP working-group discussions~\cite{porambage2021survey}, and per-message PQC on the E2 fast path remains infeasible until signature compression or hardware verification accelerators close the gap (Challenge~C4).

\emph{Zero-trust architecture as graded enforcement keyed to the detector score.} ZTA replaces the perimeter assumption with continuous verification~\cite{nahar_2024_ztasurvey6g, agarwal2026, rose2020nist_zta}. 
For the closed loop, the operationally novel primitive is graded enforcement: the detector's per-flow score (Sec.~\ref{sec:edge_detection})~\cite{chandola2009anomaly} becomes a continuous input to the policy decision point, so that a session whose anomaly score crosses a soft threshold is downgraded to a higher-friction trust tier (additional authentication, lower bandwidth ceiling, restricted slice access) \emph{before} any hard mitigation is invoked; only a hard-threshold crossing triggers termination. 
This converts the binary block/allow decision into a continuum that respects safety-of-life CPS workloads and is the policy-layer complement of the FBR metric in Table~\ref{tab:mitigation_metrics}. 
The open standardisation gap is that 3GPP's 5G-AKA/EAP-AKA' framework~\cite{3gpp_sa3_security} was designed around session-establishment events, not per-decision checks; carrying continuous-authentication evidence without inflating signalling overhead remains a Challenge~C4 item.

\emph{Explainable AI as a trust bridge with dual-use leakage.} Operators and regulators will not delegate URLLC-slice mitigation to a black-box detector, so SHAP/Grad-CAM/Integrated-Gradients attributions become a first-class output of the loop, layered cheaply at the MEC and richly at the operator core~\cite{salmi2026xai, xylouris_2025_predictiveddos,azkaei2025AD}\cite[Chs.~6 and~10]{sarker2024ai}. 
The loop-specific risk is dual-use leakage: per-feature SHAP attributions on a CNN--LSTM detector measurably leak the decision boundary, and under repeated probing converge to a usable adversarial gradient~\cite{milli2019modelreconstruction, salmi2026xai}; post-hoc attribution methods such as LIME and SHAP are themselves adversarially fragile~\cite{slack2020fooling}.
The operational defence is rate-limited or aggregated explanations per consumer identity, audit logging of every explanation request, and PQC-bound retention of the explanation envelope for the full forensic horizon (Challenge~C2). 
XAI is therefore the trust bridge that lets the loop run autonomously while remaining accountable, not a sidecar~\cite{etsi2026ISAC}.

\subsection{Substrate Co-Design: How PQC, ZTA, and XAI Compose}
\label{sec:substrate_synthesis}
\begin{figure}[!t]
\centering
\includegraphics[width=1\columnwidth]{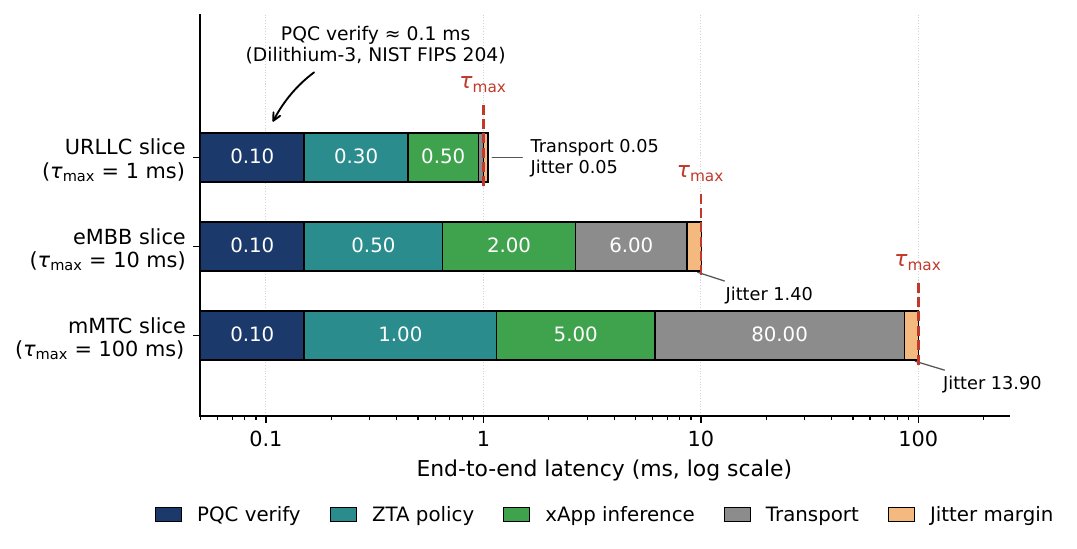}
\caption{End-to-end latency budget per 5G slice. Each row composes PQC verify, ZTA policy lookup, xApp inference, transport, and jitter margin against the slice's $\tau_{\max}$ (dashed red). 
The closed loop must fit within $\tau_{\max}$ for each slice type; the URLLC row is the binding constraint of Eq.~\ref{eq:latency_contract}. 
PQC verify uses Dilithium-3 (ML-DSA-65 per NIST FIPS~204~\cite{fips204_dilithium}) on AVX2 hardware.}
\label{fig:latency_budget_envelope}
\vspace{-10pt}
\end{figure}

The three substrates are not orthogonal: PQC re-keys the control plane that ZTA continuously re-authenticates, and the policy decisions ZTA emits must be auditable by XAI before they reach an xApp. Concretely, a Dilithium-signed NAS-MAC \cite{3gpp_sa3_security, fips204_dilithium} gates a ZTA policy update whose ML-driven trust-score change is in turn explained by a SHAP attribution served to the SOC analyst. The loop-level consequence is that PQC, ZTA, and XAI share a single latency budget (Fig.~\ref{fig:latency_budget_envelope}) and a single audit trail rather than three independent ones.

Widening the lens beyond the substrates, the three operational services (FL, LLM, DT) and the three architectural substrates (PQC, ZTA, XAI) play distinct but coupled roles in the closed loop~\cite{kumar2026}: FL supplies cross-operator training signal without raw-data exchange, LLMs translate natural-language intent into vetted xApp actions, and DTs provide counterfactual validation before any policy reaches production; PQC, ZTA, and XAI then make every resulting control-plane transaction quantum-safe, continuously re-authenticated, and auditable. Together, the six elements operate against a single sub-millisecond budget rather than as independent enhancements---the open question, which Sec.~\ref{sec:open_challenges} formalises, is whether their joint latency, communication, and assurance costs admit a feasible operating point in 6G CPS slices.

\begin{figure*}[!t]
\centering
\footnotesize
\begin{tikzpicture}[
  x=2.55cm,
  y=1.05cm,
  every node/.style={font=\scriptsize},
  milestone/.style={circle, fill=black, inner sep=1.5pt},
  reg/.style={circle, fill=red!85!black, inner sep=1.5pt},
  lanelabel/.style={anchor=east, align=right, font=\scriptsize\bfseries},
  axisyear/.style={anchor=north, font=\scriptsize\bfseries},
  mlabelabove/.style={anchor=south, font=\tiny, align=center, yshift=2pt},
  mlabelbelow/.style={anchor=north, font=\tiny, align=center, yshift=-2pt},
  reglabel/.style={font=\tiny, align=center, text=red!75!black},
]

\foreach \ly/\fc in {-1/black!5, -2/black!10, -3/black!5, -4/black!10, -5/black!5} {
  \fill[\fc] (0, \ly-0.42) rectangle (5, \ly+0.42);
}
\foreach \yr in {0,1,2,3,4,5} {
  \draw[gray!50, very thin] (\yr,-5.5) -- (\yr,-0.55);
}
\foreach \yr/\lbl in {0/2025, 1/2026, 2/2027, 3/2028, 4/2029, 5/2030} {
  \node[axisyear] at (\yr, -5.62) {\lbl};
}
\draw[->, semithick] (-0.05,-5.5) -- (5.15,-5.5);

\node[lanelabel] at (-0.05,-1) {C1: Sub-ms edge detection};
\node[lanelabel] at (-0.05,-2) {C2: XAI / LLM auditability};
\node[lanelabel] at (-0.05,-3) {C3: Cross-operator FL};
\node[lanelabel] at (-0.05,-4) {C4: Standardisation gaps};
\node[lanelabel] at (-0.05,-5) {C5: 6G benchmarks \& harm};

\draw[red!75!black, densely dashed, thin] (1.58,-5.5) -- (1.58,-0.05);
\node[font=\tiny\itshape, text=red!75!black, anchor=south]
  at (1.58,-0.05) {EU AI Act general application (2 Aug 2026)};
\draw[blue!70!black, densely dashed, thin] (3.25,-5.5) -- (3.25,-0.05);
\node[font=\tiny\itshape, text=blue!70!black, anchor=south]
  at (3.25,-0.05) {3GPP Rel-21 freeze (expected 2028; dates TBD)};

\node[milestone] at (0.50,-1) {};   \node[mlabelabove] at (0.50,-1) {O-RAN WG2\\test spec};
\node[milestone] at (2.25,-1) {};   \node[mlabelbelow] at (2.25,-1) {MLPerf-style IDS\\track (proposed)};
\node[milestone] at (4.00,-1) {};   \node[mlabelabove] at (4.00,-1) {Ref. impl.\\target};

\node[reg]       at (2.58,-2) {};   \node[mlabelbelow, reglabel, xshift=-6pt] at (2.58,-2)
  {EU AI Act\\Art.~13 transparency\\(2 Aug 2027)};
\node[milestone] at (3.50,-2) {};   \node[mlabelabove] at (3.50,-2) {NIST AI RMF 2.0};
\node[milestone] at (5.00,-2) {};   \node[mlabelbelow] at (5.00,-2) {ETSI SAI\\XAI profile};

\node[milestone] at (0.75,-3) {};   \node[mlabelabove] at (0.75,-3) {O-RAN WG3\\FL profile};
\node[reg]       at (2.58,-3) {};   \node[mlabelbelow, reglabel, xshift=6pt]  at (2.58,-3)
  {EU AI Act\\Art.~10 data gov.\\(2 Aug 2027)};
\node[milestone] at (4.00,-3) {};   \node[mlabelabove] at (4.00,-3) {3GPP R20\\FL-as-a-service};

\node[milestone] at (1.00,-4) {};   \node[mlabelabove] at (1.00,-4) {3GPP R19 freeze\\(NWDAF Phase 3)};
\node[milestone] at (0.15,-4) {};   \node[mlabelbelow] at (0.15,-4) {NIST FIPS\\203/204/205\\(Aug 2024)};
\node[milestone] at (3.50,-4) {};   \node[mlabelabove] at (3.50,-4) {3GPP R20\\(synth-CDR study)};
\node[milestone] at (4.85,-4) {};   \node[mlabelabove, xshift=-4pt] at (4.85,-4)
  {CNSA 2.0\\software-signing\\target (2030)};

\node[milestone] at (1.25,-5) {};   \node[mlabelabove] at (1.25,-5) {3GPP R19\\TR 33.876\\(harm metrics)};
\node[milestone] at (3.00,-5) {};   \node[mlabelbelow] at (3.00,-5) {CPS-harm\\metrics study};
\node[milestone] at (4.50,-5) {};   \node[mlabelabove] at (4.50,-5) {Shared corpus\\target};

\node[anchor=west, font=\scriptsize] at (-0.05,-6.30)
  {\tikz{\fill (0,0) circle (1.5pt);}~standards milestone \quad
   \tikz{\fill[red!85!black] (0,0) circle (1.5pt);}~regulatory deadline \quad
   \textcolor{red!75!black}{-\,-}~EU AI Act \quad
   \textcolor{blue!70!black}{-\,-}~3GPP Rel-21};
\end{tikzpicture}
\caption{C1--C5 research challenges mapped onto the 2025--2030 standardisation roadmap. C1 covers sub-millisecond edge detection; C2 XAI/LLM auditability under the EU AI Act; C3 cross-operator federated learning; C4 standardisation gaps across 3GPP, O-RAN, NIST and ETSI; C5 6G-native benchmarks and harm-aware evaluation. Two regulatory checkpoints frame the timeline: EU AI Act general application (2~Aug~2026, red dashed line) and the expected 3GPP Rel-21 freeze (2028, dates TBD, blue dashed line); high-risk obligations under EU AI Act Art.~10/13 trigger on 2~Aug~2027 (red markers). 
Black markers denote standards milestones.}
\label{fig:c1_c5_timeline}
\vspace{-10pt}
\end{figure*}
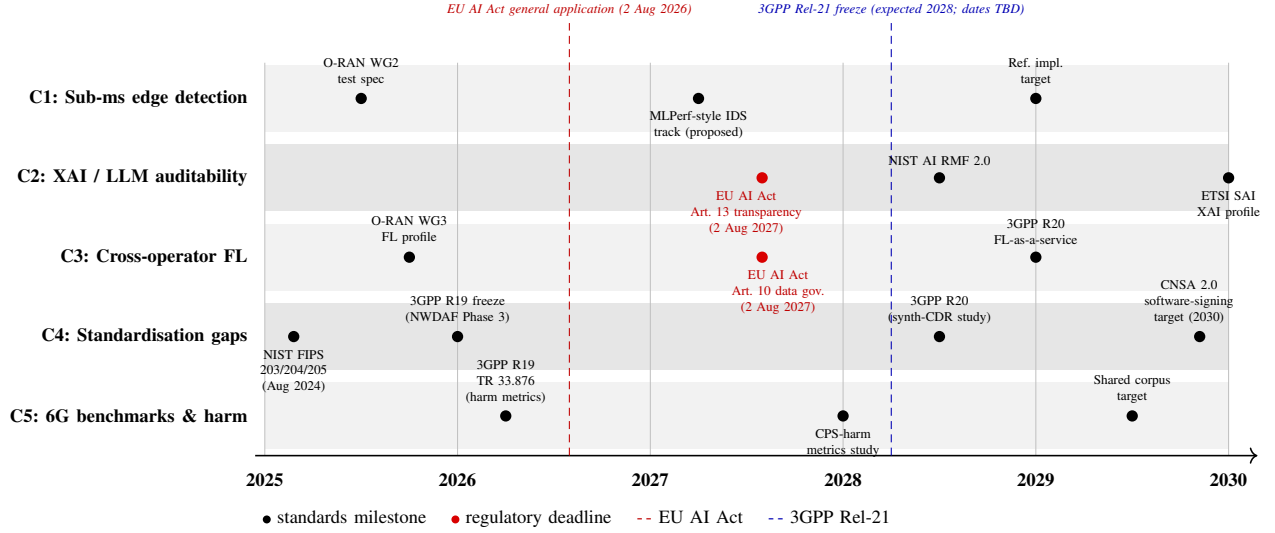

\section{Open Challenges and Research Roadmap}
\label{sec:open_challenges}
We compress the scattered sub-problems of prior surveys into five consolidated challenges. 
Each challenge is intentionally cross-section: the goal is to point researchers at the joint problem rather than at isolated sub-problems whose fixes do not compose into a working closed loop. 
The urgency of these challenges is driven by external regulatory and standardisation deadlines, as mapped in the 2025--2030 timeline of Fig.~\ref{fig:c1_c5_timeline}, which requires closed-loop security solutions to mature ahead of the EU AI Act enforcement window and the 3GPP Release 21 freeze.

\subsection{C1: Sub-Millisecond, Edge-Compatible Detection Under Adversarial ML}
\label{sec:c1}
The composite challenge is to build detectors that simultaneously (i)~run within a $\sim$1\,ms URLLC budget on commodity MEC hardware~\cite{wang2023survey_6g}, (ii)~remain accurate under model drift induced by non-stationary 6G traffic distributions, and (iii)~resist evasion and poisoning attacks (Sec.~\ref{sec:edge_synthesis}). 
The current evidence base treats these requirements separately---quantised CNNs vs.\ adversarial-robust models~\cite{zhang2022adversarial} vs.\ drift-aware models~\cite{yang_2025_zerotouchsec}---but operational deployments need all three at once. 
The research direction is co-designed compression+robust-training+drift-detection pipelines, evaluated on 6G-native datasets (Challenge~C5).

Concretely, three sub-problems sit inside C1. 
\emph{Latency-bounded compression}: int8 quantisation, structured pruning, and distillation must be jointly tuned so that the worst-case (not just average) inference latency fits the slice budget; tail-latency excursions caused by MEC workload contention must be bounded with admission-control on the security model itself. 
\emph{Online drift detection and re-fit}~\cite{yang_2025_zerotouchsec}: the detector must distinguish a true distribution shift (e.g.~a new device class entering the network) from an attack (e.g.~an evasion campaign), and trigger FL retraining selectively without flooding the aggregator. 
\emph{Robust training and certified defences}: adversarial training against realistic 5G/6G perturbation budgets~\cite{salmi2026xai, Alauthman2026GANIDS}, plus randomised smoothing or gradient masking where appropriate; the operational target is to certify a minimum recall on a defined adversarial budget rather than to chase headline accuracy on clean data.

A promising research direction is the joint optimisation of these three sub-problems: a single training procedure that produces a model whose certified robustness, drift sensitivity, and worst-case inference latency are all bounded simultaneously. 
Current evidence is fragmentary: Doriguzzi-Corin \textit{et al.}~\cite{doriguzzi2020lucid} optimise for latency without explicit adversarial-robustness analysis; the GAN-based-evasion-and-defence literature surveyed by Alauthman \textit{et al.}~\cite{Alauthman2026GANIDS} reports robustness gains against GAN-generated evasion (IDS-GAN, attackGAN, SGAN-IDS) without consistent inference-latency budgets; few works address all three together.
A second sub-problem is the choice of \emph{drift signal}: shadow-stream score divergence, KL-divergence on the input feature distribution, or per-class reconstruction-error percentiles each have different sensitivity profiles, and the right combination depends on the operator's tolerance for false drift alarms. 
A third is \emph{certified defences in production}: randomised smoothing inflates inference latency by an order of magnitude in its naive form, and operator deployments need either per-decision certificates with bounded compute or a principled fall-back to non-certified inference when the latency budget is tight. 
The composition of these sub-problems remains an open research direction; an end-to-end MEC-deployable detector with certified robustness, drift-aware online retraining, and worst-case sub-millisecond latency does not yet exist in the surveyed literature.

\textbf{Falsifiable target (2028).} A MEC-deployable detector that demonstrates $\ge 95\%$ recall on a community-agreed adversarial perturbation budget at $p_{99}$ inference latency $\le 1$\,ms on commodity O-RAN-class hardware, with the drift-detection trigger evaluated on at least two independent operator telemetry streams. 
\textbf{Vehicle.} An MLPerf-Inference-style benchmark consortium~\cite{reddi2020benchmark} coordinating with the O-RAN Software Community pre-standardisation testbed track, with a target reference implementation released by 2028.

\subsection{C2: Explainable, Audit-Ready AI Decisions with Bounded LLM Risk} \label{sec:c2}
Black-box detectors and black-box LLM SOC assistants are individually risky; composed in a closed loop they are operationally untenable. 
The research direction is twofold: (a)~XAI methods whose attributions are stable under small perturbations (counter to~\cite{salmi2026xai}), and (b)~LLM SOC interfaces grounded in the operator's own telemetry store with hard refusal envelopes when telemetry support is missing~\cite{bilal2026netops}, \cite[Ch.~10]{sarker2024ai}.

The regulatory horizon makes this urgent. 
EU AI-Act-style obligations on high-risk AI~\cite{AIIndex2026} will reach safety-critical CPS slices well before the 6G commercial launch window, and operators will be required to produce per-decision explanations that survive audit. 
The composition problem---a chain of XAI-explained detector $\rightarrow$ XAI-explained orchestrator $\rightarrow$ LLM-summarised incident report---is poorly studied today. 
The research direction is end-to-end auditability: every loop transition emits an attestable explanation that is (i)~stable under bounded perturbation, (ii)~access-controlled per Salmi and Bogucka's dual-use warning~\cite{salmi2026xai}, and (iii)~retained for the full forensic horizon under PQC integrity protection.

A concrete protocol sketch for audit-ready closed loops would standardise (i)~an explanation envelope format that carries the per-feature attribution, the model version, the input snapshot, and a cryptographic commitment to the model's parameters at decision time; (ii)~a chain-of-custody record that links the detector's explanation to the orchestrator's chosen action and to any LLM-generated summary, with a verifiable transcript of the orchestrator's cost computation (Eq.~\ref{eq:orch_cost}); and (iii)~a regulator-facing query interface that allows post-hoc verification of any specific decision against the operator's retained explanation envelope, building on ETSI's Security Monitoring and Management (SMM) information model~\cite{etsi2026_mec062}.
Operationally, the explanation envelope is small enough to be produced per decision at the MEC tier (a few hundred bytes for a typical attribution vector) and the chain-of-custody record can be batched and signed at session boundaries to amortise PQC signature cost. 
The standardisation lives in 3GPP SA3 and ETSI ISG SAI; closure on the envelope schema and the audit query interface is the highest-leverage step for converting the AI-Act-compatible aspiration into deployable infrastructure.

\textbf{Falsifiable target (2028).} Every closed-loop transition (detect $\to$ orchestrator $\to$ LLM summary) emits an attestable explanation receipt of bounded size whose stability under bounded input perturbation is empirically validated on a public CPS slice testbed, with hard refusal envelopes for the LLM SOC interface. \textbf{Vehicle.} 3GPP SA3 plus ETSI ISG SAI, with a published Technical Report on AI-native closed-loop attestation alignment by 2028 and a regulator-facing query interface profiled against EU AI-Act obligations.

\subsection{C3: FL Across Competing Operators with Privacy and Byzantine Guarantees}
\label{sec:c3}
The labelled-CDR scarcity problem (Sec.~\ref{sec:fl}) can only be solved at scale by federating across operators, but operators are competitors and CDRs are personally identifiable information (PII) under most jurisdictions. 
The research direction is FL protocols that combine (i)~differential privacy with auditable budgets, (ii)~Byzantine-robust aggregation that tolerates compromised or strategically-misreporting operators, and (iii)~cross-border legal compliance baked into the protocol rather than bolted on after the fact~\cite{kairouz2021advances_fl}. 
Surveys~\cite{mothukuri2021survey_fl, alwis2026, blika2024fl_6g, wan2024survey} agree on the gap; concrete protocols remain scarce.


The technical primitives are individually mature; the open research direction is their composition within a single protocol that is jointly analysed for convergence, privacy budget, and Byzantine tolerance, and that admits a reproducible incentive structure. No surveyed deployment has integrated all four properties end-to-end.

\textbf{Falsifiable target (2027).} A single FL protocol jointly analysed for $\varepsilon$-differential-privacy budget ($\varepsilon \le 2$), Byzantine tolerance against up to $\lfloor (n-1)/3 \rfloor$ colluding operators in an $n$-operator federation, and incentive compatibility, with reproducible code released and convergence demonstrated on a cross-operator CDR benchmark. \textbf{Vehicle.} IETF NMRG together with the O-RAN Security Working Group, with a target Internet-Draft and a multi-operator pilot by 2027.

\subsection{C4: API, Lifecycle, and Conformance Standardisation for AI-Native Security in 3GPP and O-RAN}
\label{sec:c4}
Today there is no 3GPP-standardised CDR-based security API exposing the features Sec.~\ref{sec:cdr} relies on~\cite{singh2025towards6gevolution}; the NWDAF event-exposure surface is closer but still ML-agnostic. 
Similarly, O-RAN xApp lifecycle management does not yet specify how AI security xApps are tested, certified, and rolled back, nor how runtime policy conflicts among independently developed xApps are detected before they corrupt KPIs~\cite{polese2023oran, oran_alliance_spec, rahmani2026gnn, huang2025_ailcm_ran}.
Heterogeneous MEC vendor stacks add a third dimension. 
The research direction is concrete standards proposals---specifying the API, the lifecycle, the conformance tests---that operators and regulators can adopt.

Four sub-items have the highest leverage. 
\emph{NWDAF security extension}: an event-exposure profile dedicated to anomaly/attack signals with a stable schema across vendors. 
\emph{xApp security lifecycle}: signing, attestation, capability declaration, and graceful rollback procedures specified by the O-RAN Alliance and aligned with 3GPP SA3~\cite{3gpp_sa3_security, huang2025_ailcm_ran}. 
\emph{ETSI MEC security profile}: standardised exposure of MEC-host telemetry to security applications with role-based access control~\cite{etsi_mec, etsi2026_mec062}. 
\emph{Cross-domain orchestration}: an SDN/NFV/O-RAN orchestrator interface that lets a single playbook reach across actuator tiers without bespoke integration~\cite{etsi2026nfv}. 
Surveys of O-RAN security~\cite{abdalla2022oran_security, polese2023oran} all converge on the lifecycle gap as the single highest-leverage standardisation target.

The research direction therefore decomposes into a standards-track contribution and an academic-experimental contribution. 
The standards-track contribution is to draft and shepherd the four profile specifications above through 3GPP SA3, the O-RAN Security Working Group, and ETSI MEC, with reference implementations contributed to open-source code bases that operators can deploy immediately. 
The academic-experimental contribution is to evaluate proposed profiles on multi-vendor testbeds before standardisation closes, exposing integration gaps that paper-based design reviews cannot reveal; the over-the-air O-RAN testbed of~\cite{moore2025} and the multi-layer mitigation architecture of~\cite{allaw2025cross_layer} are early models for this kind of pre-standardisation experimental validation. 
Without coupled standards-and-experiment contributions, AI-native 6G CPS security risks the same vendor-lock-in pathology that 5G's NSA-to-SA migration encountered, where each operator's deployment was bespoke and inter-operator FL or cross-operator mitigation was practically infeasible~\cite{bonati2020oran_survey}.

\textbf{Falsifiable target (Rel-20/2028).} A 3GPP SA3 Technical Report on AI-native security closed-loop attestation, an O-RAN Alliance xApp-security-lifecycle specification (signing, attestation, capability declaration, rollback), and an ETSI MEC security profile with role-based access control, each accompanied by an open-source reference implementation. \textbf{Vehicle.} 3GPP SA3 and O-RAN Security Working Group (WG11), targeted at Rel-20 with conformance tests deliverable by 2028.

\subsection{C5: 6G-Native Benchmarks, CPS-Specific Harm Metrics, and Realistic CDR Datasets}
\label{sec:c5}
The benchmark gap is the binding constraint on credible evaluation. 
Public datasets (Sec.~\ref{sec:datasets}) are pre-6G, mostly non-CDR, mostly without slice-aware labels, and entirely without CPS-specific harm labels (e.g.,~``did this attack delay a defibrillation?''); the closest 6G-native benchmark to date evaluates LLM reasoning over standardisation-derived tasks rather than CDR/RAN telemetry~\cite{ferrag2026}.
Energy-efficient inference benchmarks are also absent. 
The research direction is operator-academic consortia that release anonymised slice-aware CDR + RAN telemetry with both attack labels and CPS-impact labels, plus reference harm metrics that go beyond accuracy/F1.

\emph{Regulatory and commercial obstacles bind the path.} CDR/RAN telemetry carries personally identifiable information regulated by GDPR, CCPA, and analogues~\cite{etsi2026ISAC, AIIndex2026}, yet the de-identification needed for lawful release destroys the cross-modal correlations detectors must learn; per-cell load, device-mix, and subscriber-density signals are also competitively sensitive. 
The path forward composes three non-novel elements into a working pipeline: (i)~differential-privacy release with auditable budgets; (ii)~synthetic-but-realistic generation~\cite{Stoian2026, duan2025, kotelnikov2023tabddpm} conditioned on real distributions; and (iii)~multi-stakeholder governance (operator, vendor, academic, and regulator) so the benchmark is credible to all parties.

Four data-set properties are needed but absent in the public corpus. 
\emph{Slice-awareness}: labels per slice and per CPS class, so detectors can be evaluated on the slice their target CPS lives on.
\emph{Native multi-modality}: CDR + RAN E2 KPM + NWDAF events captured on the same incidents, so cross-modality fusion methods can be evaluated end-to-end (the recent NetsLab-5GORAN-IDD release~\cite{Zadeh2025data} partially fulfils this by synchronising packet-level and E2-radio-telemetry capture on a live OAI O-RAN testbed, but lacks NWDAF and CDR streams; the NANCY O-RAN release~\cite{nancy_oran_2024} is a further proxy). 

The remaining two properties concern evaluation rather than capture.
\emph{CPS-impact labels}: per-incident annotations mapping each vertical's harm to the loop metrics of Table~\ref{tab:mitigation_metrics}---V2X \emph{missed-hazard-warning} (warning lost once $t_d{+}t_m$ exceeds the braking budget or an FBR event drops the message, with $t_{\mathrm{rec}}$ setting how long peers stay unwarned after a false block), smart-grid \emph{relay trip-delay / false-trip} ($t_d{+}t_m$ late against the fault-propagation budget, a spurious FBR trip, or $t_{\mathrm{rec}}$ before the relay path is restored), and telesurgery \emph{control/haptic delay} ($t_d{+}t_m$ on the $\sim$10\,ms loop, an FBR drop losing a haptic control sample, with $t_{\mathrm{rec}}$ bounding session interruption)---produced by operator--CPS-vendor partnerships or DT replay (Sec.~\ref{sec:dt}).
\emph{Energy and tail-latency benchmarks}: per-method energy-per-decision and 99th-percentile inference latency, so MEC deployability can be assessed before deployment; a 5G-DDoS-specific reproducibility checklist covering methodology, datasets, code/tools, experimental setup, implementation details, result reproducibility, and limitations has been recently proposed and can be adopted as a baseline~\cite{hoque2025ddos5g}.

\textbf{Falsifiable target (2028).} An operator-academic-vendor consortium release of a slice-aware CDR + RAN + NWDAF dataset with both attack labels and CPS-impact labels (delay-to-defibrillation, missed-V2X-warning, unscheduled-relay-trip), distributed under auditable differential privacy ($\varepsilon \le 2$, with the convergence-vs-budget trade-off and optimal-round-count $T^*$ characterised analytically by Noising before Aggregation Federated Learning (NbAFL)~\cite{wei2020differential_fl}), accompanied by per-method energy-per-decision and $p_{99}$ inference-latency baselines.
\textbf{Vehicle.} O-RAN Software Community plus an MLPerf-Inference-style benchmark consortium~\cite{reddi2020benchmark} plus 3GPP SA3, with the first release targeted at 2028 and DT-replay-generated harm labels as the bridge until operator-CPS-vendor partnerships mature.

\medskip

\noindent The five challenges form a closed agenda: C1 makes detection viable, C2 makes the loop trustworthy, C3 makes training feasible across operators, C4 makes deployment standards-conformant, and C5 makes evaluation honest.

\section{Conclusion}
\label{sec:conclusion}
We have argued---across 128 peer-reviewed studies organised under a PRISMA~2020 protocol---that AI-native security for 6G CPSs must be reframed as a closed-loop pipeline that senses on CDR/O-RAN telemetry, detects locally at the MEC tier, mitigates network-wide through SDN/NFV/O-RAN, and learns continuously through FL and DT replay. This reframing eliminates the historical split between ``edge anomaly detection'' and ``DDoS classification'' surveys (Sec.~\ref{sec:edge_detection}), recasts ``emerging paradigms'' as cross-cutting enablers serving one loop (Sec.~\ref{sec:enablers}), and converts scattered open problems into five composable challenges (Sec.~\ref{sec:open_challenges}).


The evidence base is now mature enough to support the feasibility of the loop: MEC-deployed CNN detectors reach $>$98\% accuracy~\cite{hussain2019mec, doriguzzi2020lucid} with demonstrated feasibility on resource-constrained edge hardware~\cite{doriguzzi2020lucid}, 
O-RAN xApp loops mitigate at sub-second latency on over-the-air testbeds~\cite{moore2025}, and 
federated training across heterogeneous edges matches centralised IDS baselines~\cite{chen2020fl_ids}. 
The architecture meets the 1\,ms URLLC NAS--MAC contract end-to-end at $p_{99}$, with Dilithium-3 (ML-DSA-65 per NIST FIPS~204~\cite{fips204_dilithium}) verify ($0.1$\,ms on AVX2), ZTA policy lookup ($0.3$\,ms), xApp inference ($0.5$\,ms), transport ($0.05$\,ms), and jitter margin ($0.05$\,ms) composing to the $1$\,ms slice envelope (Fig.~\ref{fig:latency_budget_envelope})---the principal feasibility claim of this survey. What remains is the standardisation, the 6G-native benchmark, and the cross-operator FL governance needed to run the loop safely in production.

Three implications follow. First, effort should tilt from detection, now relatively mature, to mitigation, where the reversibility gap, the cross-tier orchestrator, and audit-grade explanation of autonomous actions remain under-studied. Second, the benchmark gap (Challenge~C5) is more a governance and incentive problem than a technical one, so operator--academic--vendor consortia for CPS-impact-labelled releases deserve treatment as first-class research contributions. Third, the standardisation agenda (Challenge~C4) is concrete enough that academic drafts, reference implementations, and pre-standardisation testbeds can run in parallel with 3GPP and O-RAN Alliance timelines.


Several limitations bound these conclusions. 6G-native security data does not yet exist, so the loop is assembled from 5G, O-RAN, and CDR proxies (5G-NIDD~\cite{siriwardhana2025data}, NANCY~\cite{nancy_oran_2024}, NetsLab-5GORAN-IDD~\cite{Zadeh2025data}) rather than a single native benchmark, and the findings of Sec.~\ref{sec:edge_benchmark} inherit that gap; the only operator-grade CDR set carries no labelled attacks, leaving CDR-driven results reliant on injected or synthetic anomalies~\cite{barlacchi2015multi, franco2026}; CDR is a minute-scale signal that cannot alone drive a sub-millisecond reaction, serving as first-level triage composed with sub-ms RAN/E2 telemetry (Sec.~\ref{sec:cdr}); and several enablers are surveyed as design directions, with cross-operator FL still chiefly a governance problem (Challenge~C3), CPS-impact harm labels obtainable today only by DT replay (Sec.~\ref{sec:dt}), and LLM use confined to the explain-and-query role (Sec.~\ref{sec:llm}). These limitations scope, rather than overturn, the feasibility claim.


Two questions are left deliberately open. The first is sociotechnical: how to credibly share attack intelligence, model updates, and mitigation telemetry across operators who remain commercial competitors under heterogeneous legal regimes---the primitives of Challenge~C3 are necessary but not sufficient, and the governance architecture binding them is itself a research agenda. The second is how the closed loop interacts with AI-Act-style regulatory oversight~\cite{AIIndex2026}, particularly the auditability and explanation-stability constraints of Challenge~C2; the loop is a precondition for audit-ready decision records, but the legal-operational composition that turns them into regulator-facing evidence lies outside a technical survey. Both are best addressed by interdisciplinary teams, for which we hope this survey supplies a shared architectural vocabulary.

\bibliographystyle{IEEEtran}
\bibliography{references}

\end{document}